\documentclass[a4paper,11pt]{article}
\pdfoutput=1 

\usepackage{jcappub} 

\usepackage[T1]{fontenc} 

\usepackage{listings}

\usepackage{graphicx}	
\usepackage{amsmath}	
\usepackage{mathtools}
\usepackage{xspace}
\usepackage{graphicx}
\usepackage{graphbox}
\usepackage{bbm}
\usepackage{soul}
\usepackage{float}
\usepackage{booktabs}
\usepackage{tabularx}
\usepackage{multirow}
\usepackage{braket}
\usepackage{threeparttable} 
\usepackage{xcolor}
\usepackage{rotating} 
\usepackage{adjustbox}
\usepackage{pdflscape}

\newcommand{\modot}{\ensuremath{\mathrm{M_\odot}}\xspace}
\newcommand{\kpc}{\ensuremath{\mathrm{kpc}}\xspace}
\newcommand{\kms}{\ensuremath{\mathrm{km / s}}\xspace}
\newcommand{\rsat}{\ensuremath{r_\mathrm{sat}}\xspace}
\newcommand{\msat}{\ensuremath{M_\mathrm{sat}}\xspace}
\newcommand{\vparallel}{\ensuremath{w_\parallel}\xspace}
\newcommand{\vperp}{\ensuremath{w_\perp}\xspace}
\newcommand{\vtotal}{\ensuremath{w_\mathrm{tot}}\xspace}
\newcommand{\asymerror}[3]{\ensuremath{#1^{+#2}_{-#3}}\xspace}

\title{Forecasting Dark Matter Subhalo Constraints from Stellar Streams using Implicit Likelihood Inference}

\author[a,b,*]{Tri Nguyen,\note[*]{Corresponding author.}}
\author[c]{Rutong Pei}
\author[d]{Zhuofu Li}
\author[d]{Nora Shipp}
\author[e,f,g]{Scott Dodelson}
\author[h]{Denis Erkal}
\author[d]{Peter S.~Ferguson}
\author[a,b,j]{Tjitske K.~Starkenburg}
\author[k]{Markus M.~Rau}
\author[l]{Alexander H.~Riley}
\author[f,i]{Alan Junzhe Zhou}
\author[]{the LSST Dark Energy Science Collaboration}

\affiliation[a]{Center for Interdisciplinary Exploration and Research in Astrophysics, Northwestern University, 1800 Sherman Ave, Evanston, IL 60201}
\affiliation[b]{NSF-Simons AI Institute for the Sky, 172 E. Chestnut St., Chicago, IL 60611, USA}
\affiliation[c]{Department of Physics, Carnegie Mellon University, Wean Hall, 5000 Forbes Ave, Pittsburgh, PA 15213, USA}
\affiliation[d]{Department of Astronomy and DiRAC Institute, University of Washington, Seattle, WA 98195, USA}
\affiliation[e]{Department of Astronomy and Astrophysics, University of Chicago, Chicago, IL 60637}
\affiliation[f]{Kavli Institute for Cosmological Physics, University of Chicago, Chicago, IL 60637, USA}
\affiliation[g]{Fermi National Accelerator Laboratory, P.O. Box 500, Batavia, IL 60510, USA}
\affiliation[h]{Department of Physics, University of Surrey, Guildford GU2 7XH, UK}
\affiliation[i]{Department of Physics, University of Chicago, Chicago, IL 60637, USA}
\affiliation[j]{Department of Physics \& Astronomy, Northwestern University, 2145 Sheridan Road, Tech F165, Evanston, IL 60208, USA}
\affiliation[k]{School of Mathematics, Statistics and Physics, Newcastle University, Newcastle upon Tyne, NE1 7RU, United Kingdom}
\affiliation[l]{Lund Observatory, Division of Astrophysics, Department of Physics, Lund University, SE-221 00 Lund, Sweden}

\emailAdd{trivtnguyen@northwestern.edu}
\emailAdd{rpei@andrew.cmu.edu}
\emailAdd{zhuofu@uw.edu}
\emailAdd{nshipp@uw.edu}
\emailAdd{dodelson@fnal.gov}
\emailAdd{d.erkal@surrey.ac.uk}
\emailAdd{pferguso@uw.edu}
\emailAdd{tjitske.starkenburg@northwestern.edu}
\emailAdd{Markus.Rau@newcastle.ac.uk}
\emailAdd{alexander.riley@fysik.lu.se}
\emailAdd{ajzhou@uchicago.edu}


\abstract{
The evidence for dark matter (DM) remains compelling, although attempts to understand its particle nature remain inconclusive. 
One promising method to study DM is detecting DM subhalos through their gravitational interactions with stellar streams.
In this study, we apply Neural Posterior Estimation (NPE) to constrain subhalo interaction parameters, including mass, scale radius, velocity, and encounter geometry, from stellar stream kinematics.
We generate particle spray simulations based on the Lagrange Cloud stripping technique, focusing on the ATLAS-Aliqa Uma stream as a test case.
We train multiple NPE models across multiple observational scenarios, quantifying how kinematic completeness affects inference and forecasting constraints from upcoming surveys including LSST, 4MOST, and 10-year Gaia data.
Our results demonstrate that NPE can produce accurate and well-calibrated posteriors.
In the idealized case with full 6D coordinates, we achieve subhalo mass uncertainties of $15-20\%$ for a $10^7 \, \modot$ subhalo, with 5D coordinates (excluding radial velocities) achieving similar performance.
Under realistic observational conditions, mass uncertainties range from $50\%$ (present-day) to $20-40\%$ (future scenarios), with comparable performance between the photometric-only LSST sample and a smaller sample that includes Gaia proper motions and 4MOST radial velocities.
Most notably, we find that velocity bimodality emerges when phase space is poorly sampled, whether due to missing kinematic information or limited stellar tracers.
Combining large photometric samples with targeted spectroscopic follow-up can effectively resolves this degeneracy.
These results demonstrate the power of implicit likelihood inference for optimizing stellar stream observational strategies and forecasting DM subhalo constraints from upcoming surveys.
}

\begin{document}
\maketitle
\flushbottom

\section{Introduction}

Stellar streams are remnants of disrupted globular clusters and dwarf galaxies orbiting galaxies like the Milky Way~\citep{2016ASSL..420.....N}. 
These streams are highly sensitive to the local mass distribution, making them powerful tools for investigating small-scale cold dark matter (DM) substructure~\citep[e.g.,][]{2002MNRAS.332..915I, johnston2002, 2009ApJ...705L.223C, 2016MNRAS.461.1590E}. 
According to the cold DM model, DM halos with masses too low to host stars are expected to exist~\citep{2008Natur.454..735D, 2008MNRAS.391.1685S}, although they cannot be detected directly. 
Their presence, however, can be inferred indirectly through their gravitational interactions with stellar streams~\citep[e.g.,][]{2002MNRAS.332..915I, johnston2002}. 
As subhalos interact with stellar streams, they may cause distinct features in stream structure, such as density gaps and kinks~\citep[e.g.,][]{2009ApJ...705L.223C, 2016MNRAS.461.1590E, Varghese2011}. 
By analyzing the signatures of subhalo perturbation in a stream’s density and velocity, we can constrain the subhalos’ properties such as mass and radius. 

To investigate stellar stream perturbation, we need precise measurements of a large number of stars along the stream. 
This has been made possible by large-scale sky surveys, such as the Dark Energy Survey~\citep[DES;][]{2016MNRAS.460.1270D}, Gaia~\citep{Prusti2016}, and Sloan Digital Sky Survey~\citep[SDSS;][]{2000AJ....120.1579Y}. 
In this work we focus on the ATLAS-Aliqa Uma (AAU) stream, a thin stellar stream from a globular cluster progenitor. 
The ATLAS stream was first discovered in the ATLAS survey~\citep{2014MNRAS.442L..85K}, and Aliqa Uma was detected in DES~\citep{2018ApJ...862..114S}. Despite a discontinuity in the stream tracks on the sky, subsequent kinematic studies found that the two components are in fact part of a single perturbed stellar stream \citep{2021ApJ...911..149L}. 
One possible explanation for this perturbation is a past interaction with a low mass, dark subhalo.

Several techniques have been proposed to infer the properties of DM subhalos from stellar streams. 
One such method is the power spectrum technique~\citep{2017MNRAS.466..628B}, which uses the power spectrum of the density of stars along a stream to measure perturbation-induced density fluctuations and thereby infer the underlying population of subhalos. 
\cite{2021MNRAS.502.2364B, 2021JCAP...10..043B} applied this approach to provide some of the strongest constraints on the particle mass of warm and fuzzy DM using the GD-1~\citep{2006ApJ...643L..17G} and Palomar-5~\citep{2001ApJ...548L.165O} streams. 

Another approach focuses on reconstructing the properties of a single subhalo impact. 
\cite{2015MNRAS.454.3542E} employed Bayesian inference with MCMC to demonstrate that subhalo interaction parameters can be recovered from stellar kinematics, even with realistic observational errors.
Using analytic forward models with the impulse approximation, they uncovered a degeneracy between subhalo mass and velocity, where similar perturbations can be produced by either high-mass or low-velocity subhalos.
\cite{2019ApJ...880...38B} fitted similar models to the GD-1 stream and reported comparable degeneracies.
More recently, \cite{hilmi24} incorporated more realistic simulations, using the Lagrange Cloud stripping technique~\citep{2014MNRAS.445.3788G, 2019MNRAS.487.2685E} (also known as particle-spray simulations), and employed Markov Chain Monte Carlo (MCMC) to infer impact properties for a mock AAU stream.
They reported projected mass constraints of 15\% (with present-day measurement uncertainties) and 10\% (with future uncertainties) for a typical $10^7 \, \modot$ subhalo encounter.

It is increasingly evident that realistic simulations are critical for the study of low mass subhalos with stellar streams. 
Simulations can incorporate detailed physical models necessary for complex systems like stream-subhalo interactions.
However, they often produce intractable likelihood functions that are poorly suited for traditional parameter inference methods.

Implicit Likelihood Inference (ILI; for a review, see \cite{2020PNAS..11730055C}) addresses these limitations by incorporating simulations directly into the inference process, bypassing the need to explicitly construct the likelihood function.
Instead of deriving an analytical likelihood, ILI methods train neural networks to learn the mapping between observed data and model parameters using large datasets of forward simulations, effectively encoding the likelihood implicitly within the network architecture.
Recent years have seen growing interest in applying ILI to astrophysical problems, with applications spanning galaxy spectral energy distribution fitting~\citep[e.g.,][]{2022ApJ...938...11H, 2022MLS&T...3dLT04K}, gravitational wave parameter estimation~\citep[e.g.,][]{2021PhRvL.127x1103D, 2023PhRvL.130q1403D}, strong gravitational lensing~\citep[e.g.,][]{2023ApJ...942...75W, 2024ApJ...975..297W, 2024arXiv241010123E}, cosmology~\citep[e.g.,][]{2022ApJ...933..236Z, 2024NatAs...8.1457H, 2025arXiv250907060S}, and dwarf galaxy studies~\citep[e.g.,][]{2023PhRvD.107d3015N, gnn2}.

Beyond addressing intractable likelihoods, ILI offers additional advantages over traditional Bayesian sampling approaches. 
Standard MCMC can struggle with efficient exploration of complex parameter spaces, particularly when the posterior distribution exhibits multimodality, though more advanced implementations can somewhat alleviate this~\citep[e.g.][]{1992EL.....19..451M, 2016MNRAS.455.1919V}. 
Nested sampling, while robust for evidence estimation, becomes computationally prohibitive in high-dimensional problems.
In contrast, ILI methods can capture arbitrary and high-dimensional posterior shapes at low computational cost once trained.

Recently, \cite{2025ApJ...987...96M} applied neural posterior estimation (NPE), a specific implementation of ILI, to constrain the subhalo interaction parameters of GD-1 using particle-spray simulations. 
Their approach employed graph neural networks and normalizing flows to directly map the particle data to subhalo interaction parameters.
However, their study was limited to two interaction parameters (the subhalo mass and total velocity), and the use of particle-level data implicitly encodes the unreliable density information presented in particle-spray simulations. 

In this work, we develop a comprehensive NPE framework for constraining six subhalo parameters: mass, scale radius, two velocity components, interaction parameter, and impact orientation angle. 
Our method employs transformer-based neural networks with normalizing flows to learn complex posterior distributions from stellar stream kinematics.
We train our model on a large dataset of particle spray simulations based on the AAU stream \citep{2016MNRAS.461.1590E, hilmi24}.
To ensure robust inference, we implement a spatial binning approach that preserves kinematic information while avoiding the unreliable density information in the simulations. 

Beyond parameter inference, ILI is particularly well-suited for observational forecasting.
Forecasts are commonly performed through full inference with MCMC or nested sampling, or using Fisher Information Matrix approach, which compute the parameter covariances by inverting the Fisher matrix. 
The latter provides faster approximation compared to MCMC and nested sampling, making it better suited for scientific forecasting applications~\citep[e.g.][]{2022ApJ...932..102W, 2023arXiv230508994C, 2023MNRAS.524.4711M, 2025arXiv250907060S}, although at a cost of assuming Gaussian likelihoods.
In comparison, ILI can capture arbitrary posterior shapes at a lower computational cost once trained.
More importantly, while ILI and machine learning methods in general can be sensitive to out-of-distribution data, this limitation is less problematic for forecasting than for real data analysis, since forecasting focuses on speed and relative performance trends rather than precise absolute constraints.

In the near future, the Vera C. Rubin Observatory Legacy Survey of Space and Time (LSST; \cite{2019ApJ...873..111I}) is expected to revolutionize the study of stellar streams with its unprecedented volume and depth of observations across half the sky. 
LSST will provide precise positions and proper motions of a large sample of AAU stars, enabling detailed modeling of its interactions with dark matter substructure. 
However, LSST does not provide radial velocity measurements and it will be infeasible to spectroscopically follow up all of the AAU stars that will be observed in LSST. 
In this work, we predict the power of future datasets to constrain DM subhalo perturbers of AAU in two example future observation scenarios. 
First, we measure the constraining power of the full sample of AAU member stars observed at 10 year LSST depths with only 2D on-sky position measurements. 
We then compare these constraints to the sample of AAU member stars that are within the magnitude limits of future spectroscopic and proper motion observations. 
In particular, we use the predicted depth and measurement uncertainties for radial velocities from the Four Metre Multi-Object Spectroscopic Telescope (4MOST; \cite{2019Msngr.175....3D}) and predicted proper motions uncertainties from the full Gaia survey~\citep{Prusti2016}. 

Given these observational capabilities, we leverage our NPE framework to conduct a comprehensive forecast study for future stellar stream surveys. 
Specifically, we assess the constraining power under different combinations of available kinematic information, quantifying how the absence of radial velocities, proper motions, and distances impact the parameter inference compared to complete 6D coordinates.
Additionally, we evaluate the performance across multiple observational scenarios spanning present-day and future survey depths, including LSST photometry, Gaia proper motions, 4MOST spectroscopy, and their combinations. 

This paper is structured as follows.
Section~\ref{section:sim} describes the particle-spray simulations and datasets.
Section~\ref{section:method} presents the NPE framework, including the data preprocessing and binning procedure (\ref{section:preprocess}) and the machine learning setup (\ref{section:ml}). 
Section~\ref{section:observable} examines the impact of different kinematic observables on inference, while Section~\ref{section:uncertainty} compares the performance across present-day and future observational scenarios.
Section~\ref{section:discussion} discusses current limitations and implications for future subhalo-stream studies.
Section~\ref{section:conclusion} summarizes the paper. 

The link to the repository can be found at: \url{https://github.com/trivnguyen/sbi_stream}.

\section{Simulation and Dataset}
\label{section:sim}

We briefly describe the particle spray simulation used to generate the training dataset.
We adopt the same simulation setup from \cite{hilmi24}, which consists of three primary steps: (1) generating the unperturbed stream through backward-forward integration with particle ejection at Lagrange points, (2) sampling the parameters of the subhalo impact, and (3) rewinding and integrating both the progenitor and subhalo forward together to generate the perturbed stream.

Adopting the AAU progenitor properties from \cite{2021ApJ...911..149L}, we model the progenitor of the unperturbed stream as a globular cluster following a Plummer sphere \cite{1911MNRAS..71..460P} with a mass and scale radius of $2 \times 10^4 \, \modot$ and $10 \, \mathrm{pc}$, respectively.
The gravitational potential in which the stream is disrupted includes both the Milky Way and the LMC, as well as the potential of the globular cluster progenitor.
The Milky Way model from \cite{2017MNRAS.465...76M} comprises six components: a bulge, a spherical Navarro-Frenk-White (NFW; \cite{1996ApJ...462..563N}) DM halo, thick and thin disks, and atomic and molecular hydrogen gas disks.
We adopt the Sun's position at $R_0 = 8.23 \, \kpc$ with velocity $(U_{\odot}, V_{\odot}, W_{\odot}) = (8.4, 12.0, 7.3) \, \kms$.
The full potential parameters are given in Table A.3 of \cite{2021ApJ...923..149S}, determined from a specific draw from the posterior chains of \cite{2017MNRAS.465...76M}, following \cite{2021ApJ...911..149L} and \cite{2021ApJ...923..149S}.
The LMC is modeled as a Hernquist profile \citep{1990ApJ...356..359H} with a mass of $1.5 \times 10^{11} \, \modot$ and a scale radius of $17.13 \, \kpc$, consistent with measurements from stellar streams and LMC satellites.

We generate the unperturbed stream using the approach of \cite{2019MNRAS.487.2685E}, which implements the modified Lagrange Cloud stripping technique \citep{2014MNRAS.445.3788G} to include the LMC.
We start with the present-day position and velocity of the progenitor and LMC, then integrate backward for $4 \, \mathrm{Gyr}$ to obtain initial conditions, accounting for the reflex motion of the Milky Way due to the LMC's influence~\citep{2019ApJ...884...51G, 2020MNRAS.498.5574E, 2020MNRAS.494L..11P, 2021NatAs...5..251P, 2021MNRAS.506.2677E}.
During the forward integration, particles are ejected from the progenitor's inner and outer Lagrange points at randomly selected time intervals drawn from a Gaussian distribution centered on each pericenter.
A mean of 2000 particles are ejected every pericenter with velocities drawn from a Gaussian distribution to match the velocity dispersion of the stream.
The perturbed stars at the outer Lagrange point drift behind the progenitor to form the trailing arm, whereas those at the inner Lagrange point drift ahead to form the leading arm, resulting in a stream of approximately 12,000 stars over the 4 Gyr integration.

\begin{figure}
    \centering
    \includegraphics[width=0.7\linewidth]{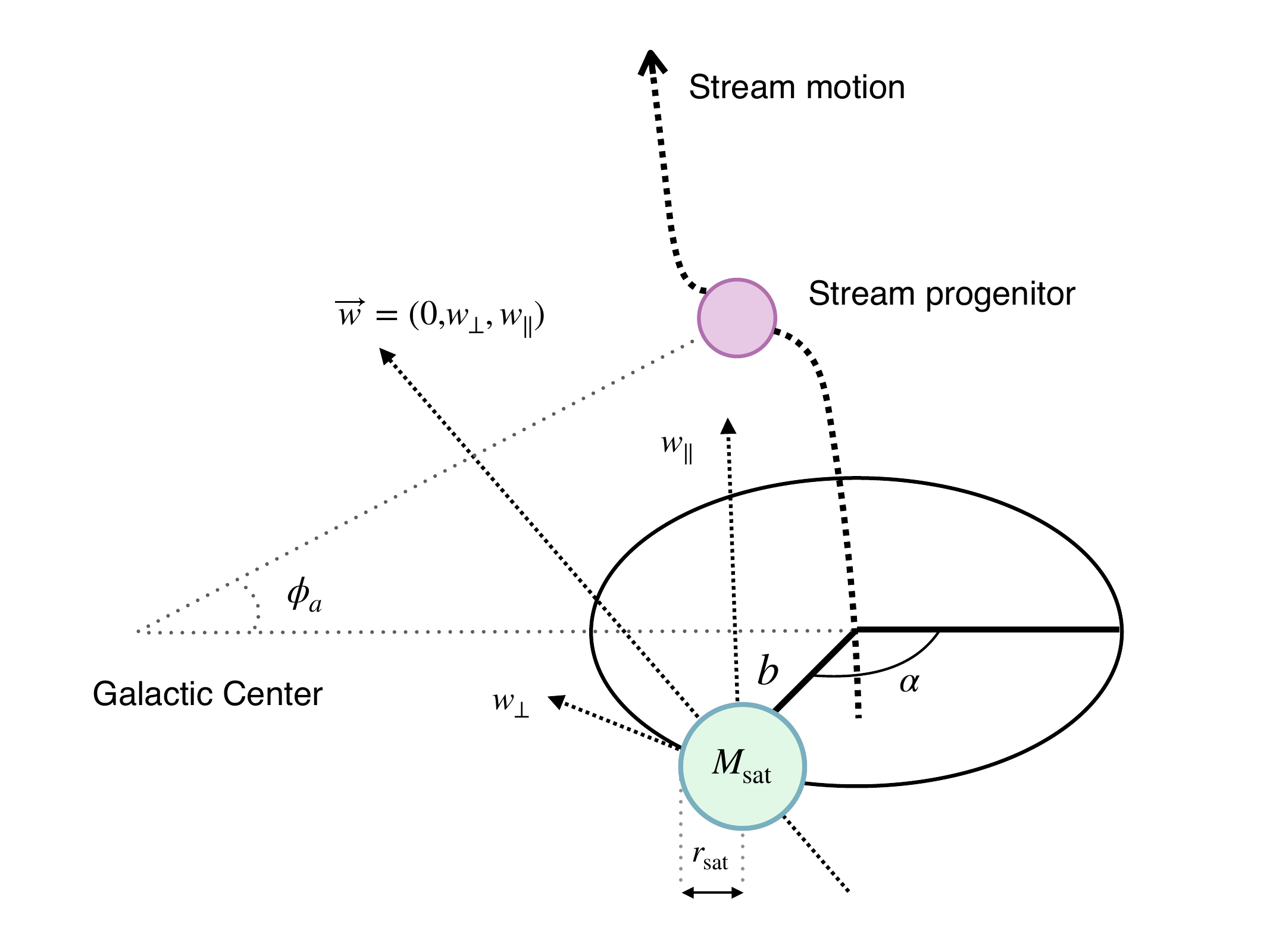}
    \caption{Geometry of the subhalo-stellar stream encounter. The impact time $T_\mathrm{a}$ is not illustrated. Adapted from Figure 1 of \cite{hilmi24}.}
    \label{fig:geometry}
\end{figure}
Next, we model the subhalo impact.
Following \cite{2015MNRAS.454.3542E}, the full encounter is characterized by eight parameters: subhalo mass $M_\mathrm{sat}$\footnote{We adopt the subscript ``sat'' instead of ``sub'' for consistency with \cite{2015MNRAS.454.3542E}. In practice, the stream could encounter either a dark subhalo or luminous satellite.}, subhalo radius $r_\mathrm{sat}$, velocity components perpendicular and parallel to the stream's progenitor ($\vperp, \vparallel$), impact parameter $b$\footnote{Throughout this work, we refer to the impact parameter $b$ as the minimum distance between the subhalo and the stream during the encounter, while interaction parameters $\vec{\theta}$ refer to the parameters of the subhalo-stream encounters.}, orientation angle $\alpha$, impact time $T_\mathrm{a}$, and angle along the stream $\phi_\mathrm{a}$.
Figure~\ref{fig:geometry} shows the diagram of the encounter geometry.  
The subhalo interaction parameters for each stream are drawn from the distributions given in Table~\ref{tab:prior}, which also serves as the prior distribution for our NPE framework.
We consider six of these parameters, excluding the impact time $T_\mathrm{a}$ and angle $\phi_\mathrm{a}$, which are fixed to $T_\mathrm{a} = -0.25 \, \mathrm{Gyr}$ and $\phi_\mathrm{a} = -6 \, \mathrm{deg}$, respectively.
We find that these parameters significantly expand the dimensionality of the parameter space and are among the most challenging to constrain from gravitational perturbations alone, leaving their inclusion for future work.

Finally, we generate the perturbed stream by rewinding both the progenitor and subhalo backward from the present day and then integrating them forward together.
This approach has two key advantages: (1) the subhalo's gravitational influence on the stream is modeled self-consistently throughout the encounter, rather than using an impulse approximation, and (2) the subhalo is guaranteed to be at the correct present-day location regardless of its properties.
The resulting perturbed stream contains the same number of particles as the unperturbed case but with modified phase-space coordinates reflecting the subhalo's gravitational perturbation.

\begin{table*}
\centering
\renewcommand{\arraystretch}{1.25}
\begin{tabular}{|lllll|}
\hline
Parameter & Description & Range & Prior & Mock AAU \\
\hline
\msat & Impact subhalo mass in $10^7 \, \modot$ & $[0.1, 10]$ & Log Uni. & $1.0$ \\
\rsat & Impact subhalo scale radius in \kpc & $[0.1, 1.0]$ & Log Uni. & $0.3$ \\ 
\vperp & Perpendicular velocity in \kms & $[-100, 100]$ & Uni. & $35$ \\ 
\vparallel & Parallel velocity in \kms & $[-100, 100]$ & Uni. & $-10$  \\ 
$b$ & Impact parameter in \kpc & $[0, 3] $& Uni. & $0.1$ \\ 
\multirow{2}{*}{$\alpha$} & Orientation of the subhalo at the & \multirow{2}{*}{$[0, 360]$} & \multirow{2}{*}{Uni.} & \multirow{2}{*}{$250$} \\ 
& point of closest approach in deg & & & \\
\hline 
\multirow{2}{*}{$\phi_\mathrm{a}$} & Angle relative to the progenitor & \multirow{2}{*}{$-6.0$} & \multirow{2}{*}{Fixed} & \multirow{2}{*}{$-6.0$}\\
& at the time of impact in deg & & & \\
$T_\mathrm{a}$ & Impact time in Gyr & $0.25$ & Fixed & $0.25$ \\
\hline
\end{tabular}
\caption{Summary of the simulation parameters, their prior distributions, and the fiducial values used for the mock AAU stream. 
The model includes six free parameters with specified priors and two parameters fixed to the fiducial AAU values for reference.
}
\label{tab:prior}
\end{table*}

We generate $150,000$ streams and partition them $80-20$ into training and validation sets.
Additionally, we generate $8,000$ streams as a hyperparameter tuning dataset for optimizing the NPE architecture and another $12,500$ streams as an independent test set, on which we present results throughout this work.
The subhalo interaction parameters for each stream are drawn from the prior distributions given in Table~\ref{tab:prior}.

\section{Machine Learning Framework}
\label{section:method}

\subsection{Data Preprocessing}
\label{section:preprocess}

For a given set of subhalo interaction parameters $\vec{\theta}$, the simulation generates 12,000 particles per stream.
To match the present observations of AAU from \cite{2021ApJ...911..149L}, we select particles within a stream longitude $\phi_1$ range of $(-20, 12) \, \mathrm{deg}$, resulting in approximately 10,000 particles per stream.
The data preprocessing pipeline consists of two primary steps: (1) incorporating observational uncertainties and (2) spatial binning along the stream track.

We calculate the expected observational uncertainties in the coordinates and velocities of each simulated particle, modeling them as Gaussian distributions.
We generate multiple realizations of each stream, each with a different random noise realization of these uncertainties, effectively augmenting our training data.
Section~\ref{section:observable} explores the idealized scenarios with no uncertainty (i.e. uncertainties set to zero), while Section~\ref{section:uncertainty} discusses the detailed uncertainty calculations for different observational scenarios and compares these different observational contexts.

Next, we divide the selected particles into bins. 
The AAU stream exhibits a distinct kink-like feature that divides it into the ATLAS half~\citep{2014MNRAS.442L..85K} and the Aliqa Uma half~\citep{2018ApJ...862..114S}.
This sharp transition complicates binning along $\phi_1$, as bins near the kink can be highly sensitive to the precise placement of the bin edges.
Additionally, within our considered parameter space, streams often exhibit complex morphology where the relationship between $\phi_1$ and $\phi_2$ is non-monotonic.

To account for more complex and diverse morphologies, one could either bin particles directly in the $(\phi_1, \phi_2)$ space or fit the stream track and bin particles along its path.
We choose the latter approach, as it maintains ordering of adjacent bins along the stream.
Additionally, compared to direct binning in the $(\phi_1, \phi_2)$ space, this method reduces the data dimensionality more effectively and more robustly handles outliers and streams with small tracer counts.

We fit the stream track using a univariate cubic spline and project the $(\phi_1, \phi_2)$ coordinates of each particle onto the track.
To determine the spline knots, we first divide the stream uniformly along $\phi_1$ into 50 bins.
For each bin that contains at least one particle, we place a knot at the bin center, thus allowing the spline to have up to 50 knots depending on the particle distribution along the stream.
The number of knots can thus vary depending on the morphology and number of tracers of the stream, and range from $50$ knots for $10,000$ member stars (Section~\ref{section:observable}) to $10-20$ knots for $96$ member stars (Section~\ref{section:uncertainty}).

We then divide the stream into 50 bins of equal arc length, i.e. each bin spans the same distance when measured along the curve of the spline. 
Since particles are not uniformly distributed along the stream, each bin contains a variable number of particles: denser regions will have more particles per bin while sparser regions will have fewer.
For each bin, we calculate the mean and standard deviation of each observable.
We ignore empty bins, which correspond to density gaps in the stream's spatial distribution.

We do not include stellar density variations along the stream as an observable in our analysis. 
While density perturbations are expected signatures of subhalo interactions~\citep[e.g.][]{johnston2002, 2009ApJ...705L.223C, 2016MNRAS.461.1590E}, the perturbations produced by particle-spray simulations, including ours, are not robust for direct comparison with real observational data.
Though this work primarily trains and tests on mock simulations, we deliberately exclude density perturbations to avoid reporting overly optimistic results that would not generalize to real data applications.
This also necessitates our binning approach, which removes the implicit density variations present in the particle-level data while preserving robust kinematic information.

\begin{figure*}
    \centering
    \includegraphics[width=0.98\linewidth]{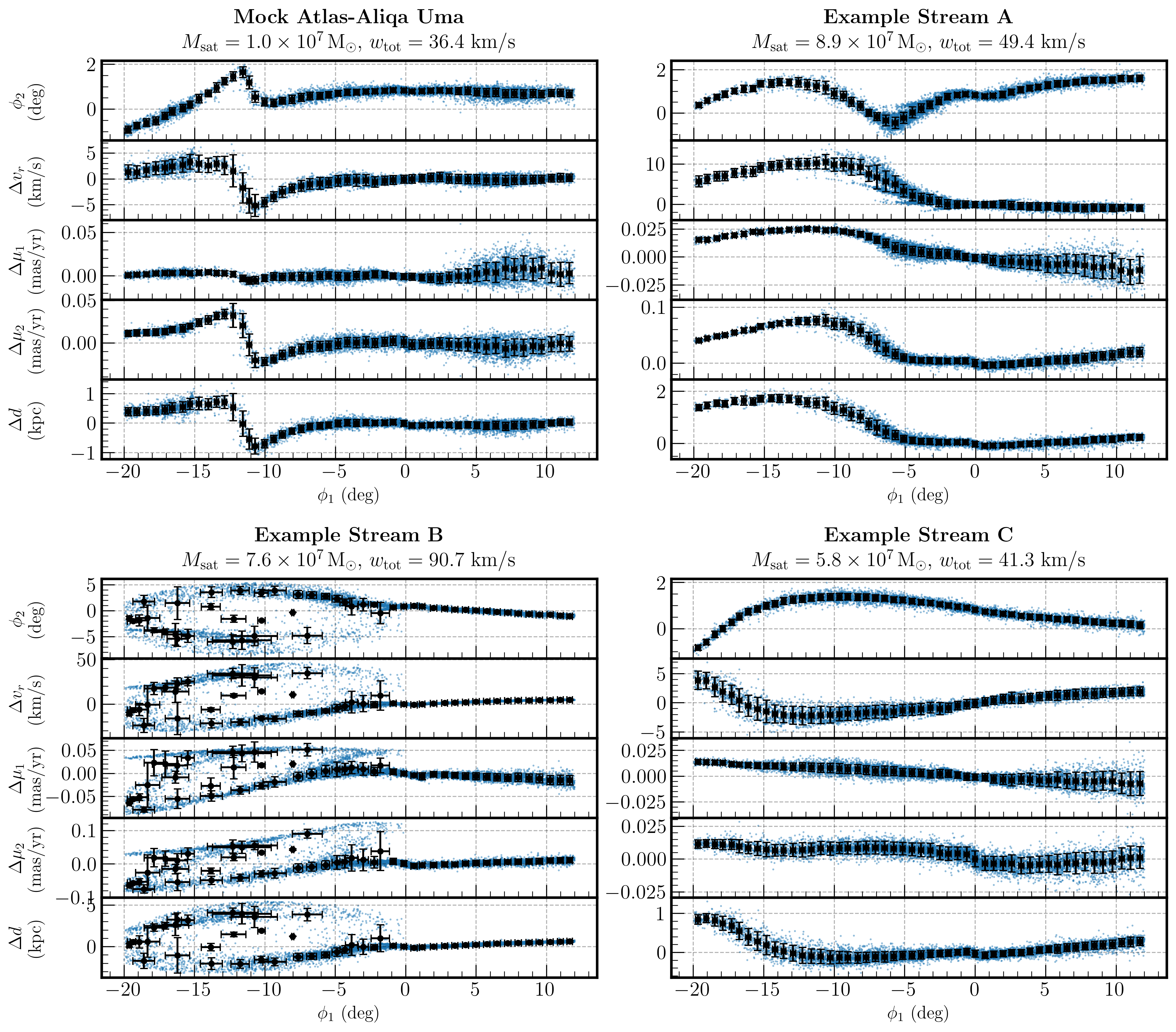}
    \caption{
    Example perturbed streams.
    Blue data points denote the raw particle data output by the simulation, while the black data points with error bars denote the preprocessed data. 
    The top left panel shows the mock AAU stream, while the rest of the panels show example streams from the training dataset.
    In each panel, each row shows a different observable as a function of the stream longitude $\phi_1$: from top to bottom, the stream latitudes $\phi_2$, radial velocities $v_r$, proper motions $\mu_1$ and $\mu_2$, and distances $d$.
    For clarity, for the velocities and distances, the differences relative to an unperturbed stream are shown, where the unperturbed stream baseline is estimated by fitting the coordinates and velocities using a fourth-order polynomial following \cite{hilmi24}.
    }
    \label{fig:example_streams}
\end{figure*}

Figure~\ref{fig:example_streams} plots the coordinates of four streams as a function of $\phi_1$, showing the mock AAU stream and three example perturbed streams from the training dataset. 
The mock AAU stream is generated using the fiducial parameters listed in Table~\ref{tab:prior} to mimic the observed AAU in \cite{2021ApJ...911..149L}.
We emphasize that this represents simulated data designed to match the properties of the real AAU stream, \textit{not actual observational measurements.}
The parameters for the other three streams (and the rest of the training dataset) are drawn from the prior distributions specified in Table~\ref{tab:prior}.

The preprocessed data in Figure~\ref{fig:example_streams} is treated as sequential input for the machine learning model, where the feature vector of each bin consists of the mean and standard deviation of the observables under consideration.
We explore different combinations of observables and present the comparative results in Section~\ref{section:observable}.
Additionally, the normalized arc length coordinate of each bin (scaled to $[0, 1]$) serves as a positional encoding that indicates the location of each bin along the stream sequence.

From Figure~\ref{fig:example_streams}, we see that binning in $\phi_1$ only does not adequately capture the morphology of streams that have undergone major perturbations (high-mass or low velocity encounters), as in Example B. 
However, although our spline fit procedure is generally more robust, it too cannot fully account for all stream morphologies. 
In Example B, some bins fall in low-density regions and deviate from the underlying particle distribution, while other denser regions of the stream are missed entirely by the binning scheme.
Additionally, in Example A, the spline fit does not capture the bifurcation in the $(\phi_1, \phi_2)$ space at approximately $\phi_1 = -5 \, \mathrm{deg}$, and the radial velocity profile shows multiple kinematic branches that are smoothed over during binning.

Through visual inspection of the preprocessed training data, we find that while small differences such as these can occur between the processed data and the particle-level data, these discrepancies typically amount to small deviations rather than systematic misrepresentation of the stream track. 
Importantly, we do not observe degraded performance in parameter recovery (Sections~\ref{section:observable} and \ref{section:uncertainty}) for streams that have undergone major perturbations (where the spline fit is likely least accurate), indicating that our preprocessing procedure remains robust even when the stream track is not fitted perfectly.
Future work could refine the binning process to better handle complex stream morphologies.
Additional discussion of these limitations is provided in Section~\ref{section:discussion}.

\subsection{Neural Posterior Estimation}
\label{section:ml}

\begin{figure*}
    \centering
    \includegraphics[width=\linewidth]{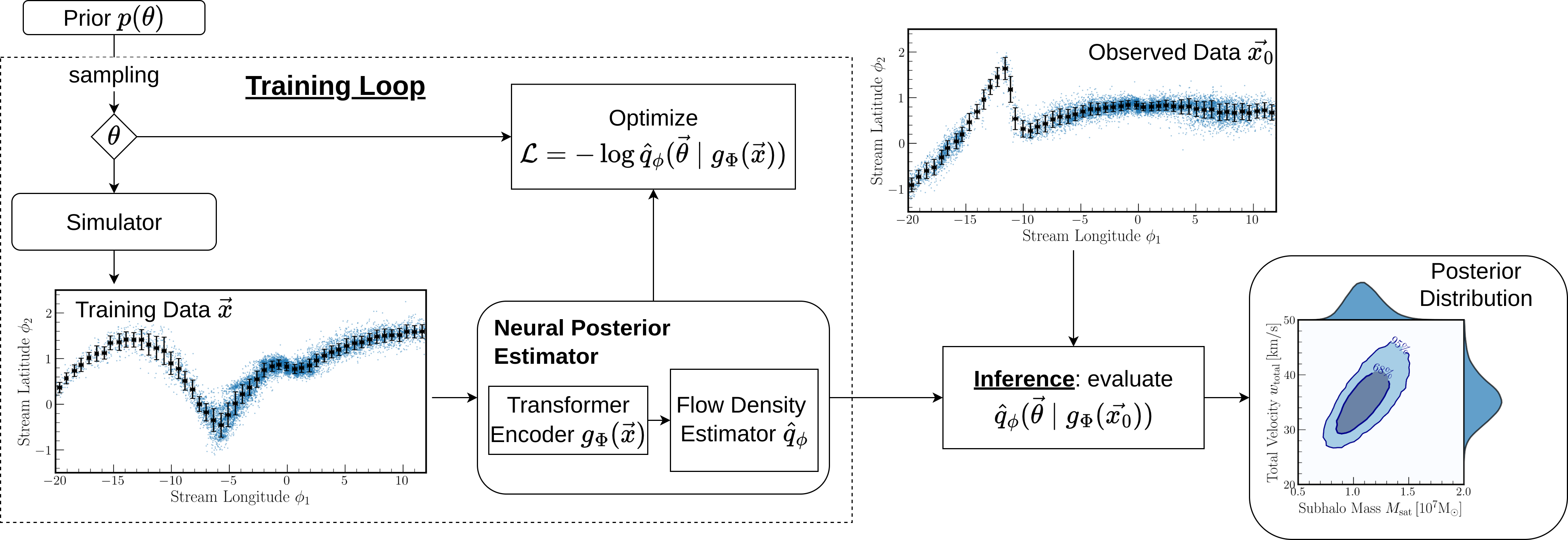}
    \caption{The flowchart of the NPE framework.
    The training data $\vec{x}$ is created by sampling the subhalo interaction parameters $\vec{\theta}$ from a prior distribution and passing them into the simulator, which implicitly encodes the likelihood function $p(\vec{x} \mid \vec{\theta})$.
    Data preprocessing steps are assumed to be incorporated within the simulator for simplicity.
    During training, the NPE model $\hat{q}_\phi(\vec{\theta} \mid \vec{x})$, which consists of a Transformer encoder and a flow density estimator, minimizes the objective $\mathcal{L}_\mathrm{NLL}$ (Equation~\ref{eq:loss_nll}).
    Once trained, during inference, given an observed stream $\vec{x}_0$, the posterior $\hat{q}_\phi(\vec{\theta} \mid \vec{x}_0)$ can be directly sampled without re-training.
    }
    \label{fig:diagram}
\end{figure*}

NPE employs neural networks to learn the direct mapping from observed data to the posterior distributions, bypassing the need to explicitly construct a likelihood function. 
Instead, the likelihood is implicitly encoded within the training simulation. 
During training, the model is exposed to pairs of simulated data and corresponding parameters, allowing it to learn the relationship between them and effectively capture the likelihood.
Figure~\ref{fig:diagram} presents the flowchart of the NPE framework. 

NPE models typically consist of a feature extraction network, which processes the input data into summary features, and a neural density estimator, which maps the summary features to the posterior distribution.
Here, we briefly outline the components of our implementation and provide a more detailed description of the architecture in Appendix~\ref{app:architecture}.

As previously discussed, the binned data is treated as a sequence with variable lengths due to empty bins. 
To handle this, we use an encoder-only transformer architecture~\citep{2017arXiv170603762V} as the feature extraction network. 
The transformer employs a self-attention mechanism to capture relationships between bins and is particularly effective for modeling sequential data.
In contrast, \cite{2025ApJ...987...96M} used a graph neural network (GNN) on particle-level data as the feature extractor.
However, as discussed in Section~\ref{section:preprocess}, using particle-level data would implicitly incorporate the unreliable density information present in particle-spray simulations.

For the density estimator, we use a conditional normalizing flow~\citep{2015arXiv150505770J, 2017arXiv170507057P, 2019arXiv191202762P}.
Normalizing flows map a simple base distribution (e.g., Gaussian) to a complex target distribution by applying a sequence of invertible transformations with tractable Jacobian.
These transformations are typically parametrized by neural networks with learnable parameters, enabling them to learn complex distributions, including multimodal or highly non-Gaussian shapes. 
In our setup, the conditional flow consists of a sequence of Neural Spline Flow transformations~\citep{2019arXiv190604032D}, which use piecewise rational quadratic splines to model the mappings between the base and target distributions. 

During forward pass, the input sequential data is first processed through the transformer, then aggregated into a single feature vector.
This feature vector serves as the condition, also known as the context, for the flow. 
Given an input data $\vec{x}$ with temporal coordinate $\vec{t}$, and parameters $\vec{\theta}$, the flow models the distribution $\hat{q}_\phi(\vec{\theta} \mid g_\Phi(\vec{x},\vec{t}))$, where $g_\Phi(\vec{x}, \vec{t})$ represents the output of the transformer. 
The learnable parameters of the flow and transformer are denoted $\phi$ and $\Phi$, respectively.
As a reminder, $\vec{x}$ denotes the binned coordinates of the stream, $\vec{t}$ denotes the arc lengths along the stream track, and $\vec{\theta}$ denotes the subhalo interaction parameters.

During training, we optimize both the transformer and the flow simultaneously using the negative log-likelihood loss function:
\begin{equation}
    \label{eq:loss_nll}
    \mathcal{L}_\mathrm{NLL} = -\log \hat{q}_\phi(\vec{\theta} \mid g_\Phi(\vec{x},\vec{t})).
\end{equation}
We use the AdamW~\citep{adamw2019, kingma2014adam} gradient descent optimizer with a peak learning rate of $9 \times 10^{-4}$ and a weight decay coefficient of $3 \times 10^{-4}$.
Following standard practice for transformer-based models, we implement a cosine annealing learning rate scheduler~\citep{2016arXiv160803983L} with $35,000$ linear warm up steps and $750,000$ decay steps. 
The training batch size is $256$.

We determine these hyperparameters, along with the model architecture parameters, using Bayesian optimization with \texttt{Optuna}~\citep{2019arXiv190710902A}.
Specifically, we train approximately 200 models with varying hyperparameters on an independent dataset of 8,000 streams (as mentioned in Section~\ref{section:sim}), and select the configuration that achieves the best performance (see Appendix~\ref{app:architecture}).
Given the computational expense of this tuning process, which requires several GPU days, we perform hyperparameter optimization only on the 6D Coordinates dataset (see Section~\ref{section:observable}) and adopt the resulting configuration for all models.
Training on the full dataset requires approximately 8 to 12 hours per model on a single NVIDIA Tesla A100 GPU.

\section{Impact of Observational Completeness}
\label{section:observable}

We first explore NPE model performance in idealized scenarios with no measurement uncertainty (i.e., we set the uncertainty of all observables to zero).
To assess the impact of different observational constraints on subhalo parameter inference, we systematically examine four sets of observables that reflect varying levels of observational completeness:
\begin{itemize}
    \item \textbf{6D Coordinates}: stream coordinates $(\phi_1, \phi_2)$, proper motions $(\mu_1, \mu_2)$, radial velocity $v_r$, distance $d$
    \item \textbf{5D Coordinates}: stream coordinates $(\phi_1, \phi_2)$, proper motions $(\mu_1, \mu_2)$, distance $d$
    \item \textbf{3D Coordinates}: stream coordinates $(\phi_1, \phi_2)$, distance $d$
    \item \textbf{2D Coordinates}: stream coordinates $(\phi_1, \phi_2)$ only
\end{itemize}
As discussed in Section~\ref{section:method}, we do not include density variations as observables since they are difficult to reliably model in particle-spray simulations.

These observable combinations are motivated by the capabilities of current and future astronomical surveys.
Radial velocities are the most resource-intensive to obtain as they require dedicated spectroscopic follow-up observations. 
Proper motions require multi-epoch observations with sufficient temporal baseline.
Distance measurements can be obtained from tracer stellar populations such as RR Lyrae and blue horizontal branch stars.
We note that this is appropriate for AAU, which has sufficient distance tracers. 
For streams that have fewer tracers and require photometric distances, proper motions may be more reliable.

For each observable set, we train a different NPE model, using the same preprocessing procedure and network architecture as in Section~\ref{section:method}.

\subsection{Diagnostic Tests}
\label{section:observable:test}
\begin{figure*}
    \centering
    \includegraphics[width=0.98\linewidth]{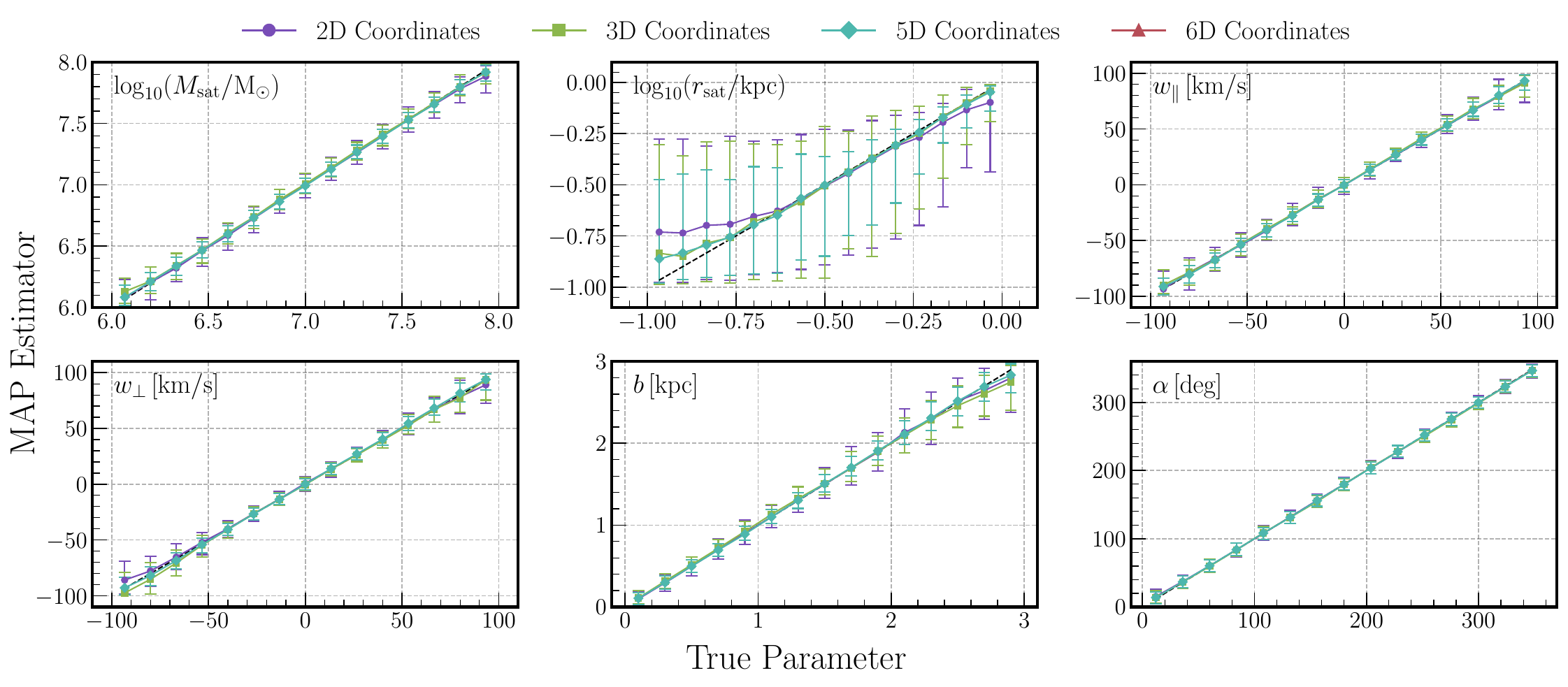}
    \caption{
    Comparison between the MAP estimators and the true parameters for the four observable sets.
    Each panel shows the MAP estimators versus the true parameters for different subhalo parameters. 
    In each panel, streams are grouped into bins based on their true parameters. 
    Data points and error bars show the mean and 68\% confidence interval of the MAP estimators within each bin, with different markers and colors corresponding to different observable sets.
    The black dashed lines show the one-to-one correspondence indicating perfect recovery.
    }
    \label{fig:predict-true-obs}
\end{figure*}

\begin{figure}
    \centering
    \includegraphics[width=0.7\linewidth]{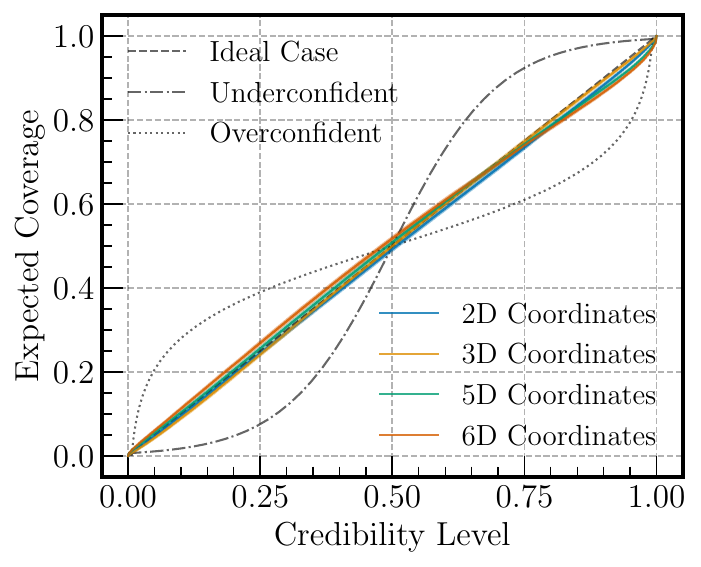}
    \caption{
    Credibility level versus expected coverage from TARP for the four observable sets.
    Each solid line and shaded band show the median and 68\% confidence interval of the TARP result across 100 bootstrap samples of a different observable set (color). 
    Bands are too narrow to be visible at this scale.
    The black dashed diagonal line represents the ideal case where posteriors are perfectly calibrated.
    The black S-shaped curves indicate under-confident and over-confident posteriors, respectively.
    }
    \label{fig:tarp-obs}
\end{figure}

We present diagnostic tests to assess each model's performance across the full test dataset.
We first evaluate the accuracy of parameter recovery by comparing the maximum a posteriori (MAP) estimators to the true subhalo parameters.
To obtain the MAP estimator of each stream, we pass the observed kinematics through the trained model and generate $1000$ posterior samples using the normalizing flow.
We then evaluate the posterior density of each sample using the flow and select the parameter set with the highest flow-estimated posterior probability as the MAP estimator.
This approach identifies the joint maximum in 6D posterior space, accounting for correlations between parameters.

Figure~\ref{fig:predict-true-obs} shows the MAP estimators plotted against the true parameters for each subhalo property.
To visualize trends and assess systematic biases, we bin the streams according to their true parameter values and compute the mean and 68\% confidence interval of the MAP estimators within each bin.

Overall, the framework accurately recovers the true parameters across all observable sets.
Performance is consistent across most parameters, with the notable exception of the subhalo scale radius $\rsat$.
Compared to other parameters, the scatter in the MAP estimates of $\rsat$ is significantly larger relative to its prior range. 
At low $\rsat$ values, all models exhibit systematic biases toward higher values, with this effect being more pronounced when fewer kinematic observables are available.

The difficulty in constraining $\rsat$ likely stems from the fact that sufficiently small subhalos produce gravitational signatures indistinguishable from point masses, regardless of their actual radius, as well as from the fact that the impact parameters is not defined in units of $\rsat$, so a denser satellite will effectively pass by the stream at a larger distance.
Additionally, although the models accurately infer the impact parameter $b$, the scatter increases toward larger values of $b$.
This is consistent with the above observations, as larger impact parameters combined with small subhalo radii approach the point-mass interaction regime.
Lastly, we do not observe significant performance differences in recovering $\rsat$ and $b$ when considering only low-mass ($\msat < 10^{6.5} \, \modot$) versus high-mass ($\msat > 10^{7.5} \, \modot$) halos in our test samples, with only a slight increase in scatter for the lower mass sample.

It is important to note that Figure~\ref{fig:predict-true-obs} evaluates only the accuracy of parameter recovery and does not assess the quality of the uncertainty quantification or the width of the posterior distributions.
Accurate point estimates do not necessarily guarantee well-calibrated posteriors with appropriate confidence intervals.
Additionally, we expect small numbers of observables (e.g., 2D) to naturally provide weaker constraints than higher-dimensional cases, even when the MAP estimators are equally accurate.
We now turn to evaluating the calibration of the posterior distributions.

We apply Tests of Accuracy with Random Points (TARP; \cite{2023PMLR..20219256L}) to assess how well the posteriors are calibrated.
TARP is a sampling-based test specifically designed for high-dimensional posterior distributions and does not require explicit likelihood evaluation.
\cite{2023PMLR..20219256L} demonstrate that TARP provides both \textit{necessary and sufficient} conditions for posterior calibration.
Alternative calibration methods have notable limitations: rank-based tests such as those presented in \cite{2025ApJ...987...96M} can only evaluate 1D marginal distributions rather than the full joint posterior, while highest posterior density tests cannot detect systematic biases in the posterior mean (as proven in \cite{2023PMLR..20219256L}).

Figure~\ref{fig:tarp-obs} shows the TARP calibration curves, plotting credibility level against expected coverage for each observable set.
To assess statistical uncertainty in the calibration assessment, we compute results across 100 bootstrap samples (shown as solid lines and shaded bands), although the confidence intervals are too narrow to be visible.
We also include reference S-shaped curves to illustrate how deviations from the diagonal (ideal case) would appear for under- and over- confident posteriors.

We find that all observable combinations produce TARP curves that closely follow the diagonal, indicating well-calibrated posteriors.
The curves show no significant differences between observable combinations, demonstrating consistent calibration quality regardless of the number of observables used.

\subsection{Results on an idealized mock ATLAS-Aliqa Uma Stream}
\label{section:observable:aau}

\begin{figure*}
    \centering
    \includegraphics[width=0.98\linewidth]{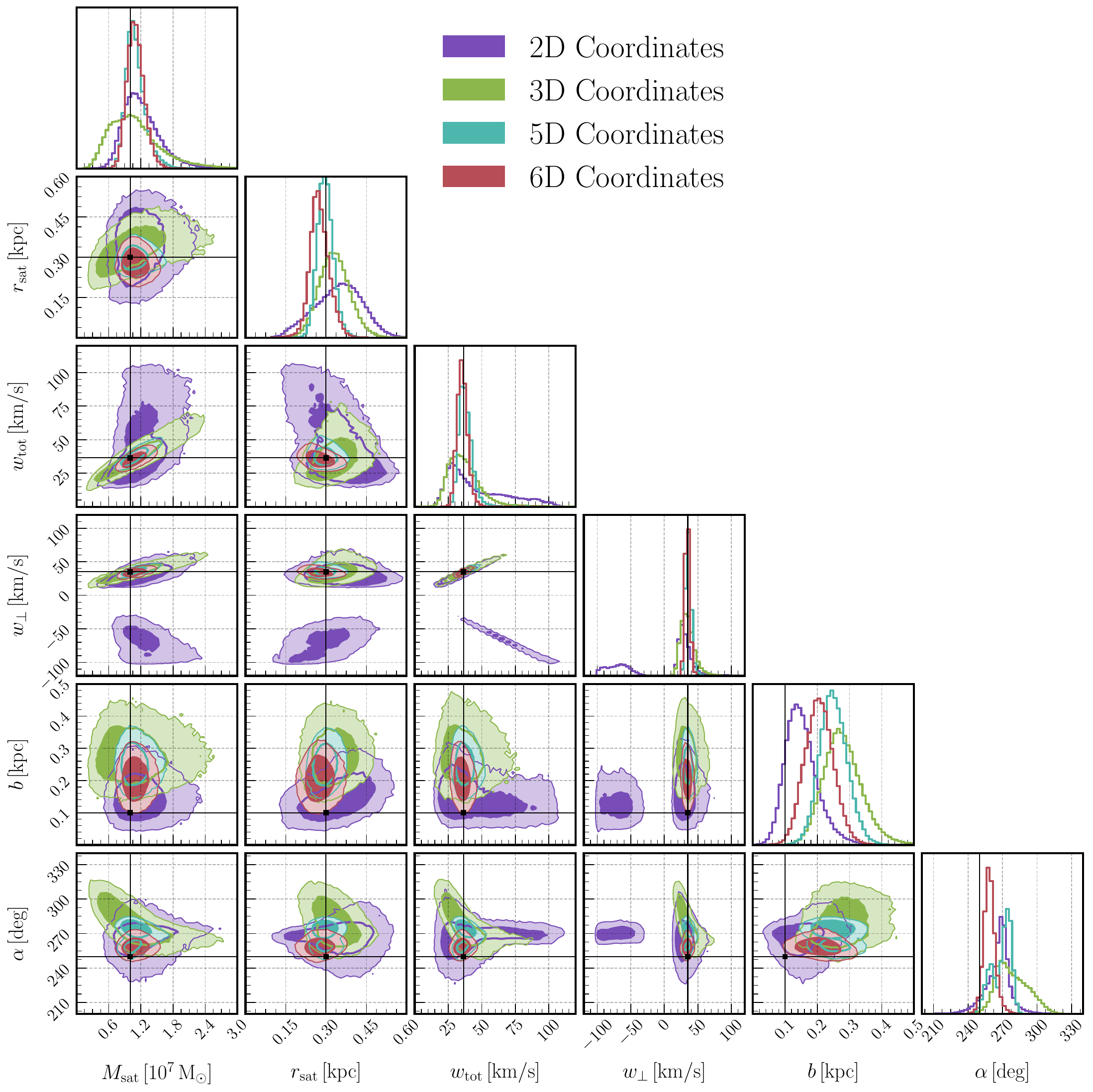}
    \caption{Recovered 6-parameter posterior distributions for the mock AAU stream for the four observable sets.
    The total velocity, $\vtotal = \sqrt{\vparallel^2 + \vperp^2}$, is shown in place of the parallel velocity \vparallel.  
    Contours show 68\% and 95\% credible regions.
    True parameter values are marked with dashed black lines.}
    \label{fig:corner_obs_mv}
\end{figure*}

\begin{landscape}
\begin{table*}[p]
\centering
\renewcommand{\arraystretch}{1.35}
\begin{tabular}{|llllllll|}
\hline
Scenario & \msat & \rsat & \vparallel & \vperp & \vtotal  & $b$ & $\alpha$  \\
& ($10^7$ \modot) & (\kpc) & (\kms) & (\kms) & (\kms) & (\kpc) & (deg) \\
\hline
6D Coordinates
& \asymerror{1.10}{0.17}{0.15} 
& \asymerror{0.27}{0.04}{0.03}
& \asymerror{-8.75}{1.88}{2.55}
& \asymerror{34.25}{3.45}{3.34}
& \asymerror{35.36}{3.92}{3.60}
& \asymerror{0.21}{0.04}{0.04}
& \asymerror{258.54}{4.86}{4.37} \\
5D Coordinates 
& \asymerror{1.08}{0.20}{0.14} 
& \asymerror{0.30}{0.03}{0.03}  
& \asymerror{-11.14}{2.73}{3.55}
& \asymerror{35.57}{4.39}{3.49}
& \asymerror{37.27}{5.18}{3.93} 
& \asymerror{0.25}{0.04}{0.04}
& \asymerror{271.37}{5.66}{12.70} \\
3D Coordinates 
& \asymerror{1.03}{0.51}{0.39} 
& \asymerror{0.33}{0.06}{0.05} 
& \asymerror{-10.73}{3.88}{7.26}
& \asymerror{33.79}{10.71}{8.79} 
& \asymerror{35.52}{12.32}{9.40}  
& \asymerror{0.28}{0.06}{0.05} 
& \asymerror{276.50}{17.12}{11.59} \\
2D Coordinates 
& \asymerror{1.17}{0.38}{0.28} 
& \asymerror{0.35}{0.08}{0.10}  
& \asymerror{-9.55}{11.93}{9.09}
& \asymerror{24.08}{12.35}{94.81}
& \asymerror{39.32}{33.96}{13.27} 
& \asymerror{0.15}{0.05}{0.04}
& \asymerror{267.72}{6.68}{11.95} \\
\hline
Present 6D
& \asymerror{1.14}{1.06}{0.63} 
& \asymerror{0.28}{0.27}{0.13}
& \asymerror{-14.07}{9.39}{20.98}
& \asymerror{35.64}{23.78}{20.15}
& \asymerror{44.27}{31.62}{20.20}
& \asymerror{0.50}{0.39}{0.23}
& \asymerror{272.54}{25.56}{21.96} \\
Future 2D 
& \asymerror{1.11}{0.38}{0.29} 
& \asymerror{0.23}{0.06}{0.05}  
& \asymerror{-12.49}{4.97}{7.41}
& \asymerror{37.04}{12.42}{7.94}
& \asymerror{40.80}{18.19}{8.61} 
& \asymerror{0.22}{0.06}{0.05}
& \asymerror{266.04}{15.33}{14.19} \\
Future 6D 
& \asymerror{1.05}{0.49}{0.31} 
& \asymerror{0.27}{0.08}{0.07} 
& \asymerror{-11.21}{4.49}{6.37}
& \asymerror{34.55}{9.80}{8.20} 
& \asymerror{37.71}{12.03}{8.45}  
& \asymerror{0.24}{0.08}{0.07} 
& \asymerror{267.29}{18.48}{18.47} \\
Future 6+2D
& \asymerror{1.00}{0.23}{0.18} 
& \asymerror{0.24}{0.05}{0.04} 
& \asymerror{-10.97}{2.85}{3.59} 
& \asymerror{34.94}{5.51}{4.41}  
& \asymerror{36.69}{6.17}{4.90}  
& \asymerror{0.22}{0.04}{0.03} 
& \asymerror{265.83}{13.65}{10.31} \\
\hline
Mock AAU & $1.00$ & $0.30$ & $-10.0$ & $35.0$ & $36.4$ & $0.10$ & $250.0$ \\
\hline
\end{tabular}
\caption{
The recovered impact parameters of the mock AAU stream for four kinematic observable sets and four observational scenarios.
The columns show the medians and 68\% credible intervals of the marginal distributions for the six impact parameters and one derived parameter (the total velocity \vtotal).
The true parameters are also shown for reference.
}
\label{tab:aau}
\end{table*}
\end{landscape}

We now present the posterior distributions for the mock AAU stream.
We again emphasize that this analysis uses a simulated stream generated with the fiducial parameters from Table~\ref{tab:prior} to represent the AAU system, rather than actual observational data.
For each kinematic observable set, we sample the posterior $100,000$ times. 
Table~\ref{tab:aau} shows the medians and 68\% credible intervals of the marginal distributions of the interaction parameters.

Figure~\ref{fig:corner_obs_mv} compares the corner plots of the posterior distributions of the four observable sets. 
We show the total velocity, $\vtotal = \sqrt{\vparallel^2 + \vperp^2}$ instead of \vparallel since \vtotal is inversely proportional to the strength of the gravitational impulse imparted to the stream during the encounter. 
Contours represent 68\% and 95\% credible regions.

From Figure~\ref{fig:corner_obs_mv}, all models recover the subhalo mass and velocity within the 95\% credible regions.
The posterior contours become progressively wider when there are fewer stellar kinematic observables, as expected.
The 6D and 5D cases provide the tightest posteriors and are remarkably comparable, with the 68\% credible intervals having widths of approximately $15-20\%$ of the best-fit value for \msat\ and about $4-5 \, \kms$ for \vtotal.
This suggests that stellar radial velocity measurements may not be essential for determining the subhalo mass and velocity, though radial velocities may still be necessary for confirming stellar membership in the stream.

Compared to the 6D and 5D cases, the 3D case shows moderately broader posteriors, with the 68\% credible intervals having widths of about $45\%$ of the best-fit value for mass and $10 \, \kms$ for velocity.
Interestingly, the 2D case yields a somewhat tighter constraint on \msat compared to the 3D case, with a credible interval width of about $30\%$ for mass.
We note that such variations between individual cases can arise from run-to-run stochasticity in ILI training, and the overall trend from Figure~\ref{fig:predict-true-obs} shows that models with fewer observables generally perform worse.
Furthermore, in this example, the 2D case still performs worse overall, exhibiting the largest uncertainties with a velocity constraint width of about $20 \, \kms$.

All models struggle with constraining the impact parameter $b$ and consistently show systematic overestimation.
The constraints on $b$ are roughly comparable across all observable sets, with no clear trend with the number of observables, suggesting that $b$ is generally challenging to constrain regardless of the available kinematic information.
The true impact parameter is recovered within the 95\% contours for the 6D, 5D, and 2D cases and within the 99.7\% contour for the 3D case.
This difficulty may arise from the true value ($b = 0.1 \, \kpc$) being very close to the lower edge of the prior range of $[0, 3] \, \kpc$.
Note that although the 2D case appears to perform better for $b$ in this specific realization, this is likely due to run-to-run variations rather than a systematic advantage, as demonstrated by the overall trends in Figure~\ref{fig:predict-true-obs}.

Most notably, the posterior of the perpendicular velocity \vperp in the 2D case is bimodal, with the two peaks roughly corresponding to the different signs of \vperp.
The distribution is not perfectly symmetric around zero: the negative peak is at approximately $-50 \, \mathrm{km \, s^{-1}}$ and more spread out compared to the positive peak, which aligns closer to the true value of $35 \, \mathrm{km \, s^{-1}}$.
The model correctly identifies the positive peak as the primary solution.
Additionally, the contours reveal a connection between $\vperp$ and $\alpha$, with the incorrect negative peak failing to encompass the true $\alpha$ value even within the 95\% contour.
We note that this bimodality appears only in the velocity parameters and not in other parameters such as mass or orientation angle.
Although not shown in Figure~\ref{fig:corner_obs_mv}, we do not observe bimodality for the parallel velocity $\vparallel$ in this example; however, the true $\vparallel$ value of $-10 \, \mathrm{km \, s^{-1}}$ is close to zero, which may mask this effect.

The velocity bimodality is not unique to the 2D kinematic case.
Although this particular mock AAU example only shows bimodality in the 2D case, we observe similar velocity bimodality (in both $\vparallel$ and $\vperp$) across different observable sets in other test cases.
In general, velocity bimodality occurs when the phase-space is not sufficiently sampled, whether due to limited kinematic information or other factors that reduce the information content of the data (see Section~\ref{section:uncertainty:aau}).

In summary, our analysis demonstrates that NPE successfully recovers DM subhalo properties even with limited observational data, though the precision of parameter constraints depends strongly on the available stellar kinematic information.
The remarkable similarity between 6D and 5D results indicates that radial velocity measurements, while valuable for confirming stellar membership, may not be critical for determining subhalo mass and velocity.
However, the degradation in performance for the 2D case, particularly the emergence of bimodality in velocity posteriors, reveals the fundamental challenges of subhalo characterization when relying solely on photometric data.

\section{Observational Scenarios}
\label{section:uncertainty}

We now examine the performance of the framework under more realistic observational scenarios. 
Specifically, we inject measurement uncertainties into the coordinates and velocities and downsample the total number of stars to match present-day observations and future projections.  
We consider four scenarios:
\begin{itemize}
   \item \textbf{Present~6D}: present-day observational scenario matching the AAU observations from \cite{2021ApJ...911..149L}.
   We consider $96$ member stars with $(\phi_1, \phi_2, \mu_1, \mu_2, v_r, d)$ down to $r < 19.9$.
   
   \item \textbf{Future~2D}: future observational scenario with LSST DR2 photometry. We consider $4096$ member stars down to $r<25.25.$
   
   \item \textbf{Future~6D}: future observational scenario with LSST photometry, Gaia proper motions, and spectroscopic measurements from 4MOST.
   We consider $396$ member stars with $(\phi_1, \phi_2, \mu_1, \mu_2, v_r, d)$ down to $r < 21$. 
   
   \item \textbf{Future~6+2D}: combination of the Future~2D and Future~6D scenarios.
   In other words, we consider $396$ member stars with $(\phi_1, \phi_2, \mu_1, \mu_2, v_r, d)$ and $4096$ member stars with $(\phi_1, \phi_2)$, for a total of $4492$ stars.
\end{itemize}

The models are described in greater detail in \ref{section:uncertainty:model}. For each of the first three scenarios (i.e. Present~6D, Future~2D, and Future~6D), we train a different NPE model, using the same data preprocessing pipeline and machine learning architecture described in Section~\ref{section:method}.
We generate 10 independent realizations of each stream, where each realization involves randomly subsampling the particles to match the number of member stars of each scenario and adding Gaussian noise to the observables\footnote{To prevent data leakage, we ensure that all realizations of the same underlying stream (i.e., same parameter set) are assigned to the same training or validation split, preventing the model from seeing different noise realizations of identical streams across splits.}.
We then apply the binning procedure to each realization.

For the Future~6+2D scenario, we do not train a new NPE model, as this would require incorporating methods to handle incomplete data.
While this is generally feasible (e.g., treating missing data as having infinite uncertainty, see \cite{2023ApJ...952L..10W}), such procedures are computationally expensive and introduce additional complexity to the framework.
Instead, we simply combine the posteriors from the Future~6D and Future~2D models.

For simplicity, we assume that the 6D and 2D data, denoted as $\vec{x}_\mathrm{6D}$ and $\vec{x}_\mathrm{2D}$, of a single stream are independent (i.e., no overlap in the stellar samples).
Under this assumption, the joint likelihood can be written as $p(\vec{x}_\mathrm{6D}, \vec{x}_\mathrm{2D} \mid \theta) = p(\vec{x}_\mathrm{6D}\mid \theta)p(\vec{x}_\mathrm{2D} \mid \theta)$, where $\theta$ represents the subhalo parameters.
The joint posterior can be obtained from Bayes' theorem:
\begin{equation}
    \hat{q}(\theta \mid \vec{x}_\mathrm{6D}, \vec{x}_\mathrm{2D}) = \frac{\hat{q}(\theta \mid \vec{x}_\mathrm{6D})\hat{q}(\theta \mid \vec{x}_\mathrm{2D})}{p(\theta)},
\end{equation}
where $\hat{q}(\theta \mid \vec{x}_\mathrm{6D})$ and $\hat{q}(\theta \mid \vec{x}_\mathrm{2D})$ are the posterior distributions from the Future~6D and Future~2D models, respectively, and $p(\theta)$ is the prior distribution, which is identical for all models in this work.
In this manner, we can estimate the posterior by resampling either $\hat{q}(\theta \mid \vec{x}_\mathrm{6D})$ or $\hat{q}(\theta \mid \vec{x}_\mathrm{2D})$ with the appropriate importance weights.
Since we use normalizing flows for density estimation, this is straightforward as we can directly evaluate the density at any point in parameter space.

We note that, in practice, the stellar sample of the 2D data is likely to be a subset of the 6D data, e.g., bright stars in the photometric sample that are selected for spectroscopic follow-up.
In this case, the distributions $p(\vec{x}_\mathrm{6D}\mid \theta)$ and $p(\vec{x}_\mathrm{2D} \mid \theta)$ are no longer independent.
Since the spectroscopic sample would contain additional information about the same stellar population, we expect this correlation to lead to less improvement in parameter constraints than our independent assumption suggests, making our results somewhat optimistic for the Future~6+2D scenario.

In Section~\ref{section:uncertainty:model}, we describe details of the uncertainty modeling, including the projected number of stars.
Section~\ref{section:uncertainty:test} and Section~\ref{section:uncertainty:aau} present the results over the entire test dataset and posteriors of the mock AAU stream, respectively. 

\subsection{Mock AAU Observations}
\label{section:uncertainty:model}


\begin{table}
\centering
\renewcommand{\arraystretch}{1.25}
\begin{tabular}{|lllll|}
\hline
Scenario & $N_\star$ & $r_\mathrm{max}$ & \multicolumn{2}{c|}{$\sigma_{\mu} (r = 19)$} \\ 
 & & (mag) & ($\mathrm{mas~yr}^{-1}$) & ($\mathrm{km~s}^{-1}$) \\
\hline
Present 6D & 96 & 19.8 & 0.23 & 25 \\
Future 2D & 4096 & 25.25 & -- & -- \\
Future 6D & 396 & 21.0 & 0.035 & 3.8 \\
\hline
\end{tabular}
\caption{Parameters for three observational scenarios, showing the number of tracers, the $r$-band magnitude limits, and proper motion uncertainty at $r= 19$. This proper motion uncertainty in $\mathrm{km~s}^{-1}$ is calculated assuming an AAU distance modulus of 16.8.}
\label{tab:obs}
\end{table}

In this section, we describe our procedure for determining the number of member stars observed (Section \ref{section:mock:nstars}) and calculating uncertainties for the proper motion ($\mu_1, \mu_2$; Section \ref{section:mock:proper_motion}), radial velocity ($v_r$; Section \ref{section:mock:radial_velocities}), and distance ($d$; Section \ref{section:mock:distances}) measurements. 
Appendix~\ref{app:uncertainty} presents additional information.
In both present-day and future scenarios, we assume negligible uncertainty in the on-sky stream coordinates $(\phi_1, \phi_2)$.
Table~\ref{tab:obs} summarizes the properties for all three observational scenarios.

Astrometric and spectroscopic uncertainties are typically parameterized as functions of survey magnitude.
We first assign apparent DECam $g$- and $r$-band magnitudes to each simulated particle according to the initial mass function and isochrone properties of the observed AAU stream, then use these magnitudes to calculate the associated uncertainties.
For simplicity, the $g$- and $r$- band magnitudes are assigned randomly to each particle, independent of the coordinates in the stream. 

\subsubsection{Member stars}
\label{section:mock:nstars}
Current observations of AAU contain 96 stars with 6D measurements \citep{2021ApJ...911..149L} and $\sim$ 2000 stars with 2D measurements \citep{2018ApJ...862..114S}. These observations include photometric measurements by DES, proper motion measurements from Gaia DR3, and radial velocities measured by the Southern Stellar Stream Spectroscopic Survey ($S^5$, \cite{2019MNRAS.490.3508L}).
For our estimates of future observations,
we start by assuming that the stellar population follows a MIST isochrone \citep{2016ApJS..222....8D} with a Chabrier initial mass function \citep[][IMF;]{2005ASSL..327...41C}. 
The exact isochrone parameters are taken from \cite{2021ApJ...911..149L}: $[\mathrm{Fe/H}] = -1.99$, $[\alpha/\mathrm{Fe}] = 0.4$, $Y = 0.4$, and $\mathrm{Age} = 11.5 \, \mathrm{Gyr}$. Following \cite{2021ApJ...911..149L}, we apply an additional empirical shift of $0.143$ and $0.188$ in the $g$- and $r$-bands, respectively, to better match the observed stellar population. 

To estimate the number of AAU member stars that will be included in future 2D and 6D observations, we sample stars from the isochrone models down to the magnitude limits of upcoming surveys. 
For our Future 2D scenario, we consider the first data release of LSST that includes homogeneous coverage across the full survey footprint (DR2\footnote{LSST DR2 will include one year of observations and is currently scheduled to be released in late 2027.}~\cite{RTN:011}). 
Simulations of the LSST survey \citep{peter_yoachim_2025_15368965} suggest that the $10\sigma$ $r$-band depth after 1 year of observations will be $r_{10}\sim25.25$. 
We can use this to estimate the number of detected AAU member stars by taking the number in DES ($r_{10}\sim23.5$, $N_\star = 2{,}000$) and integrating a Chabrier IMF to the new limiting magnitude. 
From this we estimate that $\sim4{,}096$ member stars will be detected with $\phi_1, \phi_2$ measurements in the LSST DR2 data.

For the Future 6D scenario, we consider stars to the depth of 4MOST \citep{2019Msngr.175....3D}--an ESO-led spectroscopic survey on the VISTA telescope that aims to measure radial velocities of stars across a large fraction of the southern sky to the limits of Gaia proper motion measurements ($r \sim 21$). 
We again integrate the Chabrier IMF, scaled by $N_\star = 96$ at $r = 19.8$ to match the Present 6D measurements. 
This produces an estimate of 396 member stars with 6D measurements down to $r=21$.

We note that this scaling approach assumes similar completeness between current and future observations, which is reasonable for photometry but more uncertain for spectroscopy.
Current spectroscopic observations from surveys like $S^5$ employ relatively broad selection criteria that likely capture most stream members~\citep{2019MNRAS.490.3508L, 2021ApJ...911..149L}, though future surveys may have different targeting strategies.
We expect other observational uncertainties to dominate over these selection effects.

\subsubsection{Proper Motions}
\label{section:mock:proper_motion}
The Present 6D scenario includes proper motions from Gaia Data Release 3 (DR3), whose uncertainties we model using the \texttt{PyGaia} package.
We first convert the sampled $g$- and $r-$band magnitudes to Gaia $G$-band magnitudes using the relation:
\begin{equation}
    G = -0.09 -0.71(g-r) -0.13(g-r)^2,
\end{equation}
which we derive by performing a polynomial fit using the isochrones generated in the DES and Gaia photometric systems.
The proper motion uncertainties are then calculated with \texttt{PyGaia} as follows:
\begin{align}
\label{eq:astro_error}
    \sigma_X &= c_X \sigma_\varpi, \quad \sigma_\varpi = \sqrt{40 + 800 z + 30 z^2}, \\
    \log_{10}z &= 0.4 (\mathrm{max}(G, G_\mathrm{bright}) - 15.0),
\end{align}
where $\sigma_\varpi$ is the parallax error in $\mathrm{mas}$ and $G_\mathrm{bright}=13$.
For Gaia DR3, the corresponding coefficients $c_X$ for proper motions $\mu_\alpha$ and $\mu_\delta$ are $1.03$ and $0.89$ in unit of $\mathrm{mas \, yr^{-1}}$, respectively.
Gaia proper motion uncertainties scale as $T^{-1.5}$, where $T$ is the mission duration in years \citep{2021A&A...649A...1G, 2023A&A...674A...1G}.
Therefore, proper motion uncertainties from the expected 10-year Gaia data (Gaia DR5) will be a factor of $0.15$ times those of the 5.5-year observations included in Gaia DR3~\citep{2025arXiv250301533B}. 
Lastly, we note that \texttt{PyGaia} includes a $10\%$ ``science margin'' in its uncertainty calculations (i.e., multiplies uncertainties by a factor of $1.1$), making our modeled proper motion uncertainties potentially conservative.

\subsubsection{Radial Velocities}
\label{section:mock:radial_velocities}
We model the radial velocity uncertainties for the Present 6D scenario by fitting a function to the radial velocity errors of $S^5$ measurements of AAU member stars:
\[
\sigma_{v_r} = 10.3317 - 1.8028\,r + 0.0741\,r^2
\]
where $r$ is the DECam $r$-band magnitude.  

For the Future 6D model, we use the 4MOST radial velocity uncertainties presented in \cite{2012SPIE.8446E..0TD}. We extract the data from their Figure 1 and fit the equation:
\begin{equation}
\sigma_{v_r} = a + \frac{b}{1 + \exp(-c(V - d))},
\end{equation}
where $a = 1.98 \, \kms$, $b = 211.5 \, \kms$, $c = 1.17$, and $d = 24.3$. As these uncertainties are a function of $V$-band magnitudes, we convert our sampled $g$- and $r$-band magnitudes using the relation from \cite{2005AJ....130..873J}:
\begin{equation}
    V = -0.01 - 0.59 (g - r) + g.
\end{equation}

\subsubsection{Distances}
\label{section:mock:distances}
We adopt a distance uncertainty of 10\% for all simulated particles. 
This is typical of RR Lyrae and blue horizontal branch (BHB) stars, which are commonly used as standard candles for stellar stream distance measurements \citep{2021ApJ...911..149L, 2022AJ....163...18F}.
We note that this approach is only approximate and specifically calibrated for the AAU stream, which contains a relatively high density of distance tracers~\citep{2021ApJ...911..149L}.
In practice, most stream stars are not distance tracers (as also evidenced in Figure~\ref{fig:aau_cmd}), and obtaining reliable distance measurements requires at least one RR Lyrae or BHB star per stream segment.
While AAU has sufficient coverage for most of its length, this is not guaranteed for all stream segments or for other stellar streams more generally (e.g., see Figure 6 of \cite{2021ApJ...911..149L} and Figure 4 of \cite{2022AJ....163...18F}).

An alternative approach would be to model isochrone-based photometric distances for the majority of stars, which typically have uncertainties of $\sim 20\%$.
However, we do not adopt this approach here since we are specifically modeling the AAU stream.

We note an additional caveat.
For simplicity, we assume all measurement uncertainties are uncorrelated in our simulations, while in practice distance and radial velocity uncertainties can be covariant, particularly for RR Lyrae stars.
RR Lyrae and BHB stars with spectroscopic information can achieve excellent distance precision of $\sim 5\%$ \citep{2025MNRAS.542..560B, 2025arXiv250402924M}, but obtaining reliable radial velocities from RR Lyrae is challenging due to pulsations and requires multiple spectroscopic observations \citep[e.g.][]{2020ApJ...901...23H, 2025arXiv250402924M}.
This covariance between distance and radial velocity uncertainties varies across different stellar types within the stream.

Given these various observational complexities, we emphasize that our treatment of distance uncertainties is specifically tailored to the characteristics of the AAU stream.
Future applications of this framework to other streams will require calibration to match their specific observational properties.

\subsection{Results on the test dataset}
\label{section:uncertainty:test}

\begin{figure*}
    \centering
    \includegraphics[width=0.98\linewidth]{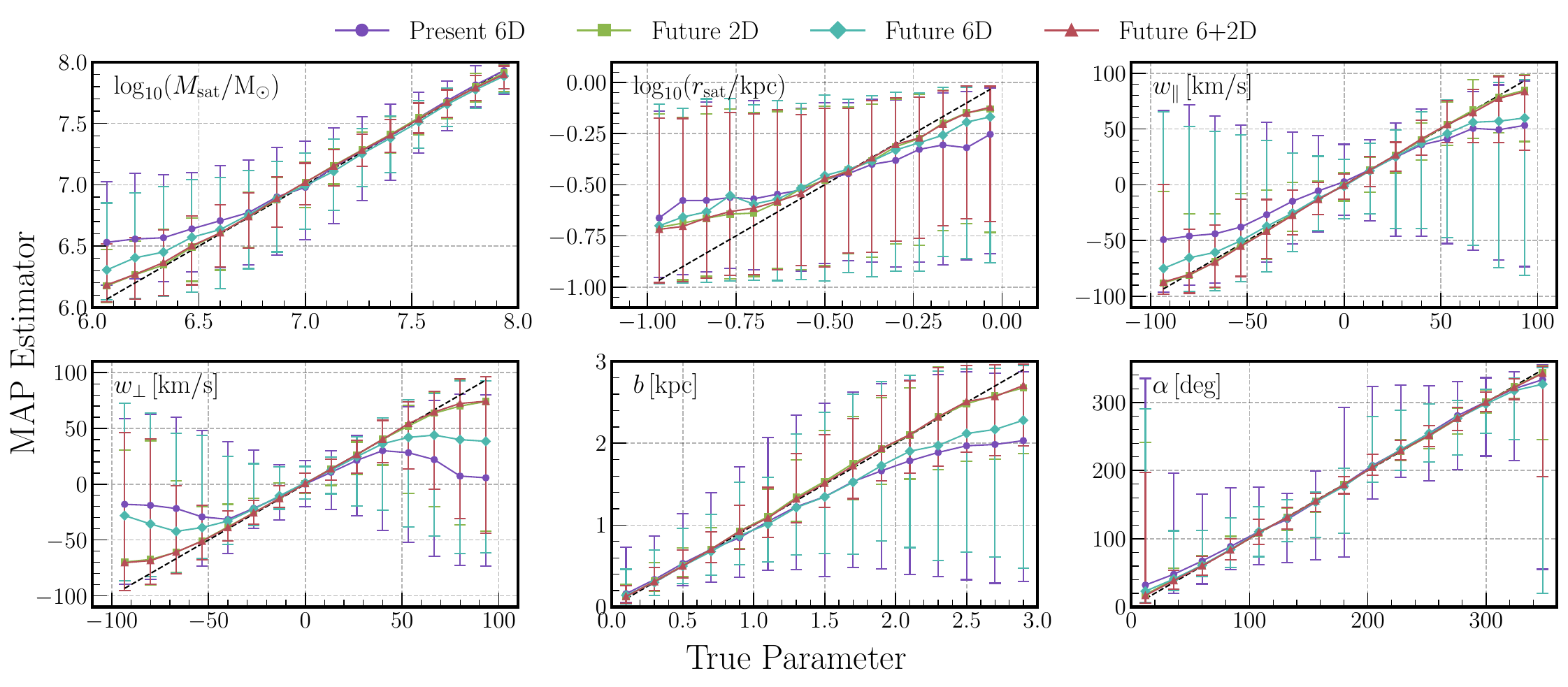}
    \caption{Comparison between the MAP estimators and the true parameters for the four observational scenarios. 
    Panel layouts are the same as Figure~\ref{fig:predict-true-obs}.
    }
    \label{fig:predict-true-err}
\end{figure*}
\begin{figure}
    \centering
    \includegraphics[width=0.7\linewidth]{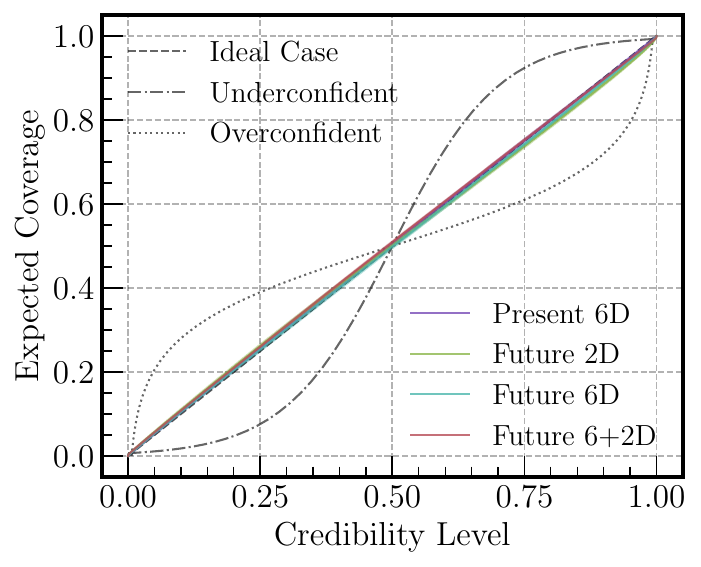}
    \caption{Credibility level versus expected coverage from TARP for the four observational scenarios.
    Panel layouts are the same as Figure~\ref{fig:tarp-obs}.
    }
    \label{fig:tarp-err}
\end{figure}

We first present results on the full test dataset.
Similar to Section~\ref{section:observable:test}, we compare the MAP estimates with true parameters and examine the TARP calibration curves.
These results show the general performance trends and provide important validation of the framework's overall performance.

Figure~\ref{fig:predict-true-err} shows the MAP estimators plotted against the true parameters for each subhalo property across the four observational scenarios. 
As before, we bin the streams according to their true parameter values and compute the mean and 68\% confidence interval of the MAP estimators within each bin to visualize trends and assess systematic biases.

Overall, the performance of all models is significantly degraded compared to Figure~\ref{fig:predict-true-obs}.
This is expected, as the introduction of observational uncertainties and the reduction in the number of available stars both reduce the information content of the observational data.

Specifically, we observe systematic biases in parameter regimes where subhalo-stream interactions are expected to be weaker.
This occurs for low subhalo mass $\msat$, high encounter velocity $\vtotal$, as well as when the subhalo radius $\rsat$ or the impact parameter $b$ is large.
To first order, these trends can be understood through the velocity kick impulse approximation imparted to stream stars during the encounter \citep{2015MNRAS.454.3542E}:
\begin{equation}
    \Delta v = \frac{G \msat \vperp}{\vtotal^2 \sqrt{b^2 + \rsat^2}}.
\end{equation}
When $\Delta v$ becomes small relative to observational uncertainties, parameter recovery becomes increasingly challenging.
While similar trends are subtly present in Figure~\ref{fig:predict-true-obs}, they become much more pronounced with the addition of observational noise.
Additionally, we again observe systematic bias at low $\rsat$, where compact subhalos become indistinguishable from point masses.

We note that for the velocity parameters, we observe large scatter in the MAP estimators, especially at high velocities.
This scatter is primarily due to bimodality in the velocity posteriors, where the MAP estimator can correspond to either peak.

Among the four observational scenarios, Present~6D performs worst overall, reflecting the limited number of available stars (96 stars) and relatively large observational uncertainties.
Interestingly, Future~2D (4,096 stars) and Future~6D (396 stars) show similar performance levels, with Future~6D performing slightly worse despite having access to full 6D kinematic information.
This suggests a trade-off between sample size and kinematic completeness in phase-space sampling, with both contributing significantly to constraining power.
Future~2D shows accuracy nearly comparable to the ideal cases in Section~\ref{section:observable}, demonstrating the power of large statistical samples.
As expected, Future~6+2D performs best overall, benefiting from the combination of both high statistical power from the 2D sample and detailed kinematic constraints from the 6D subset.

Figure~\ref{fig:tarp-err} shows the TARP calibration results, plotting credibility level versus expected coverage.
Despite the reduced accuracy in parameter recovery, we find that the models still return well-calibrated posterior distributions for all observational scenarios.

\subsection{Results on a mock ATLAS-Aliqa Uma Stream}
\label{section:uncertainty:aau}
\begin{figure*}
    \centering
    \includegraphics[width=0.98\linewidth]{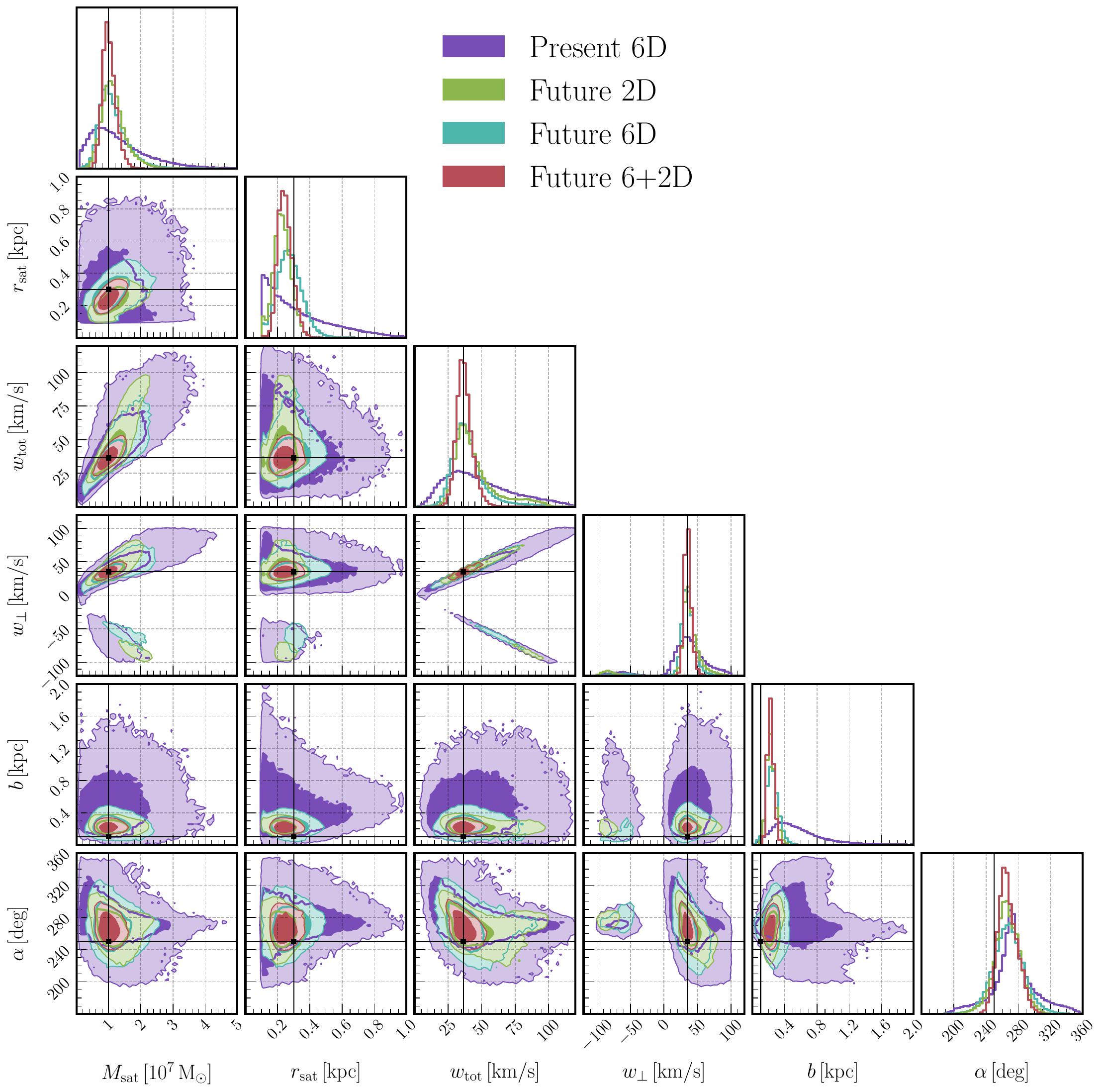}
    \caption{
    Recovered 6-parameter posterior distributions for mock AAU stream for the four observational scenarios.
    Contours show 68\% and 95\% credible regions.
    True parameter values are marked with dashed lines.}
    \label{fig:corner_err_mv}
\end{figure*}

We now examine the posteriors for the mock AAU stream under realistic observational conditions. 
Figure~\ref{fig:corner_err_mv} shows the corner plots of the posterior distributions, with each color denoting a different observational scenario.
Table~\ref{tab:aau} presents the medians and 68\% credible intervals of the marginal distributions for all impact parameters.

We find that incorporating realistic observational conditions substantially broadens the posterior contours compared to those in Figure~\ref{fig:corner_obs_mv}.
The Present~6D scenario exhibits the poorest performance, with marginal posteriors for $\msat$, $\vtotal$, and $b$ showing pronounced asymmetry and long tails extending toward higher values.
The subhalo mass uncertainty is approximately $50\%$ of the best-fit value, while velocity uncertainties reach $10-15 \, \kms$.
Most notably, the model fails to meaningfully constrain the subhalo radius $\rsat$, effectively returning the prior distribution.
The impact parameter $b$ exhibits systematic overestimation, similar to the trends observed in Section~\ref{section:observable:aau}, though the true value remains within the 95\% credible region.

The Future~2D and Future~6D scenarios demonstrate substantially better performance and produce remarkably similar posterior distributions, consistent with the trends observed in Section~\ref{section:uncertainty:test}.
Future~6D benefits from complete kinematic information while Future~2D compensates with a substantially larger sample size (4,096 vs 396 stars), resulting in comparable constraining power for both scenarios.
The subhalo mass uncertainty decreases to approximately $40\%$, with velocity uncertainties reduced to around $10-20 \, \kms$.
The subhalo radius $\rsat$ shows meaningful constraints compared to the Present~6D case, though the posterior retains a long tail toward higher values.
The impact parameter $b$ continues to show systematic overestimation but remains within the 95\% credible region.

The Future~6+2D scenario achieves the tightest constraints overall by combining the statistical power of the large 2D sample with the detailed kinematic information from the smaller 6D subset.
The performance approaches that of the idealized cases in Section~\ref{section:observable}, with subhalo mass uncertainties of approximately $20\%$ and velocity uncertainties of approximately $5-10 \, \kms$.

Most notably, the bimodality in the perpendicular velocity $\vperp$, as previously observed only in the idealized 2D coordinates case in Section~\ref{section:observable:aau}, now emerges across multiple observational scenarios.
Specifically, this bimodality appears in the Present~6D, Future~2D, and Future~6D scenarios.
This suggests that bimodality emerges when phase-space is poorly sampled, rather than being specific to missing kinematic information, whether due to limited kinematic observables (as in the 2D case) or reduced statistical power from fewer stars.
As before, this manifests as a negative and a positive peak, with all models correctly selecting the positive peak.

Importantly, the bimodality is absent in the Future~6+2D case, where the negative peak is ruled out at the 95\% confidence level.
The Future~6+2D case shows that adequate phase-space coverage through complementary datasets successfully resolves these degeneracies.

\section{Discussion}
\label{section:discussion}

\subsection{Comparison with Hilmi et al. 2024}

We compare our inference results with Hilmi et al. (2024) \cite{hilmi24} (hereafter H24), which represents the closest related work.
H24 employs identical particle-spray simulations of the AAU stream \citep[see also][]{2019MNRAS.487.2685E, 2015MNRAS.454.3542E} and applies MCMC inference using a 1D binned Gaussian likelihood approach, binning particles along $\phi_1$, to perform inference on 8 interaction parameters (our 6 parameters along with the additional $\phi_a$ and $T_a$ parameters).
They explore three scenarios with 6D kinematics: no uncertainty, present-day uncertainties, and 4MOST uncertainties.
Since their uncertainty prescriptions differ from ours for both present-day and 4MOST scenarios, we focus our comparison on the idealized case.

We report mass constraints of approximately $15\%$ uncertainties for the idealized 6D scenario.
In comparison, H24 reports tighter constraints with approximately $4\%$ uncertainties for their idealized case.
Notably, we observe bimodality in the perpendicular velocity in several scenarios, which H24 does not report despite using the same underlying simulations.

Several factors may contribute to these differences.
We note that H24 uses a uniform prior on subhalo mass while we employ a log-uniform prior, which may contribute to differences in posterior shape, particularly longer tails toward higher masses in our results. 
However, this difference is unlikely to significantly affect the comparison of uncertainty magnitudes.
Additionally, our NPE model may require additional training data to achieve tighter uncertainty estimates.
However, when tested with a reduced training dataset ($100,000$ vs. $150,000$ streams), we find roughly similar performance, suggesting our results are not primarily limited by training data size.
We will conduct further investigation to explore this systematically in future work.

However, we note that the constraints reported in H24 may be optimistic for several reasons.
Our spline fitting approach captures more complex stream morphology compared to their binning along $\phi_1$.
As illustrated in Figure~\ref{fig:example_streams}, binning along $\phi_1$ alone may not capture all relevant stream features, which could lead to different sensitivity to parameter constraints.

Additionally, H24 reports surprisingly tight constraints on the impact time $T_\mathrm{a}$ and relative angle $\phi_\mathrm{a}$.
In our work, we find that including these parameters significantly expands the parameter space, and many combinations result in the subhalo missing the stream entirely, creating scenarios with no gravitational perturbation.
It is unclear how H24 handles such unphysical regions of parameter space.
Without the amortized framework that enables systematic calibration testing, it remains unclear whether the MCMC posteriors are well-calibrated.

Furthermore, the absence of velocity bimodality in H24's results may reflect differences in sampling approaches.
Although advanced MCMC techniques such as parallel tempering can better handle such cases~\citep[e.g.][]{1992EL.....19..451M, 2016MNRAS.455.1919V}, standard MCMC algorithms can struggle to detect and adequately sample multimodal distributions.
The limited description of the MCMC implementation in H24 makes it challenging to fully understand the source of these methodological differences.

\subsection{Implications for future observations}

In Section~\ref{section:uncertainty}, we compare the inference performance between four scenarios: present-day observations matching AAU observations (96 stars with 6D coordinates), LSST DR2 photometry (4,096 stars with 2D coordinates), 4MOST spectroscopy (396 stars with 6D coordinates), and a combined LSST+4MOST approach (4,492 total stars with mixed observational completeness).

We find that the DM subhalo mass \msat can be accurately inferred, though the constraints vary across scenarios: present-day, LSST photometry, 4MOST spectroscopy, and LSST+4MOST achieve $50\%$, $30\%$, $40\%$, and $20\%$ uncertainties, respectively.
While the total velocities \vtotal are well-constrained, the individual velocity components show bimodality with peaks at different signs of the perpendicular velocity \vperp.
This bimodality emerges when phase space is poorly sampled due to limited kinematic information (e.g. large samples with 2D coordinates) or insufficient stellar tracers (e.g. small samples with 6D coordinates).
On the other hand, the impact parameter $b$ and subhalo radius \rsat remain poorly constrained across all scenarios.

The DM subhalo mass represents the most crucial measurement, as it directly probes the subhalo mass function and thus can be used for testing cold DM predictions~\citep{2008Natur.454..735D, 2008MNRAS.391.1685S} and galaxy formation physics through subhalo abundance matching~\citep[e.g.][]{2004ApJ...609...35K, 2004MNRAS.353..189V, 2006ApJ...647..201C}.
Additionally, the subhalo radius, velocities, and impact parameter provide complementary probes of DM. 
The radius can be used to determine the subhalo concentration~\citep[e.g.][]{1997ApJ...490..493N, 2017MNRAS.466.4974M, 2018PhR...730....1T, 2021MNRAS.506.4421S}.
Similarly, velocities and impact parameters can infer the present-day positions of interacting subhalos, enabling studies of DM substructure distribution within the Milky Way halo~\citep{2015MNRAS.446.1000F, 2021MNRAS.507.4826L}

Our results indicate that while the subhalo mass can be reliably inferred even with limited kinematic information and tracer counts, other secondary parameters are much more challenging.
The subhalo radius and impact parameter suffer from a degeneracy where compact subhalos produce gravitational signatures nearly indistinguishable from point masses, thus preventing reliable constraints on subhalo concentration.
Similarly, present-day subhalo positions cannot be determined due to directional ambiguity in velocity measurements and poor constraints on the impact parameter.

We identify a fundamental trade-off between sample size and kinematic completeness in constraining subhalo properties.
The comparable performance between LSST's large photometric sample (4,096 stars) and 4MOST's smaller but kinematically complete sample (396 stars) demonstrates that both statistical power and phase-space coverage contribute significantly to inference quality.
For applications focused primarily on subhalo mass functions, LSST's photometric-only approach may provide sufficient constraints at minimal observational cost.
However, the optimal strategy combines both approaches: large photometric samples with targeted spectroscopic follow-up effectively resolve velocity degeneracies while maximizing statistical power.
This has direct implications for survey planning, as modest spectroscopic investments can significantly enhance the scientific return of photometric surveys for DM substructure studies.

Finally, we note that the results in Section~\ref{section:uncertainty} are likely to be conservative since we do not utilize density information along the streams in the inference process.
Using density perturbations would help constrain subhalo parameters, but as indicated in Sections~\ref{section:sim} and \ref{section:method}, density information from particle-spray simulations is unreliable and unsuitable for direct comparison with observational data.
Additionally, the proper motion uncertainties provided by \texttt{PyGaia} include a conservative 10\% science margin, though this does not affect the LSST DR2 photometric scenarios.

\subsection{Current limitations and future improvements}

We discuss several potential areas for improvements in future work. 
As noted in Section~\ref{section:preprocess}, our spline fitting and binning procedure, while generally robust, does not fully capture complex stream morphologies, particularly for streams that have undergone major perturbations.
While this does not significantly impact parameter recovery, future developments should adopt more robust track-fitting algorithms or, more ideally, utilize particle-level data, as in \cite{2025ApJ...987...96M}.
However, particle-level approaches face a fundamental challenge: they implicitly encode density information, which is known to be unreliable in particle-spray simulations used in both this work and \cite{2025ApJ...987...96M}.
Therefore, if particle-level data is to be used effectively, future work must either focus on improving the underlying simulations or develop methods to remove or mitigate the spurious density information (e.g., by resampling particles).

Our current simulations are specifically calibrated for the AAU stream, which necessitates assumptions about the properties and orbits of the progenitor.
While the NPE framework can be readily applied to different streams, this requires updating the progenitor model in the underlying simulations.
Additionally, our uncertainty modeling is calibrated using observed properties specific to AAU, including its magnitude completeness and distance tracer availability.
Different streams may have substantially different observational characteristics, requiring recalibration of these models and potentially affecting inference performance.

Additionally, as mentioned, we fix the encounter time $T_\mathrm{a}$ and the relative angle along the stream at the time of impact $\phi_\mathrm{a}$.
Including these parameters significantly expands the parameter space and introduces computational challenges.
Moreover, some combinations of $T_\mathrm{a}$ and $\phi_\mathrm{a}$ would result in the subhalo missing the stream entirely.
This creates scenarios where no gravitational perturbation occurs and complicates the definition of other interaction parameters, as quantities like impact parameter $b$ become ill-defined when no actual encounter takes place.
Future work could address this challenge by implementing hierarchical models that first determine interaction probability or constraining the parameter space to physically meaningful encounter geometries.

Machine learning-based methods, including ILI, are more susceptible to model misspecification when the target observational data falls outside the training distribution.
This limitation is less problematic for forecasting applications, such as ours, where we focus on relative performance comparisons within the same simulation framework.
However, for applications to real observational data, future work could address this challenge through domain adaptation techniques, which have shown promise in astrophysical contexts~\citep[e.g.][]{2023MLS&T...4b5013C, 2023arXiv231101588R, 2025MLS&T...6c5032P}, though this remains an active area of research.
Ultimately, it is better to improve the underlying simulations to match the target domain as closely as possible, by both updating the physics models, e.g. by including time-varying potentials or more realistic treatments of globular cluster dynamics~\citep{2025arXiv250903599P, 2025arXiv250514792W, 2024MNRAS.532.2657B, 2025arXiv250915307W}, and accounting for more realistic observational effects.
Currently, we are addressing the latter by extending this forecasting framework to include systematic effects expected in early LSST data.
Among the effects under consideration are contamination from galaxies, contamination from foreground and background stars, and survey properties such as seeing and airmass variations.

\section{Conclusion}
\label{section:conclusion}

Interactions between DM subhalos and stellar streams offer a promising avenue for constraining the particle nature of DM.
These interactions leave detectable kinematic substructures in stellar streams such as density gaps and velocity perturbations.
By analyzing these substructures, we can infer the properties of the interacting DM subhalos. 

In this work, we present an NPE framework that maps the stellar kinematics (coordinates and velocities) to the parameters of the interacting DM subhalos. 
The framework employs a transformer-based architecture for feature extraction and a normalizing flow for density estimation. 
We train our model on particle-spray simulations based on the AAU stream from \cite{hilmi24, 2016MNRAS.461.1590E, 2019MNRAS.487.2685E}. 
We target six interaction parameters: subhalo mass $\msat$ and radius $\rsat$, velocity components ($\vperp$, $\vparallel$), impact parameter $b$, and impact orientation angle $\alpha$.
We use our NPE framework to systematically evaluate how kinematic completeness and observational uncertainties affect subhalo parameter inference under idealized conditions. 

In Section~\ref{section:observable}, we examine four idealized kinematic scenarios with 10,000 member stars and no observational uncertainties.
These include full 6D coordinates, 5D coordinates (excluding radial velocities), 3D coordinates (excluding radial velocities and proper motions), and 2D coordinates only.
For each scenario, we train separate models and evaluate both parameter recovery accuracy and posterior calibration across the entire test dataset.
We summarize our key findings below.
\begin{itemize}
    \item For all kinematic scenarios, our models accurately recover all subhalo interaction parameters and produce well-calibrated posterior distributions (Section~\ref{section:observable:test}).
    Of all interaction parameters, the subhalo radius \rsat is most challenging to constrain at small values, while the impact parameter $b$ shows reduced accuracy at larger values, due to a fundamental physical degeneracy where compact subhalos are nearly indistinguishable from point masses.
    
    \item We apply our framework to a mock AAU stream with interaction parameters from \cite{hilmi24}, which are designed to roughly match the observations in \cite{2021ApJ...911..149L}.
    The 6D and 5D scenarios achieve remarkably similar precision with mass uncertainties of $15-20$\% and velocity constraints of $4-5 \, \kms$, demonstrating that radial velocities may not be essential for accurate subhalo characterization (Section~\ref{section:observable:aau}).
    However, as we show in later sections, spectroscopic data becomes crucial for constraining subhalo velocities under realistic observational conditions with limited stellar samples and measurement uncertainties.
    
    \item The 2D case (coordinates only) shows degraded performance with mass uncertainties of about $30$\% and velocity uncertainties of about $20 \, \kms$, while exhibiting velocity bimodality with peaks corresponding to different signs of the perpendicular velocity, though the distribution is not perfectly symmetric around zero.
    The model correctly identifies the peak corresponding to the true velocity value as the primary solution.
    This reveals fundamental degeneracies that emerge when phase space is inadequately sampled due to insufficient kinematic information.
    
\end{itemize}

We then forecast performance under realistic observational scenarios incorporating measurement uncertainties and finite sample sizes characteristic of current and future surveys.
In Section~\ref{section:uncertainty}, we examine four observational scenarios: present-day observations (96 stars with full 6D coordinates matching \cite{2021ApJ...911..149L}), future LSST photometry (4,096 stars with coordinates only), future 4MOST spectroscopy (396 stars with full 6D coordinates), and a combined LSST and 4MOST strategy (4,096 total stars with mixed observational completeness).
Similarly, we train separate models for the first three scenarios with realistic uncertainty models, while the combined scenario employs joint likelihood analysis using the LSST and 4MOST posteriors.
Our findings are as follows.
\begin{itemize}
    \item Comparing to the idealized cases, the accuracy degrades significantly when incorporating realistic observational uncertainties, particularly for weak interactions involving low masses, high velocities, or large impact parameters, though models maintain well-calibrated posteriors (Section~\ref{section:uncertainty:test}).
    Present-day observations perform worst due to limited sample size and large uncertainties, while the LSST photometric and 4MOST spectroscopic scenarios show comparable performance despite the trade-off between sample size and kinematic completeness.
    The combined LSST+4MOST strategy achieves the best overall performance by leveraging both large statistical samples and detailed kinematic information.
    
    \item For the mock AAU stream under realistic conditions, we report mass uncertainties ranging from approximately $50\%$ (present-day) to $20-40\%$ (future scenarios), with velocity constraints spanning $10-45 \, \kms$ (Section~\ref{section:uncertainty:aau}).
    Similar to the idealized 2D case, we observe velocity bimodality emerging across multiple realistic scenarios (present-day, LSST photometric, and 4MOST spectroscopic).
    All models correctly identify the peak corresponding to the true velocity value as the primary solution.
    We demonstrate that the combined LSST+4MOST approach achieves optimal performance with approximately $20\%$ mass uncertainty and $5-10 \, \kms$ velocity constraints while eliminating velocity bimodality, showcasing the power of combining large photometric samples with targeted spectroscopic follow-up.
\end{itemize}

To conclude, our results demonstrate that NPE provides a powerful and flexible framework for constraining DM subhalo properties from stellar stream observations, with the ability to capture complex posterior distributions while maintaining proper calibration. 
This work establishes that combining complementary datasets—large photometric surveys with targeted spectroscopic follow-up—represents the optimal strategy for future DM studies, providing a robust forecasting framework for maximizing the scientific return of upcoming surveys in probing the nature of DM substructure.

\acknowledgments

This paper has undergone internal review in the LSST Dark Energy Science Collaboration. 

SD and NS initially proposed this project and worked with RP in the early stages and then merged efforts with TN and others. 
SD supervised the first stages of the work and compared SBI results with earlier MCMC results to help guide final decisions about the SBI architecture.	
TN and NS conceived the project, supervised all stages of the research, and wrote and edited the manuscript.	
TN and RP designed the machine learning methodology and trained the models.
DE provided expert advice and supervision on the simulations and comparison to the previous MCMC analysis.
PF provided expert advice on the observational uncertainty modeling and contributed to writing and editing that section of the manuscript.	
ZL, NS, and PSF conducted the final observational uncertainty modeling.
AZ and TS provided scientific input.
AR and MR provided comments as part of the LSST Dark Energy Science Collaboration internal review process.
All authors reviewed and approved the final manuscript.

We thank Matthew Ho for helpful discussions regarding TARP and hyperparameter tuning, and Nathaniel Starkman for providing the spline stream track fitting code.
We also thank Claude-André Faucher-Giguère, Ting S. Li, Robyn E. Sanderson, Arpit Arora, Justine Zeghal for helpful feedback. 

TN is supported by the CIERA Postdoctoral Fellowship.
TN and TS gratefully acknowledge the support of the NSF-Simons AI-Institute for the Sky (SkAI) via grants NSF AST-2421845 and Simons Foundation MPS-AI-00010513. 
TS was supported by NSF through grant AST-2510183 and by NASA through grants 22-ROMAN22-0055 and 22-ROMAN22-0013.
We acknowledge support from the DiRAC Institute in the Department of Astronomy at the University of Washington. The DiRAC Institute is supported through generous gifts from the Charles and Lisa Simonyi Fund for Arts and Sciences, Janet and Lloyd Frink, and the Washington Research Foundation.
The DESC acknowledges ongoing support from the Institut National de 
Physique Nucl\'eaire et de Physique des Particules in France; the 
Science \& Technology Facilities Council in the United Kingdom; and the
Department of Energy and the LSST Discovery Alliance
in the United States.  DESC uses resources of the IN2P3 
Computing Center (CC-IN2P3--Lyon/Villeurbanne - France) funded by the 
Centre National de la Recherche Scientifique; the National Energy 
Research Scientific Computing Center, a DOE Office of Science User 
Facility supported by the Office of Science of the U.S.\ Department of
Energy under Contract No.\ DE-AC02-05CH11231; STFC DiRAC HPC Facilities, 
funded by UK BEIS National E-infrastructure capital grants; and the UK 
particle physics grid, supported by the GridPP Collaboration.  This 
work was performed in part under DOE Contract DE-AC02-76SF00515.

This research makes use of the following packages:
\texttt{corner}~\citep{PER-GRA:2007}, 
\texttt{IPython}~\citep{PER-GRA:2007}, 
\texttt{Jupyter}~\citep{2016ppap.book...87K},
\texttt{Matplotlib}~\citep{2007CSE.....9...90H},
\texttt{NumPy}~\citep{harris2020array},
\texttt{Optuna}~\citep{2019arXiv190710902A},
\texttt{PyGaia}~\citep{pygaia},
\texttt{PyTorch}~\citep{2019arXiv191201703P}, 
\texttt{PyTorch Lightning}~\citep{william_falcon_2020_3828935},
\texttt{SciPy}~\citep{2020SciPy-NMeth},
\texttt{ugali}~\citep{ugali},
\texttt{zuko}~\citep{2023zndo...7625672R}.

\appendix

\appendix
\section{Machine learning architecture}
\label{app:architecture}

As briefly described in Section~\ref{section:ml}, our architecture consists of a transformer-based feature extractor followed by a normalizing flow composed of multiple rational quadratic spline flows transformations.
The input bin stream features and time information (representing arc length along the stream) are independently projected to $32$-dimensional embedding spaces and then concatenated.
The transformer encoder comprises $4$ layers, each with $4$ attention heads and a feedforward dimension of $256$.
We experimented with alternative encoding schemes for the time information, including sinusoidal positional encodings as described in \cite{2017arXiv170603762V}, but found similar performance with the simple linear projection.

The transformer output is then passed through a multi-layer perceptron with $3$ hidden layers of width $256$, incorporating ReLU activation functions, dropout regularization ($p=0.1$), and batch normalization.
The processed features are subsequently transformed through a normalizing flow architecture consisting of $5$ rational quadratic spline flow transformations, each utilizing $6$ hidden layers of dimension $256$ and $15$ spline bins per transformation.
In total, this architecture contains approximately $2.7\mathrm{M}$ trainable parameters.

As described in Section~\ref{section:ml}, we use \texttt{Optuna} to perform hyperparameter tuning on an independent dataset of $8,000$ streams.
We find that tuning solely based on the negative log-likelihood loss (Equation~\ref{eq:loss_nll}) sometimes yields accurate point estimates but produces poorly calibrated posterior distributions.
To achieve both accurate and well-calibrated posterior distributions, we incorporate a ``calibration metric'' into our hyperparameter optimization.
Specifically, after each tuning run, we calculate the mean squared error (MSE) of the TARP curve deviations from the diagonal (see Section~\ref{section:observable:test}) and use this as an additional optimization metric.
Note that we do not train directly on this TARP calibration loss.
We select the hyperparameter configuration that provides the best trade-off between low negative log-likelihood loss and good calibration performance, effectively choosing from the Pareto frontier (configurations where improving one metric would worsen the other).

\section{Modeling mock AAU observations}
\label{app:uncertainty}

\begin{figure}
    \centering
    \includegraphics[width=0.7\linewidth]{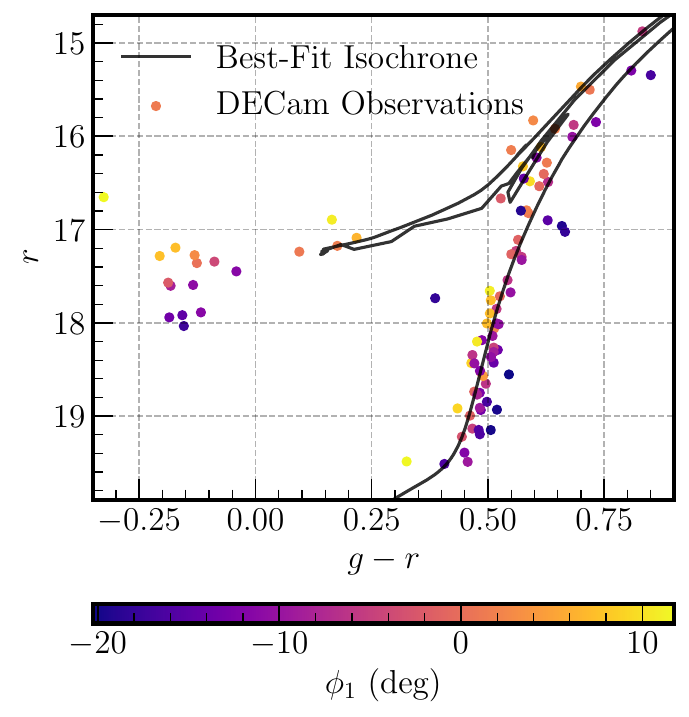} 
    \caption{Color-magnitude diagram of the 96 AAU stream member stars.
    Member stars are color-coded by their $\phi_1$ coordinates.
    The black line shows the best-fit MIST isochrone, with parameters adopted from \cite{2021ApJ...911..149L}: $[\mathrm{Fe/H}] = -1.99$, $[\alpha/\mathrm{Fe}] = 0.4$, $Y = 0.4$, $\mathrm{Age} = 11.5 \, \mathrm{Gyr}$, and distance modulus $\mu=16.8$. Note, the observed spread in stars from the isochrone is due to the distance gradient along the stream.}
    \label{fig:aau_cmd}
\end{figure}

Figure~\ref{fig:aau_cmd} shows the color-magnitude diagram of the 96 AAU member stars, along with the best-fit isochrone at $m-M$ = 16.8 \citep{2021ApJ...911..149L}.

\bibliography{references}

@ARTICLE{2025arXiv250915307W,
       author = {{Weatherford}, Newlin C. and {Bonaca}, Ana},
        title = "{Kinematics of Stellar Streams from Globular Clusters Depend on Black Hole Retention and Star Mass: A Selection Effect for Dark Matter Inference}",
      journal = {arXiv e-prints},
         year = 2025,
        month = sep,
          eid = {arXiv:2509.15307},
        pages = {arXiv:2509.15307},
          doi = {10.48550/arXiv.2509.15307},
archivePrefix = {arXiv},
       eprint = {2509.15307},
 primaryClass = {astro-ph.GA},
       adsurl = {https://ui.adsabs.harvard.edu/abs/2025arXiv250915307W}
}

@ARTICLE{2024MNRAS.532.2657B,
       author = {{Brooks}, Richard A.~N. and {Sanders}, Jason L. and {Lilleengen}, Sophia and {Petersen}, Michael S. and {Pontzen}, Andrew},
        title = "{Action and energy clustering of stellar streams in deforming Milky Way dark matter haloes}",
      journal = {\mnras},
         year = 2024,
        month = aug,
       volume = {532},
       number = {2},
        pages = {2657-2673},
          doi = {10.1093/mnras/stae1565},
archivePrefix = {arXiv},
       eprint = {2401.11990},
 primaryClass = {astro-ph.GA},
       adsurl = {https://ui.adsabs.harvard.edu/abs/2024MNRAS.532.2657B}
}

@ARTICLE{2025arXiv250514792W,
       author = {{Weerasooriya}, Sachi and {Starkenburg}, Tjitske and {Cunningham}, Emily C. and {Johnston}, Kathryn V},
        title = "{Dancing Streams In Merging Halos: Stellar Streams in a MW--LMC-like merger}",
      journal = {arXiv e-prints},
         year = 2025,
        month = may,
          eid = {arXiv:2505.14792},
        pages = {arXiv:2505.14792},
          doi = {10.48550/arXiv.2505.14792},
archivePrefix = {arXiv},
       eprint = {2505.14792},
 primaryClass = {astro-ph.GA},
       adsurl = {https://ui.adsabs.harvard.edu/abs/2025arXiv250514792W}
}

@ARTICLE{2025arXiv250903599P,
       author = {{Panithanpaisal}, Nondh and {Sanderson}, Robyn E. and {Rodriguez}, Carl L. and {Starkenburg}, Tjitske and {Pearson}, Sarah and {Bonaca}, Ana and {Hopkins}, Philip F. and {Cook}, Brian T. and {Arora}, Arpit and {Weatherford}, Newlin C.},
        title = "{Breaking Down the $\textsf{CosmoGEMS}$: Toward Modeling and Understanding Globular Cluster Stellar Streams in a Fully Cosmological Context}",
      journal = {arXiv e-prints},
         year = 2025,
        month = sep,
          eid = {arXiv:2509.03599},
        pages = {arXiv:2509.03599},
          doi = {10.48550/arXiv.2509.03599},
archivePrefix = {arXiv},
       eprint = {2509.03599},
 primaryClass = {astro-ph.GA},
       adsurl = {https://ui.adsabs.harvard.edu/abs/2025arXiv250903599P}
}

@ARTICLE{2021MNRAS.507.4826L,
       author = {{Lovell}, Mark R. and {Cautun}, Marius and {Frenk}, Carlos S. and {Hellwing}, Wojciech A. and {Newton}, Oliver},
        title = "{The spatial distribution of Milky Way satellites, gaps in streams, and the nature of dark matter}",
      journal = {\mnras},
         year = 2021,
        month = nov,
       volume = {507},
       number = {4},
        pages = {4826-4839},
          doi = {10.1093/mnras/stab2452},
archivePrefix = {arXiv},
       eprint = {2104.03322},
 primaryClass = {astro-ph.GA},
       adsurl = {https://ui.adsabs.harvard.edu/abs/2021MNRAS.507.4826L}
}

@ARTICLE{1997ApJ...490..493N,
       author = {{Navarro}, Julio F. and {Frenk}, Carlos S. and {White}, Simon D.~M.},
        title = "{A Universal Density Profile from Hierarchical Clustering}",
      journal = {\apj},
         year = 1997,
        month = dec,
       volume = {490},
       number = {2},
        pages = {493-508},
          doi = {10.1086/304888},
archivePrefix = {arXiv},
       eprint = {astro-ph/9611107},
 primaryClass = {astro-ph},
       adsurl = {https://ui.adsabs.harvard.edu/abs/1997ApJ...490..493N}
}

@ARTICLE{2018PhR...730....1T,
       author = {{Tulin}, Sean and {Yu}, Hai-Bo},
        title = "{Dark matter self-interactions and small scale structure}",
      journal = {\physrep},
         year = 2018,
        month = feb,
       volume = {730},
        pages = {1-57},
          doi = {10.1016/j.physrep.2017.11.004},
archivePrefix = {arXiv},
       eprint = {1705.02358},
 primaryClass = {hep-ph},
       adsurl = {https://ui.adsabs.harvard.edu/abs/2018PhR...730....1T}
}

@ARTICLE{2021MNRAS.506.4421S,
       author = {{Shen}, Xuejian and {Hopkins}, Philip F. and {Necib}, Lina and {Jiang}, Fangzhou and {Boylan-Kolchin}, Michael and {Wetzel}, Andrew},
        title = "{Dissipative dark matter on FIRE - I. Structural and kinematic properties of dwarf galaxies}",
      journal = {\mnras},
         year = 2021,
        month = sep,
       volume = {506},
       number = {3},
        pages = {4421-4445},
          doi = {10.1093/mnras/stab2042},
archivePrefix = {arXiv},
       eprint = {2102.09580},
 primaryClass = {astro-ph.GA},
       adsurl = {https://ui.adsabs.harvard.edu/abs/2021MNRAS.506.4421S}
}

@ARTICLE{2017MNRAS.466.4974M,
       author = {{Molin{\'e}}, {\'A}ngeles and {S{\'a}nchez-Conde}, Miguel A. and {Palomares-Ruiz}, Sergio and {Prada}, Francisco},
        title = "{Characterization of subhalo structural properties and implications for dark matter annihilation signals}",
      journal = {\mnras},
         year = 2017,
        month = apr,
       volume = {466},
       number = {4},
        pages = {4974-4990},
          doi = {10.1093/mnras/stx026},
archivePrefix = {arXiv},
       eprint = {1603.04057},
 primaryClass = {astro-ph.CO},
       adsurl = {https://ui.adsabs.harvard.edu/abs/2017MNRAS.466.4974M}
}

@ARTICLE{2015MNRAS.446.1000F,
       author = {{Feldmann}, Robert and {Spolyar}, Douglas},
        title = "{Detecting dark matter substructures around the Milky Way with Gaia}",
      journal = {\mnras},
         year = 2015,
        month = jan,
       volume = {446},
       number = {1},
        pages = {1000-1012},
          doi = {10.1093/mnras/stu2147},
archivePrefix = {arXiv},
       eprint = {1310.2243},
 primaryClass = {astro-ph.GA},
       adsurl = {https://ui.adsabs.harvard.edu/abs/2015MNRAS.446.1000F}
}

@ARTICLE{2004MNRAS.353..189V,
       author = {{Vale}, A. and {Ostriker}, J.~P.},
        title = "{Linking halo mass to galaxy luminosity}",
      journal = {\mnras},
         year = 2004,
        month = sep,
       volume = {353},
       number = {1},
        pages = {189-200},
          doi = {10.1111/j.1365-2966.2004.08059.x},
archivePrefix = {arXiv},
       eprint = {astro-ph/0402500},
 primaryClass = {astro-ph},
       adsurl = {https://ui.adsabs.harvard.edu/abs/2004MNRAS.353..189V}
}

@ARTICLE{2006ApJ...647..201C,
       author = {{Conroy}, Charlie and {Wechsler}, Risa H. and {Kravtsov}, Andrey V.},
        title = "{Modeling Luminosity-dependent Galaxy Clustering through Cosmic Time}",
      journal = {\apj},
         year = 2006,
        month = aug,
       volume = {647},
       number = {1},
        pages = {201-214},
          doi = {10.1086/503602},
archivePrefix = {arXiv},
       eprint = {astro-ph/0512234},
 primaryClass = {astro-ph},
       adsurl = {https://ui.adsabs.harvard.edu/abs/2006ApJ...647..201C}
}

@ARTICLE{2004ApJ...609...35K,
       author = {{Kravtsov}, Andrey V. and {Berlind}, Andreas A. and {Wechsler}, Risa H. and {Klypin}, Anatoly A. and {Gottl{\"o}ber}, Stefan and {Allgood}, Brandon and {Primack}, Joel R.},
        title = "{The Dark Side of the Halo Occupation Distribution}",
      journal = {\apj},
         year = 2004,
        month = jul,
       volume = {609},
       number = {1},
        pages = {35-49},
          doi = {10.1086/420959},
archivePrefix = {arXiv},
       eprint = {astro-ph/0308519},
 primaryClass = {astro-ph},
       adsurl = {https://ui.adsabs.harvard.edu/abs/2004ApJ...609...35K}
}

@ARTICLE{2023arXiv231101588R,
       author = {{Roncoli}, Andrea and {{\'C}iprijanovi{\'c}}, Aleksandra and {Voetberg}, Maggie and {Villaescusa-Navarro}, Francisco and {Nord}, Brian},
        title = "{Domain Adaptive Graph Neural Networks for Constraining Cosmological Parameters Across Multiple Data Sets}",
      journal = {arXiv e-prints},
         year = 2023,
        month = nov,
          eid = {arXiv:2311.01588},
        pages = {arXiv:2311.01588},
          doi = {10.48550/arXiv.2311.01588},
archivePrefix = {arXiv},
       eprint = {2311.01588},
 primaryClass = {astro-ph.CO},
       adsurl = {https://ui.adsabs.harvard.edu/abs/2023arXiv231101588R}
}

@ARTICLE{2023MLS&T...4b5013C,
       author = {{{\'C}iprijanovi{\'c}}, A. and {Lewis}, A. and {Pedro}, K. and {Madireddy}, S. and {Nord}, B. and {Perdue}, G.~N. and {Wild}, S.~M.},
        title = "{DeepAstroUDA: semi-supervised universal domain adaptation for cross-survey galaxy morphology classification and anomaly detection}",
      journal = {Machine Learning: Science and Technology},
         year = 2023,
        month = jun,
       volume = {4},
       number = {2},
          eid = {025013},
        pages = {025013},
          doi = {10.1088/2632-2153/acca5f},
archivePrefix = {arXiv},
       eprint = {2302.02005},
 primaryClass = {astro-ph.GA},
       adsurl = {https://ui.adsabs.harvard.edu/abs/2023MLS&T...4b5013C}
}

@ARTICLE{2025MLS&T...6c5032P,
       author = {{Pandya}, Sneh and {Patel}, Purvik and {Nord}, Brian D. and {Walmsley}, Mike and {{\'C}iprijanovi{\'c}}, Aleksandra},
        title = "{SIDDA: SInkhorn Dynamic Domain Adaptation for image classification with equivariant neural networks}",
      journal = {Machine Learning: Science and Technology},
         year = 2025,
        month = sep,
       volume = {6},
       number = {3},
          eid = {035032},
        pages = {035032},
          doi = {10.1088/2632-2153/adf701},
archivePrefix = {arXiv},
       eprint = {2501.14048},
 primaryClass = {stat.ML},
       adsurl = {https://ui.adsabs.harvard.edu/abs/2025MLS&T...6c5032P}
}

@ARTICLE{2023MNRAS.524.4711M,
       author = {{Mason}, Charlotte A. and {Mu{\~n}oz}, Julian B. and {Greig}, Bradley and {Mesinger}, Andrei and {Park}, Jaehong},
        title = "{21CMFISH: Fisher-matrix framework for fast parameter forecasts from the cosmic 21-cm signal}",
      journal = {\mnras},
         year = 2023,
        month = sep,
       volume = {524},
       number = {3},
        pages = {4711-4728},
          doi = {10.1093/mnras/stad2145},
archivePrefix = {arXiv},
       eprint = {2212.09797},
 primaryClass = {astro-ph.CO},
       adsurl = {https://ui.adsabs.harvard.edu/abs/2023MNRAS.524.4711M}
}

@ARTICLE{2023arXiv230508994C,
       author = {{Coulton}, William R. and {Wandelt}, Benjamin D.},
        title = "{How to estimate Fisher information matrices from simulations}",
      journal = {arXiv e-prints},
         year = 2023,
        month = may,
          eid = {arXiv:2305.08994},
        pages = {arXiv:2305.08994},
          doi = {10.48550/arXiv.2305.08994},
archivePrefix = {arXiv},
       eprint = {2305.08994},
 primaryClass = {stat.ME},
       adsurl = {https://ui.adsabs.harvard.edu/abs/2023arXiv230508994C}
}

@ARTICLE{2022ApJ...932..102W,
       author = {{Wang}, Ziming and {Liu}, Chang and {Zhao}, Junjie and {Shao}, Lijing},
        title = "{Extending the Fisher Information Matrix in Gravitational-wave Data Analysis}",
      journal = {\apj},
         year = 2022,
        month = jun,
       volume = {932},
       number = {2},
          eid = {102},
        pages = {102},
          doi = {10.3847/1538-4357/ac6b99},
archivePrefix = {arXiv},
       eprint = {2203.02670},
 primaryClass = {gr-qc},
       adsurl = {https://ui.adsabs.harvard.edu/abs/2022ApJ...932..102W}
}

@ARTICLE{RTN:011,
    author = {{Guy}, L.P. and {Bechtol}, K. and {Bellm}, E. and {Blum}, B},
    title = "{Rubin Observatory Plans for an Early Science Program}",
    year = 2025,
    month = jun,
    journal={},
}

@ARTICLE{2019MNRAS.490.3508L,
       author = {{Li}, T.~S. and {Koposov}, S.~E. and {Zucker}, D.~B. and {Lewis}, G.~F. and {Kuehn}, K. and {Simpson}, J.~D. and {Ji}, A.~P. and {Shipp}, N. and {Mao}, Y. -Y. and {Geha}, M. and {Pace}, A.~B. and {Mackey}, A.~D. and {Allam}, S. and {Tucker}, D.~L. and {Da Costa}, G.~S. and {Erkal}, D. and {Simon}, J.~D. and {Mould}, J.~R. and {Martell}, S.~L. and {Wan}, Z. and {De Silva}, G.~M. and {Bechtol}, K. and {Balbinot}, E. and {Belokurov}, V. and {Bland-Hawthorn}, J. and {Casey}, A.~R. and {Cullinane}, L. and {Drlica-Wagner}, A. and {Sharma}, S. and {Vivas}, A.~K. and {Wechsler}, R.~H. and {Yanny}, B. and {S5 Collaboration}},
        title = "{The southern stellar stream spectroscopic survey (S$^{5}$): Overview, target selection, data reduction, validation, and early science}",
      journal = {\mnras},
         year = 2019,
        month = dec,
       volume = {490},
       number = {3},
        pages = {3508-3531},
          doi = {10.1093/mnras/stz2731},
archivePrefix = {arXiv},
       eprint = {1907.09481},
 primaryClass = {astro-ph.GA},
       adsurl = {https://ui.adsabs.harvard.edu/abs/2019MNRAS.490.3508L}
}

@ARTICLE{2020ApJ...901...23H,
       author = {{Hansen}, T.~T. and {Riley}, A.~H. and {Strigari}, L.~E. and {Marshall}, J.~L. and {Ferguson}, P.~S. and {Zepeda}, J. and {Sneden}, C.},
        title = "{A Chemo-dynamical Link between the Gj{\"o}ll Stream and NGC 3201}",
      journal = {\apj},
         year = 2020,
        month = sep,
       volume = {901},
       number = {1},
          eid = {23},
        pages = {23},
          doi = {10.3847/1538-4357/ababa5},
archivePrefix = {arXiv},
       eprint = {2007.12165},
 primaryClass = {astro-ph.SR},
       adsurl = {https://ui.adsabs.harvard.edu/abs/2020ApJ...901...23H}
}

@ARTICLE{2025MNRAS.542..560B,
       author = {{Bystr{\"o}m}, Amanda and {Koposov}, Sergey E. and {Lilleengen}, Sophia and {Li}, Ting S. and {Bell}, Eric and {Beraldo e Silva}, Leandro and {Carrillo}, Andreia and {Chandra}, Vedant and {Gnedin}, Oleg Y. and {Han}, Jiwon Jesse and {Medina}, Gustavo E. and {Najita}, Joan and {Riley}, Alexander H. and {Thomas}, Guillaume and {Valluri}, Monica and {Aguilar}, Jessica N. and {Ahlen}, Steven and {Prieto}, Carlos Allende and {Brooks}, David and {Claybaugh}, Todd and {Cole}, Shaun and {Dawson}, Kyle and {de la Macorra}, Axel and {Font-Ribera}, Andreu and {Forero-Romero}, Jaime E. and {Gazta{\~n}aga}, Enrique and {Gontcho}, Satya Gontcho A. and {Kremin}, Anthony and {Lambert}, Andrew and {Landriau}, Martin and {Le Guillou}, Laurent and {Levi}, Michael E. and {Meisner}, Aaron and {Miquel}, Ramon and {Moustakas}, John and {Prada}, Francisco and {P{\'e}rez-R{\`a}fols}, Ignasi and {Rossi}, Graziano and {Sanchez}, Eusebio and {Schlegel}, David and {Schubnell}, Michael and {Sprayberry}, David and {Tarl{\'e}}, Gregory and {Weaver}, Benjamin A. and {Zou}, Hu},
        title = "{Exploring the interaction between the MW and LMC with a large sample of blue horizontal branch stars from the DESI survey}",
      journal = {\mnras},
         year = 2025,
        month = sep,
       volume = {542},
       number = {2},
        pages = {560-582},
          doi = {10.1093/mnras/staf1219},
archivePrefix = {arXiv},
       eprint = {2410.09149},
 primaryClass = {astro-ph.GA},
       adsurl = {https://ui.adsabs.harvard.edu/abs/2025MNRAS.542..560B}
}

@ARTICLE{2025arXiv250402924M,
       author = {{Medina}, Gustavo E. and {Li}, Ting S. and {Koposov}, Sergey E. and {Riley}, A.~H. and {Beraldo e Silva}, L. and {Valluri}, M. and {Wang}, W. and {Bystr{\"o}m}, A. and {Gnedin}, O.~Y. and {Carlberg}, R.~G. and {Kizhuprakkat}, N. and {Weaver}, B.~A. and {Aguilar}, J. and {Ahlen}, S. and {Bianchi}, D. and {Brooks}, D. and {Claybaugh}, T. and {Cooper}, A.~P. and {de la Macorra}, A. and {Dey}, A. and {Doel}, P. and {Font-Ribera}, A. and {Forero-Romero}, J.~E. and {Gazta{\~n}aga}, E. and {Gontcho}, S. Gontcho A and {Gutierrez}, G. and {Guy}, J. and {Honscheid}, K. and {Ishak}, M. and {Kisner}, T. and {Landriau}, M. and {Le Guillou}, L. and {Meisner}, A. and {Miquel}, R. and {Myers}, A.~D. and {Nadathur}, S. and {Poppett}, C. and {Prada}, F. and {P{\'e}rez-R{\`a}fols}, I. and {Rossi}, G. and {Sanchez}, E. and {Seo}, H. and {Sprayberry}, D. and {Tarl{\'e}}, G. and {Wechsler}, R.~H. and {Zhou}, R. and {Zou}, H.},
        title = "{The DESI Y1 RR Lyrae catalog I: Empirical modeling of the cyclic variation of spectroscopic properties and a chemodynamical analysis of the outer halo}",
      journal = {arXiv e-prints},
         year = 2025,
        month = apr,
          eid = {arXiv:2504.02924},
        pages = {arXiv:2504.02924},
          doi = {10.48550/arXiv.2504.02924},
archivePrefix = {arXiv},
       eprint = {2504.02924},
 primaryClass = {astro-ph.GA},
       adsurl = {https://ui.adsabs.harvard.edu/abs/2025arXiv250402924M}
}

@ARTICLE{2019ApJ...884...51G,
       author = {{Garavito-Camargo}, Nicolas and {Besla}, Gurtina and {Laporte}, Chervin F.~P. and {Johnston}, Kathryn V. and {G{\'o}mez}, Facundo A. and {Watkins}, Laura L.},
        title = "{Hunting for the Dark Matter Wake Induced by the Large Magellanic Cloud}",
      journal = {\apj},
         year = 2019,
        month = oct,
       volume = {884},
       number = {1},
          eid = {51},
        pages = {51},
          doi = {10.3847/1538-4357/ab32eb},
archivePrefix = {arXiv},
       eprint = {1902.05089},
 primaryClass = {astro-ph.GA},
       adsurl = {https://ui.adsabs.harvard.edu/abs/2019ApJ...884...51G}
}

@ARTICLE{2020MNRAS.498.5574E,
       author = {{Erkal}, Denis and {Belokurov}, Vasily A. and {Parkin}, Daniel L.},
        title = "{Equilibrium models of the Milky Way mass are biased high by the LMC}",
      journal = {\mnras},
         year = 2020,
        month = nov,
       volume = {498},
       number = {4},
        pages = {5574-5580},
          doi = {10.1093/mnras/staa2840},
archivePrefix = {arXiv},
       eprint = {2001.11030},
 primaryClass = {astro-ph.GA},
       adsurl = {https://ui.adsabs.harvard.edu/abs/2020MNRAS.498.5574E}
}

@ARTICLE{2020MNRAS.494L..11P,
       author = {{Petersen}, Michael S. and {Pe{\~n}arrubia}, Jorge},
        title = "{Reflex motion in the Milky Way stellar halo resulting from the Large Magellanic Cloud infall}",
      journal = {\mnras},
         year = 2020,
        month = may,
       volume = {494},
       number = {1},
        pages = {L11-L16},
          doi = {10.1093/mnrasl/slaa029},
archivePrefix = {arXiv},
       eprint = {2001.09142},
 primaryClass = {astro-ph.GA},
       adsurl = {https://ui.adsabs.harvard.edu/abs/2020MNRAS.494L..11P}
}

@ARTICLE{2021MNRAS.506.2677E,
       author = {{Erkal}, Denis and {Deason}, Alis J. and {Belokurov}, Vasily and {Xue}, Xiang-Xiang and {Koposov}, Sergey E. and {Bird}, Sarah A. and {Liu}, Chao and {Simion}, Iulia T. and {Yang}, Chengqun and {Zhang}, Lan and {Zhao}, Gang},
        title = "{Detection of the LMC-induced sloshing of the Galactic halo}",
      journal = {\mnras},
         year = 2021,
        month = sep,
       volume = {506},
       number = {2},
        pages = {2677-2684},
          doi = {10.1093/mnras/stab1828},
archivePrefix = {arXiv},
       eprint = {2010.13789},
 primaryClass = {astro-ph.GA},
       adsurl = {https://ui.adsabs.harvard.edu/abs/2021MNRAS.506.2677E}
}

@ARTICLE{2021NatAs...5..251P,
       author = {{Petersen}, Michael S. and {Pe{\~n}arrubia}, Jorge},
        title = "{Detection of the Milky Way reflex motion due to the Large Magellanic Cloud infall}",
      journal = {Nature Astronomy},
         year = 2021,
        month = jan,
       volume = {5},
        pages = {251-255},
          doi = {10.1038/s41550-020-01254-3},
archivePrefix = {arXiv},
       eprint = {2011.10581},
 primaryClass = {astro-ph.GA},
       adsurl = {https://ui.adsabs.harvard.edu/abs/2021NatAs...5..251P}
}

@ARTICLE{2025arXiv250301533B,
       author = {{Brown}, Anthony G.~A.},
        title = "{Gaia: Ten Years of Surveying the Milky Way and Beyond}",
      journal = {arXiv e-prints},
         year = 2025,
        month = mar,
          eid = {arXiv:2503.01533},
        pages = {arXiv:2503.01533},
          doi = {10.48550/arXiv.2503.01533},
archivePrefix = {arXiv},
       eprint = {2503.01533},
 primaryClass = {astro-ph.GA},
       adsurl = {https://ui.adsabs.harvard.edu/abs/2025arXiv250301533B}
}

@software{peter_yoachim_2025_15368965,
  author       = {Peter Yoachim and
                  Lynne Jones and
                  Eric H. Neilsen, Jr. and
                  Tiago and
                  John Parejko and
                  Eric Bellm and
                  Rachel Street and
                  Jeff Carlin and
                  Humna and
                  Matthew R. Becker and
                  pgris and
                  erykoff and
                  Loredana Prisinzano and
                  Erik Dennihy and
                  Giovanni A. Gollotti and
                  Jonathan Sick and
                  lmptc and
                  LI and
                  Natasha Abrams and
                  Roberto J. Assef and
                  Leanne Guy and
                  Ross and
                  Katja Bricman and
                  Johan Bregeon and
                  Kian-Tat Lim and
                  Michael Kelley and
                  Igor Andreoni},
  title        = {lsst/rubin\_sim: v2.2.4},
  month        = may,
  year         = 2025,
  publisher    = {Zenodo},
  version      = {v2.2.4},
  doi          = {10.5281/zenodo.15368965},
  url          = {https://doi.org/10.5281/zenodo.15368965},
  swhid        = {swh:1:dir:5af5919756f84d2bf9470ccb114582e3aa92bfe3
                   ;origin=https://doi.org/10.5281/zenodo.7087822;vis
                   it=swh:1:snp:2e8977767b6c389a3a28789861b70c066f73b
                   734;anchor=swh:1:rel:d906fa8a449db09c551889053d5f4
                   9bba84e1646;path=lsst-rubin\_sim-6eda8b6
                  },
}

@ARTICLE{2023ApJ...952L..10W,
       author = {{Wang}, Bingjie and {Leja}, Joel and {Villar}, V. Ashley and {Speagle}, Joshua S.},
        title = "{SBI$^{++}$: Flexible, Ultra-fast Likelihood-free Inference Customized for Astronomical Applications}",
      journal = {\apjl},
         year = 2023,
        month = jul,
       volume = {952},
       number = {1},
          eid = {L10},
        pages = {L10},
          doi = {10.3847/2041-8213/ace361},
archivePrefix = {arXiv},
       eprint = {2304.05281},
 primaryClass = {astro-ph.IM},
       adsurl = {https://ui.adsabs.harvard.edu/abs/2023ApJ...952L..10W}
}

@ARTICLE{2019arXiv191201703P,
       author = {{Paszke}, Adam and {Gross}, Sam and {Massa}, Francisco and {Lerer}, Adam and {Bradbury}, James and {Chanan}, Gregory and {Killeen}, Trevor and {Lin}, Zeming and {Gimelshein}, Natalia and {Antiga}, Luca and {Desmaison}, Alban and {K{\"o}pf}, Andreas and {Yang}, Edward and {DeVito}, Zach and {Raison}, Martin and {Tejani}, Alykhan and {Chilamkurthy}, Sasank and {Steiner}, Benoit and {Fang}, Lu and {Bai}, Junjie and {Chintala}, Soumith},
        title = "{PyTorch: An Imperative Style, High-Performance Deep Learning Library}",
      journal = {arXiv e-prints},
         year = 2019,
        month = dec,
          eid = {arXiv:1912.01703},
        pages = {arXiv:1912.01703},
          doi = {10.48550/arXiv.1912.01703},
archivePrefix = {arXiv},
       eprint = {1912.01703},
 primaryClass = {cs.LG},
       adsurl = {https://ui.adsabs.harvard.edu/abs/2019arXiv191201703P}
}

@ARTICLE{2022AJ....163...18F,
       author = {{Ferguson}, P.~S. and {Shipp}, N. and {Drlica-Wagner}, A. and {Li}, T.~S. and {Cerny}, W. and {Tavangar}, K. and {Pace}, A.~B. and {Marshall}, J.~L. and {Riley}, A.~H. and {Adam{\'o}w}, M. and {Carlin}, J.~L. and {Choi}, Y. and {Erkal}, D. and {James}, D.~J. and {Koposov}, Sergey E. and {Kuropatkin}, N. and {Mart{\'\i}nez-V{\'a}zquez}, C.~E. and {Mau}, S. and {Mutlu-Pakdil}, B. and {Olsen}, K.~A.~G. and {Sakowska}, J.~D. and {Stringfellow}, G.~S. and {Yanny}, B. and {Yanny}, B.},
        title = "{DELVE-ing into the Jet: A Thin Stellar Stream on a Retrograde Orbit at 30 kpc}",
      journal = {\aj},
         year = 2022,
        month = jan,
       volume = {163},
       number = {1},
          eid = {18},
        pages = {18},
          doi = {10.3847/1538-3881/ac3492},
archivePrefix = {arXiv},
       eprint = {2104.11755},
 primaryClass = {astro-ph.GA},
       adsurl = {https://ui.adsabs.harvard.edu/abs/2022AJ....163...18F}
}

@ARTICLE{2007CSE.....9...90H,
       author = {{Hunter}, John D.},
        title = "{Matplotlib: A 2D Graphics Environment}",
      journal = {Computing in Science and Engineering},
         year = 2007,
        month = may,
       volume = {9},
       number = {3},
        pages = {90-95},
          doi = {10.1109/MCSE.2007.55},
       adsurl = {https://ui.adsabs.harvard.edu/abs/2007CSE.....9...90H}
}

@article{harris2020array,
  title={Array programming with NumPy},
  author={Harris, Charles R and Millman, K Jarrod and Van Der Walt, St{\'e}fan J and Gommers, Ralf and Virtanen, Pauli and Cournapeau, David and Wieser, Eric and Taylor, Julian and Berg, Sebastian and Smith, Nathaniel J and others},
  journal={Nature},
  volume={585},
  number={7825},
  pages={357--362},
  year={2020},
  publisher={Nature Publishing Group}
}

@INCOLLECTION{2016ppap.book...87K,
       author = {{Kluyver}, Thomas and {Ragan-Kelley}, Benjain and {P{\'e}rez}, Fernando and {Granger}, Brian and {Bussonnier}, Matthias and {Frederic}, Jonathan and {Kelley}, Kyle and {Hamrick}, Jessica and {Grout}, Jason and {Corlay}, Sylvain and {Ivanov}, Paul and {Avila}, Dami{\'a}n and {Abdalla}, Safia and {Willing}, Carol and {Jupyter Development Team}},
        title = "{Jupyter Notebooks{\textemdash}a publishing format for reproducible computational workflows}",
    booktitle = {IOS Press},
         year = 2016,
        pages = {87-90},
          doi = {10.3233/978-1-61499-649-1-87},
       adsurl = {https://ui.adsabs.harvard.edu/abs/2016ppap.book...87K}
}

@software{2023zndo...7625672R,
       author = {{Rozet}, Fran{\c{c}}ois and {Divo}, Felix and {Schnake}, Simon},
        title = "{probabilists/zuko: Zuko 1.1.0}",
         year = 2024,
        month = jan,
          eid = {10.5281/zenodo.7625672},
          doi = {10.5281/zenodo.7625672},
      version = {1.1.0},
    publisher = {Zenodo},
       adsurl = {https://ui.adsabs.harvard.edu/abs/2023zndo...7625672R}
}

@software{william_falcon_2020_3828935,
	Author = {William Falcon and others},
	Doi = {10.5281/zenodo.3828935},
	Month = may,
	Publisher = {Zenodo},
	Title = {PyTorchLightning/pytorch-lightning: 0.7.6 release},
	Url = {https://doi.org/10.5281/zenodo.3828935},
	Version = {0.7.6},
	Year = 2020
}

@article{2020SciPy-NMeth,
	Adsurl = {https://rdcu.be/b08Wh},
	Author = {{Virtanen}, Pauli and others},
	Doi = {https://doi.org/10.1038/s41592-019-0686-2},
	Journal = {Nature Methods},
	Title = {{SciPy 1.0: Fundamental Algorithms for Scientific Computing in Python}},
	Year = {2020}}

@article{PER-GRA:2007,
	Adsurl = {https://ui.adsabs.harvard.edu/abs/2007CSE.....9c..21P},
	Author = {{Perez}, Fernando and {Granger}, Brian E.},
	Doi = {10.1109/MCSE.2007.53},
	Journal = {Computing in Science and Engineering},
	Month = {Jan},
	Number = {3},
	Pages = {21-29},
	Title = {{IPython: A System for Interactive Scientific Computing}},
	Volume = {9},
	Year = {2007}}

@ARTICLE{2019MNRAS.487.2685E,
       author = {{Erkal}, D. and {Belokurov}, V. and {Laporte}, C.~F.~P. and {Koposov}, S.~E. and {Li}, T.~S. and {Grillmair}, C.~J. and {Kallivayalil}, N. and {Price-Whelan}, A.~M. and {Evans}, N.~W. and {Hawkins}, K. and {Hendel}, D. and {Mateu}, C. and {Navarro}, J.~F. and {del Pino}, A. and {Slater}, C.~T. and {Sohn}, S.~T. and {Orphan Aspen Treasury Collaboration}},
        title = "{The total mass of the Large Magellanic Cloud from its perturbation on the Orphan stream}",
      journal = {\mnras},
         year = 2019,
        month = aug,
       volume = {487},
       number = {2},
        pages = {2685-2700},
          doi = {10.1093/mnras/stz1371},
archivePrefix = {arXiv},
       eprint = {1812.08192},
 primaryClass = {astro-ph.GA},
       adsurl = {https://ui.adsabs.harvard.edu/abs/2019MNRAS.487.2685E}
}

@ARTICLE{1996ApJ...462..563N,
       author = {{Navarro}, Julio F. and {Frenk}, Carlos S. and {White}, Simon D.~M.},
        title = "{The Structure of Cold Dark Matter Halos}",
      journal = {\apj},
         year = 1996,
        month = may,
       volume = {462},
        pages = {563},
          doi = {10.1086/177173},
archivePrefix = {arXiv},
       eprint = {astro-ph/9508025},
 primaryClass = {astro-ph},
       adsurl = {https://ui.adsabs.harvard.edu/abs/1996ApJ...462..563N}
}

@ARTICLE{1990ApJ...356..359H,
       author = {{Hernquist}, Lars},
        title = "{An Analytical Model for Spherical Galaxies and Bulges}",
      journal = {\apj},
         year = 1990,
        month = jun,
       volume = {356},
        pages = {359},
          doi = {10.1086/168845},
       adsurl = {https://ui.adsabs.harvard.edu/abs/1990ApJ...356..359H}
}

@ARTICLE{2017MNRAS.465...76M,
       author = {{McMillan}, Paul J.},
        title = "{The mass distribution and gravitational potential of the Milky Way}",
      journal = {\mnras},
         year = 2017,
        month = feb,
       volume = {465},
       number = {1},
        pages = {76-94},
          doi = {10.1093/mnras/stw2759},
archivePrefix = {arXiv},
       eprint = {1608.00971},
 primaryClass = {astro-ph.GA},
       adsurl = {https://ui.adsabs.harvard.edu/abs/2017MNRAS.465...76M}
}

@ARTICLE{2021ApJ...923..149S,
       author = {{Shipp}, Nora and {Erkal}, Denis and {Drlica-Wagner}, Alex and {Li}, Ting S. and {Pace}, Andrew B. and {Koposov}, Sergey E. and {Cullinane}, Lara R. and {Da Costa}, Gary S. and {Ji}, Alexander P. and {Kuehn}, Kyler and {Lewis}, Geraint F. and {Mackey}, Dougal and {Simpson}, Jeffrey D. and {Wan}, Zhen and {Zucker}, Daniel B. and {Bland-Hawthorn}, Joss and {Ferguson}, Peter S. and {Lilleengen}, Sophia and {Lilleengen}, Sophia},
        title = "{Measuring the Mass of the Large Magellanic Cloud with Stellar Streams Observed by S $^{5}$}",
      journal = {\apj},
         year = 2021,
        month = dec,
       volume = {923},
       number = {2},
          eid = {149},
        pages = {149},
          doi = {10.3847/1538-4357/ac2e93},
archivePrefix = {arXiv},
       eprint = {2107.13004},
 primaryClass = {astro-ph.GA},
       adsurl = {https://ui.adsabs.harvard.edu/abs/2021ApJ...923..149S}
}

@ARTICLE{1911MNRAS..71..460P,
       author = {{Plummer}, H.~C.},
        title = "{On the problem of distribution in globular star clusters}",
      journal = {\mnras},
         year = 1911,
        month = mar,
       volume = {71},
        pages = {460-470},
          doi = {10.1093/mnras/71.5.460},
       adsurl = {https://ui.adsabs.harvard.edu/abs/1911MNRAS..71..460P}
}

@ARTICLE{2019arXiv190710902A,
       author = {{Akiba}, Takuya and {Sano}, Shotaro and {Yanase}, Toshihiko and {Ohta}, Takeru and {Koyama}, Masanori},
        title = "{Optuna: A Next-generation Hyperparameter Optimization Framework}",
      journal = {arXiv e-prints},
         year = 2019,
        month = jul,
          eid = {arXiv:1907.10902},
        pages = {arXiv:1907.10902},
          doi = {10.48550/arXiv.1907.10902},
archivePrefix = {arXiv},
       eprint = {1907.10902},
 primaryClass = {cs.LG},
       adsurl = {https://ui.adsabs.harvard.edu/abs/2019arXiv190710902A}
}

@ARTICLE{2005AJ....130..873J,
       author = {{Jester}, Sebastian and {Schneider}, Donald P. and {Richards}, Gordon T. and {Green}, Richard F. and {Schmidt}, Maarten and {Hall}, Patrick B. and {Strauss}, Michael A. and {Vanden Berk}, Daniel E. and {Stoughton}, Chris and {Gunn}, James E. and {Brinkmann}, Jon and {Kent}, Stephen M. and {Smith}, J. Allyn and {Tucker}, Douglas L. and {Yanny}, Brian},
        title = "{The Sloan Digital Sky Survey View of the Palomar-Green Bright Quasar Survey}",
      journal = {\aj},
         year = 2005,
        month = sep,
       volume = {130},
       number = {3},
        pages = {873-895},
          doi = {10.1086/432466},
archivePrefix = {arXiv},
       eprint = {astro-ph/0506022},
 primaryClass = {astro-ph},
       adsurl = {https://ui.adsabs.harvard.edu/abs/2005AJ....130..873J}
}

@ARTICLE{2025ApJ...987...96M,
       author = {{Ma}, Peter Xiangyuan and {Rogers}, Keir K. and {Li}, Ting S. and {Hlo{\v{z}}ek}, Ren{\'e}e and {Webb}, Jeremy J. and {Huang}, Ruth and {Meunier}, Julian},
        title = "{Toward Characterizing Dark Matter Subhalo Perturbations in Stellar Streams with Graph Neural Networks}",
      journal = {\apj},
         year = 2025,
        month = jul,
       volume = {987},
       number = {1},
          eid = {96},
        pages = {96},
          doi = {10.3847/1538-4357/add698},
       adsurl = {https://ui.adsabs.harvard.edu/abs/2025ApJ...987...96M}
}

@ARTICLE{2016arXiv160803983L,
       author = {{Loshchilov}, Ilya and {Hutter}, Frank},
        title = "{SGDR: Stochastic Gradient Descent with Warm Restarts}",
      journal = {arXiv e-prints},
         year = 2016,
        month = aug,
          eid = {arXiv:1608.03983},
        pages = {arXiv:1608.03983},
          doi = {10.48550/arXiv.1608.03983},
archivePrefix = {arXiv},
       eprint = {1608.03983},
 primaryClass = {cs.LG},
       adsurl = {https://ui.adsabs.harvard.edu/abs/2016arXiv160803983L}
}

@inproceedings{adamw2019,
	Author = {Ilya Loshchilov and Frank Hutter},
	Booktitle = {International Conference on Learning Representations},
	Title = {Decoupled Weight Decay Regularization},
	Url = {https://openreview.net/forum?id=Bkg6RiCqY7},
	Year = {2019}}

@article{kingma2014adam,
	Author = {Kingma, Diederik P and Ba, Jimmy},
	Journal = {arXiv preprint arXiv:1412.6980},
	Title = {Adam: A method for stochastic optimization},
	Year = {2014}}

@ARTICLE{gnn2,
       author = {{Nguyen}, Tri and {Read}, Justin and {Necib}, Lina and {Mishra-Sharma}, Siddharth and {Faucher-Gigu{\`e}re}, Claude-Andr{\'e} and {Wetzel}, Andrew and {Starkenburg}, Tjitske K.},
        title = "{Trial by FIRE: Probing the dark matter density profile of dwarf galaxies with GraphNPE}",
      journal = {\mnras},
         year = 2025,
        month = jul,
          doi = {10.1093/mnras/staf1118},
archivePrefix = {arXiv},
       eprint = {2503.03812},
 primaryClass = {astro-ph.GA},
       adsurl = {https://ui.adsabs.harvard.edu/abs/2025MNRAS.tmp.1080N}
}

@ARTICLE{2023PMLR..20219256L,
       author = {{Lemos}, Pablo and {Coogan}, Adam and {Hezaveh}, Yashar and {Perreault-Levasseur}, Laurence},
        title = "{Sampling-Based Accuracy Testing of Posterior Estimators for General Inference}",
      journal = {40th International Conference on Machine Learning},
         year = 2023,
        month = jan,
       volume = {202},
        pages = {19256-19273},
          doi = {10.48550/arXiv.2302.03026},
archivePrefix = {arXiv},
       eprint = {2302.03026},
 primaryClass = {stat.ML},
       adsurl = {https://ui.adsabs.harvard.edu/abs/2023PMLR..20219256L}
}

@ARTICLE{2019Msngr.175....3D,
       author = {{de Jong}, R.~S. and {Agertz}, O. and {Berbel}, A.~A. and {Aird}, J. and {Alexander}, D.~A. and {Amarsi}, A. and {Anders}, F. and {Andrae}, R. and {Ansarinejad}, B. and {Ansorge}, W. and {Antilogus}, P. and {Anwand-Heerwart}, H. and {Arentsen}, A. and {Arnadottir}, A. and {Asplund}, M. and {Auger}, M. and {Azais}, N. and {Baade}, D. and {Baker}, G. and {Baker}, S. and {Balbinot}, E. and {Baldry}, I.~K. and {Banerji}, M. and {Barden}, S. and {Barklem}, P. and {Barth{\'e}l{\'e}my-Mazot}, E. and {Battistini}, C. and {Bauer}, S. and {Bell}, C.~P.~M. and {Bellido-Tirado}, O. and {Bellstedt}, S. and {Belokurov}, V. and {Bensby}, T. and {Bergemann}, M. and {Bestenlehner}, J.~M. and {Bielby}, R. and {Bilicki}, M. and {Blake}, C. and {Bland-Hawthorn}, J. and {Boeche}, C. and {Boland}, W. and {Boller}, T. and {Bongard}, S. and {Bongiorno}, A. and {Bonifacio}, P. and {Boudon}, D. and {Brooks}, D. and {Brown}, M.~J.~I. and {Brown}, R. and {Br{\"u}ggen}, M. and {Brynnel}, J. and {Brzeski}, J. and {Buchert}, T. and {Buschkamp}, P. and {Caffau}, E. and {Caillier}, P. and {Carrick}, J. and {Casagrande}, L. and {Case}, S. and {Casey}, A. and {Cesarini}, I. and {Cescutti}, G. and {Chapuis}, D. and {Chiappini}, C. and {Childress}, M. and {Christlieb}, N. and {Church}, R. and {Cioni}, M. -R.~L. and {Cluver}, M. and {Colless}, M. and {Collett}, T. and {Comparat}, J. and {Cooper}, A. and {Couch}, W. and {Courbin}, F. and {Croom}, S. and {Croton}, D. and {Daguis{\'e}}, E. and {Dalton}, G. and {Davies}, L.~J.~M. and {Davis}, T. and {de Laverny}, P. and {Deason}, A. and {Dionies}, F. and {Disseau}, K. and {Doel}, P. and {D{\"o}scher}, D. and {Driver}, S.~P. and {Dwelly}, T. and {Eckert}, D. and {Edge}, A. and {Edvardsson}, B. and {Youssoufi}, D.~E. and {Elhaddad}, A. and {Enke}, H. and {Erfanianfar}, G. and {Farrell}, T. and {Fechner}, T. and {Feiz}, C. and {Feltzing}, S. and {Ferreras}, I. and {Feuerstein}, D. and {Feuillet}, D. and {Finoguenov}, A. and {Ford}, D. and {Fotopoulou}, S. and {Fouesneau}, M. and {Frenk}, C. and {Frey}, S. and {Gaessler}, W. and {Geier}, S. and {Gentile Fusillo}, N. and {Gerhard}, O. and {Giannantonio}, T. and {Giannone}, D. and {Gibson}, B. and {Gillingham}, P. and {Gonz{\'a}lez-Fern{\'a}ndez}, C. and {Gonzalez-Solares}, E. and {Gottloeber}, S. and {Gould}, A. and {Grebel}, E.~K. and {Gueguen}, A. and {Guiglion}, G. and {Haehnelt}, M. and {Hahn}, T. and {Hansen}, C.~J. and {Hartman}, H. and {Hauptner}, K. and {Hawkins}, K. and {Haynes}, D. and {Haynes}, R. and {Heiter}, U. and {Helmi}, A. and {Aguayo}, C.~H. and {Hewett}, P. and {Hinton}, S. and {Hobbs}, D. and {Hoenig}, S. and {Hofman}, D. and {Hook}, I. and {Hopgood}, J. and {Hopkins}, A. and {Hourihane}, A. and {Howes}, L. and {Howlett}, C. and {Huet}, T. and {Irwin}, M. and {Iwert}, O. and {Jablonka}, P. and {Jahn}, T. and {Jahnke}, K. and {Jarno}, A. and {Jin}, S. and {Jofre}, P. and {Johl}, D. and {Jones}, D. and {J{\"o}nsson}, H. and {Jordan}, C. and {Karovicova}, I. and {Khalatyan}, A. and {Kelz}, A. and {Kennicutt}, R. and {King}, D. and {Kitaura}, F. and {Klar}, J. and {Klauser}, U. and {Kneib}, J. -P. and {Koch}, A. and {Koposov}, S. and {Kordopatis}, G. and {Korn}, A. and {Kosmalski}, J. and {Kotak}, R. and {Kovalev}, M. and {Kreckel}, K. and {Kripak}, Y. and {Krumpe}, M. and {Kuijken}, K. and {Kunder}, A. and {Kushniruk}, I. and {Lam}, M.~I. and {Lamer}, G. and {Laurent}, F. and {Lawrence}, J. and {Lehmitz}, M. and {Lemasle}, B. and {Lewis}, J. and {Li}, B. and {Lidman}, C. and {Lind}, K. and {Liske}, J. and {Lizon}, J. -L. and {Loveday}, J. and {Ludwig}, H. -G. and {McDermid}, R.~M. and {Maguire}, K. and {Mainieri}, V. and {Mali}, S. and {Mandel}, H.},
        title = "{4MOST: Project overview and information for the First Call for Proposals}",
      journal = {The Messenger},
         year = 2019,
        month = mar,
       volume = {175},
        pages = {3-11},
          doi = {10.18727/0722-6691/5117},
archivePrefix = {arXiv},
       eprint = {1903.02464},
 primaryClass = {astro-ph.IM},
       adsurl = {https://ui.adsabs.harvard.edu/abs/2019Msngr.175....3D}
}

@ARTICLE{2019ApJ...873..111I,
       author = {{Ivezi{\'c}}, {\v{Z}}eljko and {Kahn}, Steven M. and {Tyson}, J. Anthony and {Abel}, Bob and {Acosta}, Emily and {Allsman}, Robyn and {Alonso}, David and {AlSayyad}, Yusra and {Anderson}, Scott F. and {Andrew}, John and {Angel}, James Roger P. and {Angeli}, George Z. and {Ansari}, Reza and {Antilogus}, Pierre and {Araujo}, Constanza and {Armstrong}, Robert and {Arndt}, Kirk T. and {Astier}, Pierre and {Aubourg}, {\'E}ric and {Auza}, Nicole and {Axelrod}, Tim S. and {Bard}, Deborah J. and {Barr}, Jeff D. and {Barrau}, Aurelian and {Bartlett}, James G. and {Bauer}, Amanda E. and {Bauman}, Brian J. and {Baumont}, Sylvain and {Bechtol}, Ellen and {Bechtol}, Keith and {Becker}, Andrew C. and {Becla}, Jacek and {Beldica}, Cristina and {Bellavia}, Steve and {Bianco}, Federica B. and {Biswas}, Rahul and {Blanc}, Guillaume and {Blazek}, Jonathan and {Blandford}, Roger D. and {Bloom}, Josh S. and {Bogart}, Joanne and {Bond}, Tim W. and {Booth}, Michael T. and {Borgland}, Anders W. and {Borne}, Kirk and {Bosch}, James F. and {Boutigny}, Dominique and {Brackett}, Craig A. and {Bradshaw}, Andrew and {Brandt}, William Nielsen and {Brown}, Michael E. and {Bullock}, James S. and {Burchat}, Patricia and {Burke}, David L. and {Cagnoli}, Gianpietro and {Calabrese}, Daniel and {Callahan}, Shawn and {Callen}, Alice L. and {Carlin}, Jeffrey L. and {Carlson}, Erin L. and {Chandrasekharan}, Srinivasan and {Charles-Emerson}, Glenaver and {Chesley}, Steve and {Cheu}, Elliott C. and {Chiang}, Hsin-Fang and {Chiang}, James and {Chirino}, Carol and {Chow}, Derek and {Ciardi}, David R. and {Claver}, Charles F. and {Cohen-Tanugi}, Johann and {Cockrum}, Joseph J. and {Coles}, Rebecca and {Connolly}, Andrew J. and {Cook}, Kem H. and {Cooray}, Asantha and {Covey}, Kevin R. and {Cribbs}, Chris and {Cui}, Wei and {Cutri}, Roc and {Daly}, Philip N. and {Daniel}, Scott F. and {Daruich}, Felipe and {Daubard}, Guillaume and {Daues}, Greg and {Dawson}, William and {Delgado}, Francisco and {Dellapenna}, Alfred and {de Peyster}, Robert and {de Val-Borro}, Miguel and {Digel}, Seth W. and {Doherty}, Peter and {Dubois}, Richard and {Dubois-Felsmann}, Gregory P. and {Durech}, Josef and {Economou}, Frossie and {Eifler}, Tim and {Eracleous}, Michael and {Emmons}, Benjamin L. and {Fausti Neto}, Angelo and {Ferguson}, Henry and {Figueroa}, Enrique and {Fisher-Levine}, Merlin and {Focke}, Warren and {Foss}, Michael D. and {Frank}, James and {Freemon}, Michael D. and {Gangler}, Emmanuel and {Gawiser}, Eric and {Geary}, John C. and {Gee}, Perry and {Geha}, Marla and {Gessner}, Charles J.~B. and {Gibson}, Robert R. and {Gilmore}, D. Kirk and {Glanzman}, Thomas and {Glick}, William and {Goldina}, Tatiana and {Goldstein}, Daniel A. and {Goodenow}, Iain and {Graham}, Melissa L. and {Gressler}, William J. and {Gris}, Philippe and {Guy}, Leanne P. and {Guyonnet}, Augustin and {Haller}, Gunther and {Harris}, Ron and {Hascall}, Patrick A. and {Haupt}, Justine and {Hernandez}, Fabio and {Herrmann}, Sven and {Hileman}, Edward and {Hoblitt}, Joshua and {Hodgson}, John A. and {Hogan}, Craig and {Howard}, James D. and {Huang}, Dajun and {Huffer}, Michael E. and {Ingraham}, Patrick and {Innes}, Walter R. and {Jacoby}, Suzanne H. and {Jain}, Bhuvnesh and {Jammes}, Fabrice and {Jee}, M. James and {Jenness}, Tim and {Jernigan}, Garrett and {Jevremovi{\'c}}, Darko and {Johns}, Kenneth and {Johnson}, Anthony S. and {Johnson}, Margaret W.~G. and {Jones}, R. Lynne and {Juramy-Gilles}, Claire and {Juri{\'c}}, Mario and {Kalirai}, Jason S. and {Kallivayalil}, Nitya J. and {Kalmbach}, Bryce and {Kantor}, Jeffrey P. and {Karst}, Pierre and {Kasliwal}, Mansi M. and {Kelly}, Heather and {Kessler}, Richard and {Kinnison}, Veronica and {Kirkby}, David and {Knox}, Lloyd and {Kotov}, Ivan V. and {Krabbendam}, Victor L. and {Krughoff}, K. Simon and {Kub{\'a}nek}, Petr and {Kuczewski}, John and {Kulkarni}, Shri and {Ku}, John and {Kurita}, Nadine R. and {Lage}, Craig S. and {Lambert}, Ron and {Lange}, Travis and {Langton}, J. Brian and {Le Guillou}, Laurent and {Levine}, Deborah and {Liang}, Ming and {Lim}, Kian-Tat and {Lintott}, Chris J. and {Long}, Kevin E. and {Lopez}, Margaux and {Lotz}, Paul J. and {Lupton}, Robert H. and {Lust}, Nate B. and {MacArthur}, Lauren A. and {Mahabal}, Ashish and {Mandelbaum}, Rachel and {Markiewicz}, Thomas W. and {Marsh}, Darren S. and {Marshall}, Philip J. and {Marshall}, Stuart and {May}, Morgan and {McKercher}, Robert and {McQueen}, Michelle and {Meyers}, Joshua and {Migliore}, Myriam and {Miller}, Michelle and {Mills}, David J.},
        title = "{LSST: From Science Drivers to Reference Design and Anticipated Data Products}",
      journal = {\apj},
         year = 2019,
        month = mar,
       volume = {873},
       number = {2},
          eid = {111},
        pages = {111},
          doi = {10.3847/1538-4357/ab042c},
archivePrefix = {arXiv},
       eprint = {0805.2366},
 primaryClass = {astro-ph},
       adsurl = {https://ui.adsabs.harvard.edu/abs/2019ApJ...873..111I}
}

@article{Prusti2016,
  author = {{Gaia Collaboration} and {Prusti, T.} and {de Bruijne, J. H. J.} and {Brown, A. G. A.} and {Vallenari, A.} and {Babusiaux, C.} and {Bailer-Jones, C. A. L.} and {Bastian, U.} and {Biermann, M.} and {Evans, D. W.} and {Eyer, L.} and {Jansen, F.} and {Jordi, C.} and {Klioner, S. A.} and {Lammers, U.} and {Lindegren, L.} and {Luri, X.} and {Mignard, F.} and {Milligan, D. J.} and {Panem, C.} and {Poinsignon, V.} and {Pourbaix, D.} and {Randich, S.} and {Sarri, G.} and {Sartoretti, P.} and {Siddiqui, H. I.} and {Soubiran, C.} and {Valette, V.} and {van Leeuwen, F.} and {Walton, N. A.} and {Aerts, C.} and {Arenou, F.} and {Cropper, M.} and {Drimmel, R.} and {Høg, E.} and {Katz, D.} and {Lattanzi, M. G.} and {O’Mullane, W.} and {Grebel, E. K.} and {Holland, A. D.} and {Huc, C.} and {Passot, X.} and {Bramante, L.} and {Cacciari, C.} and {Castañeda, J.} and {Chaoul, L.} and {Cheek, N.} and {De Angeli, F.} and {Fabricius, C.} and {Guerra, R.} and {Hernández, J.} and {Jean-Antoine-Piccolo, A.} and {Masana, E.} and {Messineo, R.} and {Mowlavi, N.} and {Nienartowicz, K.} and {Ordóñez-Blanco, D.} and {Panuzzo, P.} and {Portell, J.} and {Richards, P. J.} and {Riello, M.} and {Seabroke, G. M.} and {Tanga, P.} and {Thévenin, F.} and {Torra, J.} and {Els, S. G.} and {Gracia-Abril, G.} and {Comoretto, G.} and {Garcia-Reinaldos, M.} and {Lock, T.} and {Mercier, E.} and {Altmann, M.} and {Andrae, R.} and {Astraatmadja, T. L.} and {Bellas-Velidis, I.} and {Benson, K.} and {Berthier, J.} and {Blomme, R.} and {Busso, G.} and {Carry, B.} and {Cellino, A.} and {Clementini, G.} and {Cowell, S.} and {Creevey, O.} and {Cuypers, J.} and {Davidson, M.} and {De Ridder, J.} and {de Torres, A.} and {Delchambre, L.} and {Dell’Oro, A.} and {Ducourant, C.} and {Frémat, Y.} and {García-Torres, M.} and {Gosset, E.} and {Halbwachs, J.-L.} and {Hambly, N. C.} and {Harrison, D. L.} and {Hauser, M.} and {Hestroffer, D.} and {Hodgkin, S. T.} and {Huckle, H. E.} and {Hutton, A.} and {Jasniewicz, G.} and {Jordan, S.} and {Kontizas, M.} and {Korn, A. J.} and {Lanzafame, A. C.} and {Manteiga, M.} and {Moitinho, A.} and {Muinonen, K.} and {Osinde, J.} and {Pancino, E.} and {Pauwels, T.} and {Petit, J.-M.} and {Recio-Blanco, A.} and {Robin, A. C.} and {Sarro, L. M.} and {Siopis, C.} and {Smith, M.} and {Smith, K. W.} and {Sozzetti, A.} and {Thuillot, W.} and {van Reeven, W.} and {Viala, Y.} and {Abbas, U.} and {Abreu Aramburu, A.} and {Accart, S.} and {Aguado, J. J.} and {Allan, P. M.} and {Allasia, W.} and {Altavilla, G.} and {Álvarez, M. A.} and {Alves, J.} and {Anderson, R. I.} and {Andrei, A. H.} and {Anglada Varela, E.} and {Antiche, E.} and {Antoja, T.} and {Antón, S.} and {Arcay, B.} and {Atzei, A.} and {Ayache, L.} and {Bach, N.} and {Baker, S. G.} and {Balaguer-Núñez, L.} and {Barache, C.} and {Barata, C.} and {Barbier, A.} and {Barblan, F.} and {Baroni, M.} and {Barrado y Navascués, D.} and {Barros, M.} and {Barstow, M. A.} and {Becciani, U.} and {Bellazzini, M.} and {Bellei, G.} and {Bello García, A.} and {Belokurov, V.} and {Bendjoya, P.} and {Berihuete, A.} and {Bianchi, L.} and {Bienaymé, O.} and {Billebaud, F.} and {Blagorodnova, N.} and {Blanco-Cuaresma, S.} and {Boch, T.} and {Bombrun, A.} and {Borrachero, R.} and {Bouquillon, S.} and {Bourda, G.} and {Bouy, H.} and {Bragaglia, A.} and {Breddels, M. A.} and {Brouillet, N.} and {Brüsemeister, T.} and {Bucciarelli, B.} and {Budnik, F.} and {Burgess, P.} and {Burgon, R.} and {Burlacu, A.} and {Busonero, D.} and {Buzzi, R.} and {Caffau, E.} and {Cambras, J.} and {Campbell, H.} and {Cancelliere, R.} and {Cantat-Gaudin, T.} and {Carlucci, T.} and {Carrasco, J. M.} and {Castellani, M.} and {Charlot, P.} and {Charnas, J.} and {Charvet, P.} and {Chassat, F.} and {Chiavassa, A.} and {Clotet, M.} and {Cocozza, G.} and {Collins, R. S.} and {Collins, P.} and {Costigan, G.} and {Crifo, F.} and {Cross, N. J. G.} and {Crosta, M.} and {Crowley, C.} and {Dafonte, C.} and {Damerdji, Y.} and {Dapergolas, A.} and {David, P.} and {David, M.} and {De Cat, P.} and {de Felice, F.} and {de Laverny, P.} and {De Luise, F.} and {De March, R.} and {de Martino, D.} and {de Souza, R.} and {Debosscher, J.} and {del Pozo, E.} and {Delbo, M.} and {Delgado, A.} and {Delgado, H. E.} and {di Marco, F.} and {Di Matteo, P.} and {Diakite, S.} and {Distefano, E.} and {Dolding, C.} and {Dos Anjos, S.} and {Drazinos, P.} and {Durán, J.} and {Dzigan, Y.} and {Ecale, E.} and {Edvardsson, B.} and {Enke, H.} and {Erdmann, M.} and {Escolar, D.} and {Espina, M.} and {Evans, N. W.} and {Eynard Bontemps, G.} and {Fabre, C.} and {Fabrizio, M.} and {Faigler, S.} and {Falcão, A. J.} and {Farràs Casas, M.} and {Faye, F.} and {Federici, L.} and {Fedorets, G.} and {Fernández-Hernández, J.} and {Fernique, P.} and {Fienga, A.} and {Figueras, F.} and {Filippi, F.} and {Findeisen, K.} and {Fonti, A.} and {Fouesneau, M.} and {Fraile, E.} and {Fraser, M.} and {Fuchs, J.} and {Furnell, R.} and {Gai, M.} and {Galleti, S.} and {Galluccio, L.} and {Garabato, D.} and {García-Sedano, F.} and {Garé, P.} and {Garofalo, A.} and {Garralda, N.} and {Gavras, P.} and {Gerssen, J.} and {Geyer, R.} and {Gilmore, G.} and {Girona, S.} and {Giuffrida, G.} and {Gomes, M.} and {González-Marcos, A.} and {González-Núñez, J.} and {González-Vidal, J. J.} and {Granvik, M.} and {Guerrier, A.} and {Guillout, P.} and {Guiraud, J.} and {Gúrpide, A.} and {Gutiérrez-Sánchez, R.} and {Guy, L. P.} and {Haigron, R.} and {Hatzidimitriou, D.} and {Haywood, M.} and {Heiter, U.} and {Helmi, A.} and {Hobbs, D.} and {Hofmann, W.} and {Holl, B.} and {Holland, G.} and {Hunt, J. A. S.} and {Hypki, A.} and {Icardi, V.} and {Irwin, M.} and {Jevardat de Fombelle, G.} and {Jofré, P.} and {Jonker, P. G.} and {Jorissen, A.} and {Julbe, F.} and {Karampelas, A.} and {Kochoska, A.} and {Kohley, R.} and {Kolenberg, K.} and {Kontizas, E.} and {Koposov, S. E.} and {Kordopatis, G.} and {Koubsky, P.} and {Kowalczyk, A.} and {Krone-Martins, A.} and {Kudryashova, M.} and {Kull, I.} and {Bachchan, R. K.} and {Lacoste-Seris, F.} and {Lanza, A. F.} and {Lavigne, J.-B.} and {Le Poncin-Lafitte, C.} and {Lebreton, Y.} and {Lebzelter, T.} and {Leccia, S.} and {Leclerc, N.} and {Lecoeur-Taibi, I.} and {Lemaitre, V.} and {Lenhardt, H.} and {Leroux, F.} and {Liao, S.} and {Licata, E.} and {Lindstrøm, H. E. P.} and {Lister, T. A.} and {Livanou, E.} and {Lobel, A.} and {Löffler, W.} and {López, M.} and {Lopez-Lozano, A.} and {Lorenz, D.} and {Loureiro, T.} and {MacDonald, I.} and {Magalhães Fernandes, T.} and {Managau, S.} and {Mann, R. G.} and {Mantelet, G.} and {Marchal, O.} and {Marchant, J. M.} and {Marconi, M.} and {Marie, J.} and {Marinoni, S.} and {Marrese, P. M.} and {Marschalkó, G.} and {Marshall, D. J.} and {Martín-Fleitas, J. M.} and {Martino, M.} and {Mary, N.} and {Matijevič, G.} and {Mazeh, T.} and {McMillan, P. J.} and {Messina, S.} and {Mestre, A.} and {Michalik, D.} and {Millar, N. R.} and {Miranda, B. M. H.} and {Molina, D.} and {Molinaro, R.} and {Molinaro, M.} and {Molnár, L.} and {Moniez, M.} and {Montegriffo, P.} and {Monteiro, D.} and {Mor, R.} and {Mora, A.} and {Morbidelli, R.} and {Morel, T.} and {Morgenthaler, S.} and {Morley, T.} and {Morris, D.} and {Mulone, A. F.} and {Muraveva, T.} and {Musella, I.} and {Narbonne, J.} and {Nelemans, G.} and {Nicastro, L.} and {Noval, L.} and {Ordénovic, C.} and {Ordieres-Meré, J.} and {Osborne, P.} and {Pagani, C.} and {Pagano, I.} and {Pailler, F.} and {Palacin, H.} and {Palaversa, L.} and {Parsons, P.} and {Paulsen, T.} and {Pecoraro, M.} and {Pedrosa, R.} and {Pentikäinen, H.} and {Pereira, J.} and {Pichon, B.} and {Piersimoni, A. M.} and {Pineau, F.-X.} and {Plachy, E.} and {Plum, G.} and {Poujoulet, E.} and {Prša, A.} and {Pulone, L.} and {Ragaini, S.} and {Rago, S.} and {Rambaux, N.} and {Ramos-Lerate, M.} and {Ranalli, P.} and {Rauw, G.} and {Read, A.} and {Regibo, S.} and {Renk, F.} and {Reylé, C.} and {Ribeiro, R. A.} and {Rimoldini, L.} and {Ripepi, V.} and {Riva, A.} and {Rixon, G.} and {Roelens, M.} and {Romero-Gómez, M.} and {Rowell, N.} and {Royer, F.} and {Rudolph, A.} and {Ruiz-Dern, L.} and {Sadowski, G.} and {Sagristà Sellés, T.} and {Sahlmann, J.} and {Salgado, J.} and {Salguero, E.} and {Sarasso, M.} and {Savietto, H.} and {Schnorhk, A.} and {Schultheis, M.} and {Sciacca, E.} and {Segol, M.} and {Segovia, J. C.} and {Segransan, D.} and {Serpell, E.} and {Shih, I-C.} and {Smareglia, R.} and {Smart, R. L.} and {Smith, C.} and {Solano, E.} and {Solitro, F.} and {Sordo, R.} and {Soria Nieto, S.} and {Souchay, J.} and {Spagna, A.} and {Spoto, F.} and {Stampa, U.} and {Steele, I. A.} and {Steidelmüller, H.} and {Stephenson, C. A.} and {Stoev, H.} and {Suess, F. F.} and {Süveges, M.} and {Surdej, J.} and {Szabados, L.} and {Szegedi-Elek, E.} and {Tapiador, D.} and {Taris, F.} and {Tauran, G.} and {Taylor, M. B.} and {Teixeira, R.} and {Terrett, D.} and {Tingley, B.} and {Trager, S. C.} and {Turon, C.} and {Ulla, A.} and {Utrilla, E.} and {Valentini, G.} and {van Elteren, A.} and {Van Hemelryck, E.} and {van Leeuwen, M.} and {Varadi, M.} and {Vecchiato, A.} and {Veljanoski, J.} and {Via, T.} and {Vicente, D.} and {Vogt, S.} and {Voss, H.} and {Votruba, V.} and {Voutsinas, S.} and {Walmsley, G.} and {Weiler, M.} and {Weingrill, K.} and {Werner, D.} and {Wevers, T.} and {Whitehead, G.} and {Wyrzykowski, Ł.} and {Yoldas, A.} and {Žerjal, M.} and {Zucker, S.} and {Zurbach, C.} and {Zwitter, T.} and {Alecu, A.} and {Allen, M.} and {Allende Prieto, C.} and {Amorim, A.} and {Anglada-Escudé, G.} and {Arsenijevic, V.} and {Azaz, S.} and {Balm, P.} and {Beck, M.} and {Bernstein, H.-H.} and {Bigot, L.} and {Bijaoui, A.} and {Blasco, C.} and {Bonfigli, M.} and {Bono, G.} and {Boudreault, S.} and {Bressan, A.} and {Brown, S.} and {Brunet, P.-M.} and {Bunclark, P.} and {Buonanno, R.} and {Butkevich, A. G.} and {Carret, C.} and {Carrion, C.} and {Chemin, L.} and {Chéreau, F.} and {Corcione, L.} and {Darmigny, E.} and {de Boer, K. S.} and {de Teodoro, P.} and {de Zeeuw, P. T.} and {Delle Luche, C.} and {Domingues, C. D.} and {Dubath, P.} and {Fodor, F.} and {Frézouls, B.} and {Fries, A.} and {Fustes, D.} and {Fyfe, D.} and {Gallardo, E.} and {Gallegos, J.} and {Gardiol, D.} and {Gebran, M.} and {Gomboc, A.} and {Gómez, A.} and {Grux, E.} and {Gueguen, A.} and {Heyrovsky, A.} and {Hoar, J.} and {Iannicola, G.} and {Isasi Parache, Y.} and {Janotto, A.-M.} and {Joliet, E.} and {Jonckheere, A.} and {Keil, R.} and {Kim, D.-W.} and {Klagyivik, P.} and {Klar, J.} and {Knude, J.} and {Kochukhov, O.} and {Kolka, I.} and {Kos, J.} and {Kutka, A.} and {Lainey, V.} and {LeBouquin, D.} and {Liu, C.} and {Loreggia, D.} and {Makarov, V. V.} and {Marseille, M. G.} and {Martayan, C.} and {Martinez-Rubi, O.} and {Massart, B.} and {Meynadier, F.} and {Mignot, S.} and {Munari, U.} and {Nguyen, A.-T.} and {Nordlander, T.} and {Ocvirk, P.} and {O’Flaherty, K. S.} and {Olias Sanz, A.} and {Ortiz, P.} and {Osorio, J.} and {Oszkiewicz, D.} and {Ouzounis, A.} and {Palmer, M.} and {Park, P.} and {Pasquato, E.} and {Peltzer, C.} and {Peralta, J.} and {Péturaud, F.} and {Pieniluoma, T.} and {Pigozzi, E.} and {Poels, J.} and {Prat, G.} and {Prod’homme, T.} and {Raison, F.} and {Rebordao, J. M.} and {Risquez, D.} and {Rocca-Volmerange, B.} and {Rosen, S.} and {Ruiz-Fuertes, M. I.} and {Russo, F.} and {Sembay, S.} and {Serraller Vizcaino, I.} and {Short, A.} and {Siebert, A.} and {Silva, H.} and {Sinachopoulos, D.} and {Slezak, E.} and {Soffel, M.} and {Sosnowska, D.} and {Straižys, V.} and {ter Linden, M.} and {Terrell, D.} and {Theil, S.} and {Tiede, C.} and {Troisi, L.} and {Tsalmantza, P.} and {Tur, D.} and {Vaccari, M.} and {Vachier, F.} and {Valles, P.} and {Van Hamme, W.} and {Veltz, L.} and {Virtanen, J.} and {Wallut, J.-M.} and {Wichmann, R.} and {Wilkinson, M. I.} and {Ziaeepour, H.} and {Zschocke, S.}},
  title = {The Gaia mission},
  journal = {\aap},
  volume = {595},
  pages = {A1},
  year = {2016},
  doi = {10.1051/0004-6361/201629272}
}

@ARTICLE{2016MNRAS.460.1270D,
       author = {{Dark Energy Survey Collaboration} and {Abbott}, T. and {Abdalla}, F.~B. and {Aleksi{\'c}}, J. and {Allam}, S. and {Amara}, A. and {Bacon}, D. and {Balbinot}, E. and {Banerji}, M. and {Bechtol}, K. and {Benoit-L{\'e}vy}, A. and {Bernstein}, G.~M. and {Bertin}, E. and {Blazek}, J. and {Bonnett}, C. and {Bridle}, S. and {Brooks}, D. and {Brunner}, R.~J. and {Buckley-Geer}, E. and {Burke}, D.~L. and {Caminha}, G.~B. and {Capozzi}, D. and {Carlsen}, J. and {Carnero-Rosell}, A. and {Carollo}, M. and {Carrasco-Kind}, M. and {Carretero}, J. and {Castander}, F.~J. and {Clerkin}, L. and {Collett}, T. and {Conselice}, C. and {Crocce}, M. and {Cunha}, C.~E. and {D'Andrea}, C.~B. and {da Costa}, L.~N. and {Davis}, T.~M. and {Desai}, S. and {Diehl}, H.~T. and {Dietrich}, J.~P. and {Dodelson}, S. and {Doel}, P. and {Drlica-Wagner}, A. and {Estrada}, J. and {Etherington}, J. and {Evrard}, A.~E. and {Fabbri}, J. and {Finley}, D.~A. and {Flaugher}, B. and {Foley}, R.~J. and {Fosalba}, P. and {Frieman}, J. and {Garc{\'\i}a-Bellido}, J. and {Gaztanaga}, E. and {Gerdes}, D.~W. and {Giannantonio}, T. and {Goldstein}, D.~A. and {Gruen}, D. and {Gruendl}, R.~A. and {Guarnieri}, P. and {Gutierrez}, G. and {Hartley}, W. and {Honscheid}, K. and {Jain}, B. and {James}, D.~J. and {Jeltema}, T. and {Jouvel}, S. and {Kessler}, R. and {King}, A. and {Kirk}, D. and {Kron}, R. and {Kuehn}, K. and {Kuropatkin}, N. and {Lahav}, O. and {Li}, T.~S. and {Lima}, M. and {Lin}, H. and {Maia}, M.~A.~G. and {Makler}, M. and {Manera}, M. and {Maraston}, C. and {Marshall}, J.~L. and {Martini}, P. and {McMahon}, R.~G. and {Melchior}, P. and {Merson}, A. and {Miller}, C.~J. and {Miquel}, R. and {Mohr}, J.~J. and {Morice-Atkinson}, X. and {Naidoo}, K. and {Neilsen}, E. and {Nichol}, R.~C. and {Nord}, B. and {Ogando}, R. and {Ostrovski}, F. and {Palmese}, A. and {Papadopoulos}, A. and {Peiris}, H.~V. and {Peoples}, J. and {Percival}, W.~J. and {Plazas}, A.~A. and {Reed}, S.~L. and {Refregier}, A. and {Romer}, A.~K. and {Roodman}, A. and {Ross}, A. and {Rozo}, E. and {Rykoff}, E.~S. and {Sadeh}, I. and {Sako}, M. and {S{\'a}nchez}, C. and {Sanchez}, E. and {Santiago}, B. and {Scarpine}, V. and {Schubnell}, M. and {Sevilla-Noarbe}, I. and {Sheldon}, E. and {Smith}, M. and {Smith}, R.~C. and {Soares-Santos}, M. and {Sobreira}, F. and {Soumagnac}, M. and {Suchyta}, E. and {Sullivan}, M. and {Swanson}, M. and {Tarle}, G. and {Thaler}, J. and {Thomas}, D. and {Thomas}, R.~C. and {Tucker}, D. and {Vieira}, J.~D. and {Vikram}, V. and {Walker}, A.~R. and {Wechsler}, R.~H. and {Weller}, J. and {Wester}, W. and {Whiteway}, L. and {Wilcox}, H. and {Yanny}, B. and {Zhang}, Y. and {Zuntz}, J.},
        title = "{The Dark Energy Survey: more than dark energy - an overview}",
      journal = {\mnras},
         year = 2016,
        month = aug,
       volume = {460},
       number = {2},
        pages = {1270-1299},
          doi = {10.1093/mnras/stw641},
archivePrefix = {arXiv},
       eprint = {1601.00329},
 primaryClass = {astro-ph.CO},
       adsurl = {https://ui.adsabs.harvard.edu/abs/2016MNRAS.460.1270D}
}

@ARTICLE{2000AJ....120.1579Y,
       author = {{York}, Donald G. and {Adelman}, J. and {Anderson}, Jr., John E. and {Anderson}, Scott F. and {Annis}, James and {Bahcall}, Neta A. and {Bakken}, J.~A. and {Barkhouser}, Robert and {Bastian}, Steven and {Berman}, Eileen and {Boroski}, William N. and {Bracker}, Steve and {Briegel}, Charlie and {Briggs}, John W. and {Brinkmann}, J. and {Brunner}, Robert and {Burles}, Scott and {Carey}, Larry and {Carr}, Michael A. and {Castander}, Francisco J. and {Chen}, Bing and {Colestock}, Patrick L. and {Connolly}, A.~J. and {Crocker}, J.~H. and {Csabai}, Istv{\'a}n and {Czarapata}, Paul C. and {Davis}, John Eric and {Doi}, Mamoru and {Dombeck}, Tom and {Eisenstein}, Daniel and {Ellman}, Nancy and {Elms}, Brian R. and {Evans}, Michael L. and {Fan}, Xiaohui and {Federwitz}, Glenn R. and {Fiscelli}, Larry and {Friedman}, Scott and {Frieman}, Joshua A. and {Fukugita}, Masataka and {Gillespie}, Bruce and {Gunn}, James E. and {Gurbani}, Vijay K. and {de Haas}, Ernst and {Haldeman}, Merle and {Harris}, Frederick H. and {Hayes}, J. and {Heckman}, Timothy M. and {Hennessy}, G.~S. and {Hindsley}, Robert B. and {Holm}, Scott and {Holmgren}, Donald J. and {Huang}, Chi-hao and {Hull}, Charles and {Husby}, Don and {Ichikawa}, Shin-Ichi and {Ichikawa}, Takashi and {Ivezi{\'c}}, {\v{Z}}eljko and {Kent}, Stephen and {Kim}, Rita S.~J. and {Kinney}, E. and {Klaene}, Mark and {Kleinman}, A.~N. and {Kleinman}, S. and {Knapp}, G.~R. and {Korienek}, John and {Kron}, Richard G. and {Kunszt}, Peter Z. and {Lamb}, D.~Q. and {Lee}, B. and {Leger}, R. French and {Limmongkol}, Siriluk and {Lindenmeyer}, Carl and {Long}, Daniel C. and {Loomis}, Craig and {Loveday}, Jon and {Lucinio}, Rich and {Lupton}, Robert H. and {MacKinnon}, Bryan and {Mannery}, Edward J. and {Mantsch}, P.~M. and {Margon}, Bruce and {McGehee}, Peregrine and {McKay}, Timothy A. and {Meiksin}, Avery and {Merelli}, Aronne and {Monet}, David G. and {Munn}, Jeffrey A. and {Narayanan}, Vijay K. and {Nash}, Thomas and {Neilsen}, Eric and {Neswold}, Rich and {Newberg}, Heidi Jo and {Nichol}, R.~C. and {Nicinski}, Tom and {Nonino}, Mario and {Okada}, Norio and {Okamura}, Sadanori and {Ostriker}, Jeremiah P. and {Owen}, Russell and {Pauls}, A. George and {Peoples}, John and {Peterson}, R.~L. and {Petravick}, Donald and {Pier}, Jeffrey R. and {Pope}, Adrian and {Pordes}, Ruth and {Prosapio}, Angela and {Rechenmacher}, Ron and {Quinn}, Thomas R. and {Richards}, Gordon T. and {Richmond}, Michael W. and {Rivetta}, Claudio H. and {Rockosi}, Constance M. and {Ruthmansdorfer}, Kurt and {Sandford}, Dale and {Schlegel}, David J. and {Schneider}, Donald P. and {Sekiguchi}, Maki and {Sergey}, Gary and {Shimasaku}, Kazuhiro and {Siegmund}, Walter A. and {Smee}, Stephen and {Smith}, J. Allyn and {Snedden}, S. and {Stone}, R. and {Stoughton}, Chris and {Strauss}, Michael A. and {Stubbs}, Christopher and {SubbaRao}, Mark and {Szalay}, Alexander S. and {Szapudi}, Istvan and {Szokoly}, Gyula P. and {Thakar}, Anirudda R. and {Tremonti}, Christy and {Tucker}, Douglas L. and {Uomoto}, Alan and {Vanden Berk}, Dan and {Vogeley}, Michael S. and {Waddell}, Patrick and {Wang}, Shu-i. and {Watanabe}, Masaru and {Weinberg}, David H. and {Yanny}, Brian and {Yasuda}, Naoki and {SDSS Collaboration}},
        title = "{The Sloan Digital Sky Survey: Technical Summary}",
      journal = {\aj},
         year = 2000,
        month = sep,
       volume = {120},
       number = {3},
        pages = {1579-1587},
          doi = {10.1086/301513},
archivePrefix = {arXiv},
       eprint = {astro-ph/0006396},
 primaryClass = {astro-ph},
       adsurl = {https://ui.adsabs.harvard.edu/abs/2000AJ....120.1579Y}
}

@article{Varghese2011,
  author = {Varghese, A. and Ibata, R. and Lewis, G. F.},
  title = {Stellar streams as probes of dark halo mass and morphology: a Bayesian reconstruction},
  journal = {Monthly Notices of the Royal Astronomical Society},
  volume = {417},
  issue = {1},
  pages = {198-215},
  year = {2011},
  doi = {10.1111/j.1365-2966.2011.19097.x}
}

@ARTICLE{2009ApJ...705L.223C,
       author = {{Carlberg}, R.~G.},
        title = "{Star Stream Folding by Dark Galactic Subhalos}",
      journal = {\apjl},
         year = 2009,
        month = nov,
       volume = {705},
       number = {2},
        pages = {L223-L226},
          doi = {10.1088/0004-637X/705/2/L223},
archivePrefix = {arXiv},
       eprint = {0908.4345},
 primaryClass = {astro-ph.CO},
       adsurl = {https://ui.adsabs.harvard.edu/abs/2009ApJ...705L.223C}
}

@article{johnston2002,
  author = {Johnston, Kathryn V. and Spergel, David N. and Haydn, Christian},
  title = {How Lumpy is the Milky Way’s Dark Matter Halo?},
  journal = {The Astrophysical Journal},
  year = {2002},
  volume = {570},
  number = {2},
  pages = {656},
  month = {may},
  doi = {10.1086/339791},
  url = {https://dx.doi.org/10.1086/339791}
}

@ARTICLE{2008MNRAS.391.1685S,
       author = {{Springel}, V. and {Wang}, J. and {Vogelsberger}, M. and {Ludlow}, A. and {Jenkins}, A. and {Helmi}, A. and {Navarro}, J.~F. and {Frenk}, C.~S. and {White}, S.~D.~M.},
        title = "{The Aquarius Project: the subhaloes of galactic haloes}",
      journal = {\mnras},
         year = 2008,
        month = dec,
       volume = {391},
       number = {4},
        pages = {1685-1711},
          doi = {10.1111/j.1365-2966.2008.14066.x},
archivePrefix = {arXiv},
       eprint = {0809.0898},
 primaryClass = {astro-ph},
       adsurl = {https://ui.adsabs.harvard.edu/abs/2008MNRAS.391.1685S}
}

@ARTICLE{2008Natur.454..735D,
       author = {{Diemand}, J. and {Kuhlen}, M. and {Madau}, P. and {Zemp}, M. and {Moore}, B. and {Potter}, D. and {Stadel}, J.},
        title = "{Clumps and streams in the local dark matter distribution}",
      journal = {\nat},
         year = 2008,
        month = aug,
       volume = {454},
       number = {7205},
        pages = {735-738},
          doi = {10.1038/nature07153},
archivePrefix = {arXiv},
       eprint = {0805.1244},
 primaryClass = {astro-ph},
       adsurl = {https://ui.adsabs.harvard.edu/abs/2008Natur.454..735D}
}

@ARTICLE{2002MNRAS.332..915I,
       author = {{Ibata}, R.~A. and {Lewis}, G.~F. and {Irwin}, M.~J. and {Quinn}, T.},
        title = "{Uncovering cold dark matter halo substructure with tidal streams}",
      journal = {\mnras},
         year = 2002,
        month = jun,
       volume = {332},
       number = {4},
        pages = {915-920},
          doi = {10.1046/j.1365-8711.2002.05358.x},
archivePrefix = {arXiv},
       eprint = {astro-ph/0110690},
 primaryClass = {astro-ph},
       adsurl = {https://ui.adsabs.harvard.edu/abs/2002MNRAS.332..915I}
}

@PROCEEDINGS{2016ASSL..420.....N,
        title = "{Tidal Streams in the Local Group and Beyond}",
    booktitle = {Tidal Streams in the Local Group and Beyond},
         year = 2016,
       editor = {{Newberg}, Heidi Jo and {Carlin}, Jeffrey L.},
       series = {Astrophysics and Space Science Library},
       volume = {420},
        month = jan,
          doi = {10.1007/978-3-319-19336-6},
       adsurl = {https://ui.adsabs.harvard.edu/abs/2016ASSL..420.....N}
}

@ARTICLE{2023PhRvD.107d3015N,
       author = {{Nguyen}, Tri and {Mishra-Sharma}, Siddharth and {Williams}, Reuel and {Necib}, Lina},
        title = "{Uncovering dark matter density profiles in dwarf galaxies with graph neural networks}",
      journal = {\prd},
         year = 2023,
        month = feb,
       volume = {107},
       number = {4},
          eid = {043015},
        pages = {043015},
          doi = {10.1103/PhysRevD.107.043015},
archivePrefix = {arXiv},
       eprint = {2208.12825},
 primaryClass = {astro-ph.CO},
       adsurl = {https://ui.adsabs.harvard.edu/abs/2023PhRvD.107d3015N}
}

@ARTICLE{2024ApJ...975..297W,
       author = {{Wagner-Carena}, Sebastian and {Lee}, Jaehoon and {Pennington}, Jeffrey and {Aalbers}, Jelle and {Birrer}, Simon and {Wechsler}, Risa H.},
        title = "{A Strong Gravitational Lens Is Worth a Thousand Dark Matter Halos: Inference on Small-scale Structure Using Sequential Methods}",
      journal = {\apj},
         year = 2024,
        month = nov,
       volume = {975},
       number = {2},
          eid = {297},
        pages = {297},
          doi = {10.3847/1538-4357/ad6e70},
archivePrefix = {arXiv},
       eprint = {2404.14487},
 primaryClass = {astro-ph.CO},
       adsurl = {https://ui.adsabs.harvard.edu/abs/2024ApJ...975..297W}
}

@ARTICLE{2024arXiv241010123E,
       author = {{Erickson}, Sydney and {Wagner-Carena}, Sebastian and {Marshall}, Phil and {Millon}, Martin and {Birrer}, Simon and {Roodman}, Aaron and {Schmidt}, Thomas and {Treu}, Tommaso and {Schuldt}, Stefan and {Shajib}, Anowar and {Venkatraman}, Padma and {The LSST Dark Energy Science Collaboration}},
        title = "{Lens Modeling of STRIDES Strongly Lensed Quasars using Neural Posterior Estimation}",
      journal = {arXiv e-prints},
         year = 2024,
        month = oct,
          eid = {arXiv:2410.10123},
        pages = {arXiv:2410.10123},
          doi = {10.48550/arXiv.2410.10123},
archivePrefix = {arXiv},
       eprint = {2410.10123},
 primaryClass = {astro-ph.IM},
       adsurl = {https://ui.adsabs.harvard.edu/abs/2024arXiv241010123E}
}

@ARTICLE{2023ApJ...942...75W,
       author = {{Wagner-Carena}, Sebastian and {Aalbers}, Jelle and {Birrer}, Simon and {Nadler}, Ethan O. and {Darragh-Ford}, Elise and {Marshall}, Philip J. and {Wechsler}, Risa H.},
        title = "{From Images to Dark Matter: End-to-end Inference of Substructure from Hundreds of Strong Gravitational Lenses}",
      journal = {\apj},
         year = 2023,
        month = jan,
       volume = {942},
       number = {2},
          eid = {75},
        pages = {75},
          doi = {10.3847/1538-4357/aca525},
archivePrefix = {arXiv},
       eprint = {2203.00690},
 primaryClass = {astro-ph.CO},
       adsurl = {https://ui.adsabs.harvard.edu/abs/2023ApJ...942...75W}
}

@ARTICLE{2021PhRvL.127x1103D,
       author = {{Dax}, Maximilian and {Green}, Stephen R. and {Gair}, Jonathan and {Macke}, Jakob H. and {Buonanno}, Alessandra and {Sch{\"o}lkopf}, Bernhard},
        title = "{Real-Time Gravitational Wave Science with Neural Posterior Estimation}",
      journal = {\prl},
         year = 2021,
        month = dec,
       volume = {127},
       number = {24},
          eid = {241103},
        pages = {241103},
          doi = {10.1103/PhysRevLett.127.241103},
archivePrefix = {arXiv},
       eprint = {2106.12594},
 primaryClass = {gr-qc},
       adsurl = {https://ui.adsabs.harvard.edu/abs/2021PhRvL.127x1103D}
}

@ARTICLE{2023PhRvL.130q1403D,
       author = {{Dax}, Maximilian and {Green}, Stephen R. and {Gair}, Jonathan and {P{\"u}rrer}, Michael and {Wildberger}, Jonas and {Macke}, Jakob H. and {Buonanno}, Alessandra and {Sch{\"o}lkopf}, Bernhard},
        title = "{Neural Importance Sampling for Rapid and Reliable Gravitational-Wave Inference}",
      journal = {\prl},
         year = 2023,
        month = apr,
       volume = {130},
       number = {17},
          eid = {171403},
        pages = {171403},
          doi = {10.1103/PhysRevLett.130.171403},
archivePrefix = {arXiv},
       eprint = {2210.05686},
 primaryClass = {gr-qc},
       adsurl = {https://ui.adsabs.harvard.edu/abs/2023PhRvL.130q1403D}
}

@ARTICLE{2022ApJ...938...11H,
       author = {{Hahn}, ChangHoon and {Melchior}, Peter},
        title = "{Accelerated Bayesian SED Modeling Using Amortized Neural Posterior Estimation}",
      journal = {\apj},
         year = 2022,
        month = oct,
       volume = {938},
       number = {1},
          eid = {11},
        pages = {11},
          doi = {10.3847/1538-4357/ac7b84},
archivePrefix = {arXiv},
       eprint = {2203.07391},
 primaryClass = {astro-ph.GA},
       adsurl = {https://ui.adsabs.harvard.edu/abs/2022ApJ...938...11H}
}

@ARTICLE{2022MLS&T...3dLT04K,
       author = {{Khullar}, Gourav and {Nord}, Brian and {{\'C}iprijanovi{\'c}}, Aleksandra and {Poh}, Jason and {Xu}, Fei},
        title = "{DIGS: deep inference of galaxy spectra with neural posterior estimation}",
      journal = {Machine Learning: Science and Technology},
         year = 2022,
        month = dec,
       volume = {3},
       number = {4},
          eid = {04LT04},
        pages = {04LT04},
          doi = {10.1088/2632-2153/ac98f4},
archivePrefix = {arXiv},
       eprint = {2211.09126},
 primaryClass = {astro-ph.GA},
       adsurl = {https://ui.adsabs.harvard.edu/abs/2022MLS&T...3dLT04K}
}

@ARTICLE{2014MNRAS.445.3788G,
       author = {{Gibbons}, S.~L.~J. and {Belokurov}, V. and {Evans}, N.~W.},
        title = "{`Skinny Milky Way please', says Sagittarius}",
      journal = {\mnras},
         year = 2014,
        month = dec,
       volume = {445},
       number = {4},
        pages = {3788-3802},
          doi = {10.1093/mnras/stu1986},
archivePrefix = {arXiv},
       eprint = {1406.2243},
 primaryClass = {astro-ph.GA},
       adsurl = {https://ui.adsabs.harvard.edu/abs/2014MNRAS.445.3788G}
}

@ARTICLE{2006ApJ...643L..17G,
       author = {{Grillmair}, C.~J. and {Dionatos}, O.},
        title = "{Detection of a 63{\textdegree} Cold Stellar Stream in the Sloan Digital Sky Survey}",
      journal = {\apjl},
         year = 2006,
        month = may,
       volume = {643},
       number = {1},
        pages = {L17-L20},
          doi = {10.1086/505111},
archivePrefix = {arXiv},
       eprint = {astro-ph/0604332},
 primaryClass = {astro-ph},
       adsurl = {https://ui.adsabs.harvard.edu/abs/2006ApJ...643L..17G}
}

@ARTICLE{2019ApJ...880...38B,
       author = {{Bonaca}, Ana and {Hogg}, David W. and {Price-Whelan}, Adrian M. and {Conroy}, Charlie},
        title = "{The Spur and the Gap in GD-1: Dynamical Evidence for a Dark Substructure in the Milky Way Halo}",
      journal = {\apj},
         year = 2019,
        month = jul,
       volume = {880},
       number = {1},
          eid = {38},
        pages = {38},
          doi = {10.3847/1538-4357/ab2873},
archivePrefix = {arXiv},
       eprint = {1811.03631},
 primaryClass = {astro-ph.GA},
       adsurl = {https://ui.adsabs.harvard.edu/abs/2019ApJ...880...38B}
}

@ARTICLE{2015MNRAS.454.3542E,
       author = {{Erkal}, Denis and {Belokurov}, Vasily},
        title = "{Properties of dark subhaloes from gaps in tidal streams}",
      journal = {\mnras},
         year = 2015,
        month = dec,
       volume = {454},
       number = {4},
        pages = {3542-3558},
          doi = {10.1093/mnras/stv2122},
archivePrefix = {arXiv},
       eprint = {1507.05625},
 primaryClass = {astro-ph.GA},
       adsurl = {https://ui.adsabs.harvard.edu/abs/2015MNRAS.454.3542E}
}

@ARTICLE{2021MNRAS.502.2364B,
       author = {{Banik}, Nilanjan and {Bovy}, Jo and {Bertone}, Gianfranco and {Erkal}, Denis and {de Boer}, T.~J.~L.},
        title = "{Evidence of a population of dark subhaloes from Gaia and Pan-STARRS observations of the GD-1 stream}",
      journal = {\mnras},
         year = 2021,
        month = apr,
       volume = {502},
       number = {2},
        pages = {2364-2380},
          doi = {10.1093/mnras/stab210},
archivePrefix = {arXiv},
       eprint = {1911.02662},
 primaryClass = {astro-ph.GA},
       adsurl = {https://ui.adsabs.harvard.edu/abs/2021MNRAS.502.2364B}
}

@ARTICLE{2001ApJ...548L.165O,
       author = {{Odenkirchen}, Michael and {Grebel}, Eva K. and {Rockosi}, Constance M. and {Dehnen}, Walter and {Ibata}, Rodrigo and {Rix}, Hans-Walter and {Stolte}, Andrea and {Wolf}, Christian and {Anderson}, Jr., John E. and {Bahcall}, Neta A. and {Brinkmann}, Jon and {Csabai}, Istv{\'a}n and {Hennessy}, G. and {Hindsley}, Robert B. and {Ivezi{\'c}}, {\v{Z}}eljko and {Lupton}, Robert H. and {Munn}, Jeffrey A. and {Pier}, Jeffrey R. and {Stoughton}, Chris and {York}, Donald G.},
        title = "{Detection of Massive Tidal Tails around the Globular Cluster Palomar 5 with Sloan Digital Sky Survey Commissioning Data}",
      journal = {\apjl},
         year = 2001,
        month = feb,
       volume = {548},
       number = {2},
        pages = {L165-L169},
          doi = {10.1086/319095},
archivePrefix = {arXiv},
       eprint = {astro-ph/0012311},
 primaryClass = {astro-ph},
       adsurl = {https://ui.adsabs.harvard.edu/abs/2001ApJ...548L.165O}
}

@ARTICLE{2021JCAP...10..043B,
       author = {{Banik}, Nilanjan and {Bovy}, Jo and {Bertone}, Gianfranco and {Erkal}, Denis and {de Boer}, T.~J.~L.},
        title = "{Novel constraints on the particle nature of dark matter from stellar streams}",
      journal = {\jcap},
         year = 2021,
        month = oct,
       volume = {2021},
       number = {10},
          eid = {043},
        pages = {043},
          doi = {10.1088/1475-7516/2021/10/043},
archivePrefix = {arXiv},
       eprint = {1911.02663},
 primaryClass = {astro-ph.GA},
       adsurl = {https://ui.adsabs.harvard.edu/abs/2021JCAP...10..043B}
}

@ARTICLE{2017MNRAS.466..628B,
       author = {{Bovy}, Jo and {Erkal}, Denis and {Sanders}, Jason L.},
        title = "{Linear perturbation theory for tidal streams and the small-scale CDM power spectrum}",
      journal = {\mnras},
         year = 2017,
        month = apr,
       volume = {466},
       number = {1},
        pages = {628-668},
          doi = {10.1093/mnras/stw3067},
archivePrefix = {arXiv},
       eprint = {1606.03470},
 primaryClass = {astro-ph.GA},
       adsurl = {https://ui.adsabs.harvard.edu/abs/2017MNRAS.466..628B}
}

@ARTICLE{2017arXiv170507057P,
       author = {{Papamakarios}, George and {Pavlakou}, Theo and {Murray}, Iain},
        title = "{Masked Autoregressive Flow for Density Estimation}",
      journal = {arXiv e-prints},
         year = 2017,
        month = may,
          eid = {arXiv:1705.07057},
        pages = {arXiv:1705.07057},
          doi = {10.48550/arXiv.1705.07057},
archivePrefix = {arXiv},
       eprint = {1705.07057},
 primaryClass = {stat.ML},
       adsurl = {https://ui.adsabs.harvard.edu/abs/2017arXiv170507057P}
}

@ARTICLE{2019arXiv191202762P,
       author = {{Papamakarios}, George and {Nalisnick}, Eric and {Jimenez Rezende}, Danilo and {Mohamed}, Shakir and {Lakshminarayanan}, Balaji},
        title = "{Normalizing Flows for Probabilistic Modeling and Inference}",
      journal = {arXiv e-prints},
         year = 2019,
        month = dec,
          eid = {arXiv:1912.02762},
        pages = {arXiv:1912.02762},
          doi = {10.48550/arXiv.1912.02762},
archivePrefix = {arXiv},
       eprint = {1912.02762},
 primaryClass = {stat.ML},
       adsurl = {https://ui.adsabs.harvard.edu/abs/2019arXiv191202762P}
}

@ARTICLE{2015arXiv150505770J,
       author = {{Jimenez Rezende}, Danilo and {Mohamed}, Shakir},
        title = "{Variational Inference with Normalizing Flows}",
      journal = {arXiv e-prints},
         year = 2015,
        month = may,
          eid = {arXiv:1505.05770},
        pages = {arXiv:1505.05770},
          doi = {10.48550/arXiv.1505.05770},
archivePrefix = {arXiv},
       eprint = {1505.05770},
 primaryClass = {stat.ML},
       adsurl = {https://ui.adsabs.harvard.edu/abs/2015arXiv150505770J}
}

@ARTICLE{2019arXiv190604032D,
       author = {{Durkan}, Conor and {Bekasov}, Artur and {Murray}, Iain and {Papamakarios}, George},
        title = "{Neural Spline Flows}",
      journal = {arXiv e-prints},
         year = 2019,
        month = jun,
          eid = {arXiv:1906.04032},
        pages = {arXiv:1906.04032},
          doi = {10.48550/arXiv.1906.04032},
archivePrefix = {arXiv},
       eprint = {1906.04032},
 primaryClass = {stat.ML},
       adsurl = {https://ui.adsabs.harvard.edu/abs/2019arXiv190604032D}
}

@ARTICLE{2017arXiv170603762V,
       author = {{Vaswani}, Ashish and {Shazeer}, Noam and {Parmar}, Niki and {Uszkoreit}, Jakob and {Jones}, Llion and {Gomez}, Aidan N. and {Kaiser}, Lukasz and {Polosukhin}, Illia},
        title = "{Attention Is All You Need}",
      journal = {arXiv e-prints},
         year = 2017,
        month = jun,
          eid = {arXiv:1706.03762},
        pages = {arXiv:1706.03762},
          doi = {10.48550/arXiv.1706.03762},
archivePrefix = {arXiv},
       eprint = {1706.03762},
 primaryClass = {cs.CL},
       adsurl = {https://ui.adsabs.harvard.edu/abs/2017arXiv170603762V}
}

@ARTICLE{2020PNAS..11730055C,
       author = {{Cranmer}, Kyle and {Brehmer}, Johann and {Louppe}, Gilles},
        title = "{The frontier of simulation-based inference}",
      journal = {Proceedings of the National Academy of Science},
         year = 2020,
        month = dec,
       volume = {117},
       number = {48},
        pages = {30055-30062},
          doi = {10.1073/pnas.1912789117},
archivePrefix = {arXiv},
       eprint = {1911.01429},
 primaryClass = {stat.ML},
       adsurl = {https://ui.adsabs.harvard.edu/abs/2020PNAS..11730055C}
}

@ARTICLE{2018ApJ...862..114S,
       author = {{Shipp}, N. and {Drlica-Wagner}, A. and {Balbinot}, E. and {Ferguson}, P. and {Erkal}, D. and {Li}, T.~S. and {Bechtol}, K. and {Belokurov}, V. and {Buncher}, B. and {Carollo}, D. and {Carrasco Kind}, M. and {Kuehn}, K. and {Marshall}, J.~L. and {Pace}, A.~B. and {Rykoff}, E.~S. and {Sevilla-Noarbe}, I. and {Sheldon}, E. and {Strigari}, L. and {Vivas}, A.~K. and {Yanny}, B. and {Zenteno}, A. and {Abbott}, T.~M.~C. and {Abdalla}, F.~B. and {Allam}, S. and {Avila}, S. and {Bertin}, E. and {Brooks}, D. and {Burke}, D.~L. and {Carretero}, J. and {Castander}, F.~J. and {Cawthon}, R. and {Crocce}, M. and {Cunha}, C.~E. and {D'Andrea}, C.~B. and {da Costa}, L.~N. and {Davis}, C. and {De Vicente}, J. and {Desai}, S. and {Diehl}, H.~T. and {Doel}, P. and {Evrard}, A.~E. and {Flaugher}, B. and {Fosalba}, P. and {Frieman}, J. and {Garc{\'\i}a-Bellido}, J. and {Gaztanaga}, E. and {Gerdes}, D.~W. and {Gruen}, D. and {Gruendl}, R.~A. and {Gschwend}, J. and {Gutierrez}, G. and {Hartley}, W. and {Honscheid}, K. and {Hoyle}, B. and {James}, D.~J. and {Johnson}, M.~D. and {Krause}, E. and {Kuropatkin}, N. and {Lahav}, O. and {Lin}, H. and {Maia}, M.~A.~G. and {March}, M. and {Martini}, P. and {Menanteau}, F. and {Miller}, C.~J. and {Miquel}, R. and {Nichol}, R.~C. and {Plazas}, A.~A. and {Romer}, A.~K. and {Sako}, M. and {Sanchez}, E. and {Santiago}, B. and {Scarpine}, V. and {Schindler}, R. and {Schubnell}, M. and {Smith}, M. and {Smith}, R.~C. and {Sobreira}, F. and {Suchyta}, E. and {Swanson}, M.~E.~C. and {Tarle}, G. and {Thomas}, D. and {Tucker}, D.~L. and {Walker}, A.~R. and {Wechsler}, R.~H. and {DES Collaboration}},
        title = "{Stellar Streams Discovered in the Dark Energy Survey}",
      journal = {\apj},
         year = 2018,
        month = aug,
       volume = {862},
       number = {2},
          eid = {114},
        pages = {114},
          doi = {10.3847/1538-4357/aacdab},
archivePrefix = {arXiv},
       eprint = {1801.03097},
 primaryClass = {astro-ph.GA},
       adsurl = {https://ui.adsabs.harvard.edu/abs/2018ApJ...862..114S}
}

@ARTICLE{2014MNRAS.442L..85K,
       author = {{Koposov}, S.~E. and {Irwin}, M. and {Belokurov}, V. and {Gonzalez-Solares}, E. and {Yoldas}, A.~K. and {Lewis}, J. and {Metcalfe}, N. and {Shanks}, T.},
        title = "{Discovery of a cold stellar stream in the ATLAS DR1 data.}",
      journal = {\mnras},
         year = 2014,
        month = jul,
       volume = {442},
        pages = {L85-L89},
          doi = {10.1093/mnrasl/slu060},
archivePrefix = {arXiv},
       eprint = {1403.3409},
 primaryClass = {astro-ph.GA},
       adsurl = {https://ui.adsabs.harvard.edu/abs/2014MNRAS.442L..85K}
}

@ARTICLE{2021ApJ...911..149L,
       author = {{Li}, Ting S. and {Koposov}, Sergey E. and {Erkal}, Denis and {Ji}, Alexander P. and {Shipp}, Nora and {Pace}, Andrew B. and {Hilmi}, Tariq and {Kuehn}, Kyler and {Lewis}, Geraint F. and {Mackey}, Dougal and {Simpson}, Jeffrey D. and {Wan}, Zhen and {Zucker}, Daniel B. and {Bland-Hawthorn}, Joss and {Cullinane}, Lara R. and {Da Costa}, Gary S. and {Drlica-Wagner}, Alex and {Hattori}, Kohei and {Martell}, Sarah L. and {Sharma}, Sanjib and {S5 Collaboration}},
        title = "{Broken into Pieces: ATLAS and Aliqa Uma as One Single Stream}",
      journal = {\apj},
         year = 2021,
        month = apr,
       volume = {911},
       number = {2},
          eid = {149},
        pages = {149},
          doi = {10.3847/1538-4357/abeb18},
archivePrefix = {arXiv},
       eprint = {2006.10763},
 primaryClass = {astro-ph.GA},
       adsurl = {https://ui.adsabs.harvard.edu/abs/2021ApJ...911..149L}
}

@ARTICLE{hilmi24,
       author = {{Hilmi}, Tariq and {Erkal}, Denis and {Koposov}, Sergey E. and {Li}, Ting S. and {Lilleengen}, Sophia and {Ji}, Alexander P. and {Lewis}, Geraint F. and {Shipp}, Nora and {Pace}, Andrew B. and {Zucker}, Daniel B. and {Limberg}, Guilherme and {Usman}, Sam A.},
        title = "{Inferring dark matter subhalo properties from simulated subhalo-stream encounters}",
      journal = {arXiv e-prints},
         year = 2024,
        month = apr,
          eid = {arXiv:2404.02953},
        pages = {arXiv:2404.02953},
          doi = {10.48550/arXiv.2404.02953},
archivePrefix = {arXiv},
       eprint = {2404.02953},
 primaryClass = {astro-ph.GA},
       adsurl = {https://ui.adsabs.harvard.edu/abs/2024arXiv240402953H}
}

@ARTICLE{2016MNRAS.461.1590E,
       author = {{Erkal}, Denis and {Sanders}, Jason L. and {Belokurov}, Vasily},
        title = "{Stray, swing and scatter: angular momentum evolution of orbits and streams in aspherical potentials}",
      journal = {\mnras},
         year = 2016,
        month = sep,
       volume = {461},
       number = {2},
        pages = {1590-1604},
          doi = {10.1093/mnras/stw1400},
archivePrefix = {arXiv},
       eprint = {1603.08922},
 primaryClass = {astro-ph.GA},
       adsurl = {https://ui.adsabs.harvard.edu/abs/2016MNRAS.461.1590E}
}

@misc{pygaia,
  author = {Anthony Brown and the Gaia Data Processing and Analysis Consortium},
  title = {PyGaia: Python modules for the simulation and basic manipulation of Gaia catalogue data and uncertainties},
  year = {2012--2024},
  howpublished = {\url{https://gaia.esac.esa.int/documentation/GaiaArchive/}},
  note = {Accessed: [Insert date here]}
}

@ARTICLE{2023A&A...674A...1G,
       author = {{Gaia Collaboration} and {Vallenari}, A. and {Brown}, A.~G.~A. and {Prusti}, T. and {de Bruijne}, J.~H.~J. and {Arenou}, F. and {Babusiaux}, C. and {Biermann}, M. and {Creevey}, O.~L. and {Ducourant}, C. and {Evans}, D.~W. and {Eyer}, L. and {Guerra}, R. and {Hutton}, A. and {Jordi}, C. and {Klioner}, S.~A. and {Lammers}, U.~L. and {Lindegren}, L. and {Luri}, X. and {Mignard}, F. and {Panem}, C. and {Pourbaix}, D. and {Randich}, S. and {Sartoretti}, P. and {Soubiran}, C. and {Tanga}, P. and {Walton}, N.~A. and {Bailer-Jones}, C.~A.~L. and {Bastian}, U. and {Drimmel}, R. and {Jansen}, F. and {Katz}, D. and {Lattanzi}, M.~G. and {van Leeuwen}, F. and {Bakker}, J. and {Cacciari}, C. and {Casta{\~n}eda}, J. and {De Angeli}, F. and {Fabricius}, C. and {Fouesneau}, M. and {Fr{\'e}mat}, Y. and {Galluccio}, L. and {Guerrier}, A. and {Heiter}, U. and {Masana}, E. and {Messineo}, R. and {Mowlavi}, N. and {Nicolas}, C. and {Nienartowicz}, K. and {Pailler}, F. and {Panuzzo}, P. and {Riclet}, F. and {Roux}, W. and {Seabroke}, G.~M. and {Sordo}, R. and {Th{\'e}venin}, F. and {Gracia-Abril}, G. and {Portell}, J. and {Teyssier}, D. and {Altmann}, M. and {Andrae}, R. and {Audard}, M. and {Bellas-Velidis}, I. and {Benson}, K. and {Berthier}, J. and {Blomme}, R. and {Burgess}, P.~W. and {Busonero}, D. and {Busso}, G. and {C{\'a}novas}, H. and {Carry}, B. and {Cellino}, A. and {Cheek}, N. and {Clementini}, G. and {Damerdji}, Y. and {Davidson}, M. and {de Teodoro}, P. and {Nu{\~n}ez Campos}, M. and {Delchambre}, L. and {Dell'Oro}, A. and {Esquej}, P. and {Fern{\'a}ndez-Hern{\'a}ndez}, J. and {Fraile}, E. and {Garabato}, D. and {Garc{\'\i}a-Lario}, P. and {Gosset}, E. and {Haigron}, R. and {Halbwachs}, J. -L. and {Hambly}, N.~C. and {Harrison}, D.~L. and {Hern{\'a}ndez}, J. and {Hestroffer}, D. and {Hodgkin}, S.~T. and {Holl}, B. and {Jan{\ss}en}, K. and {Jevardat de Fombelle}, G. and {Jordan}, S. and {Krone-Martins}, A. and {Lanzafame}, A.~C. and {L{\"o}ffler}, W. and {Marchal}, O. and {Marrese}, P.~M. and {Moitinho}, A. and {Muinonen}, K. and {Osborne}, P. and {Pancino}, E. and {Pauwels}, T. and {Recio-Blanco}, A. and {Reyl{\'e}}, C. and {Riello}, M. and {Rimoldini}, L. and {Roegiers}, T. and {Rybizki}, J. and {Sarro}, L.~M. and {Siopis}, C. and {Smith}, M. and {Sozzetti}, A. and {Utrilla}, E. and {van Leeuwen}, M. and {Abbas}, U. and {{\'A}brah{\'a}m}, P. and {Abreu Aramburu}, A. and {Aerts}, C. and {Aguado}, J.~J. and {Ajaj}, M. and {Aldea-Montero}, F. and {Altavilla}, G. and {{\'A}lvarez}, M.~A. and {Alves}, J. and {Anders}, F. and {Anderson}, R.~I. and {Anglada Varela}, E. and {Antoja}, T. and {Baines}, D. and {Baker}, S.~G. and {Balaguer-N{\'u}{\~n}ez}, L. and {Balbinot}, E. and {Balog}, Z. and {Barache}, C. and {Barbato}, D. and {Barros}, M. and {Barstow}, M.~A. and {Bartolom{\'e}}, S. and {Bassilana}, J. -L. and {Bauchet}, N. and {Becciani}, U. and {Bellazzini}, M. and {Berihuete}, A. and {Bernet}, M. and {Bertone}, S. and {Bianchi}, L. and {Binnenfeld}, A. and {Blanco-Cuaresma}, S. and {Blazere}, A. and {Boch}, T. and {Bombrun}, A. and {Bossini}, D. and {Bouquillon}, S. and {Bragaglia}, A. and {Bramante}, L. and {Breedt}, E. and {Bressan}, A. and {Brouillet}, N. and {Brugaletta}, E. and {Bucciarelli}, B. and {Burlacu}, A. and {Butkevich}, A.~G. and {Buzzi}, R. and {Caffau}, E. and {Cancelliere}, R. and {Cantat-Gaudin}, T. and {Carballo}, R. and {Carlucci}, T. and {Carnerero}, M.~I. and {Carrasco}, J.~M. and {Casamiquela}, L. and {Castellani}, M. and {Castro-Ginard}, A. and {Chaoul}, L. and {Charlot}, P. and {Chemin}, L. and {Chiaramida}, V. and {Chiavassa}, A. and {Chornay}, N. and {Comoretto}, G. and {Contursi}, G. and {Cooper}, W.~J. and {Cornez}, T. and {Cowell}, S. and {Crifo}, F. and {Cropper}, M. and {Crosta}, M. and {Crowley}, C. and {Dafonte}, C. and {Dapergolas}, A. and {David}, M. and {David}, P. and {de Laverny}, P. and {De Luise}, F. and {De March}, R.},
        title = "{Gaia Data Release 3. Summary of the content and survey properties}",
      journal = {\aap},
         year = 2023,
        month = jun,
       volume = {674},
          eid = {A1},
        pages = {A1},
          doi = {10.1051/0004-6361/202243940},
archivePrefix = {arXiv},
       eprint = {2208.00211},
 primaryClass = {astro-ph.GA},
       adsurl = {https://ui.adsabs.harvard.edu/abs/2023A&A...674A...1G}
}

@ARTICLE{2016ApJS..222....8D,
       author = {{Dotter}, Aaron},
        title = "{MESA Isochrones and Stellar Tracks (MIST) 0: Methods for the Construction of Stellar Isochrones}",
      journal = {\apjs},
         year = 2016,
        month = jan,
       volume = {222},
       number = {1},
          eid = {8},
        pages = {8},
          doi = {10.3847/0067-0049/222/1/8},
archivePrefix = {arXiv},
       eprint = {1601.05144},
 primaryClass = {astro-ph.SR},
       adsurl = {https://ui.adsabs.harvard.edu/abs/2016ApJS..222....8D}
}

@misc{ugali,
  author       = {Keith Bechtol and Alex Drlica-Wagner and contributors},
  title        = {Ultra-faint galaxy likelihood toolkit (ugali)},
  year         = {2021},
  howpublished = {\url{https://github.com/DarkEnergySurvey/ugali}},
  version      = {v1.8.0},
  note         = {Accessed: 2025-01-21},
}

@INPROCEEDINGS{2012SPIE.8446E..0TD,
       author = {{de Jong}, Roelof S. and {Bellido-Tirado}, Olga and {Chiappini}, Cristina and {Depagne}, {\'E}ric and {Haynes}, Roger and {Johl}, Diana and {Schnurr}, Olivier and {Schwope}, Axel and {Walcher}, Jakob and {Dionies}, Frank and {Haynes}, Dionne and {Kelz}, Andreas and {Kitaura}, Francisco S. and {Lamer}, Georg and {Minchev}, Ivan and {M{\"u}ller}, Volker and {Nuza}, Sebasti{\'a}n. E. and {Olaya}, Jean-Christophe and {Piffl}, Tilmann and {Popow}, Emil and {Steinmetz}, Matthias and {Ural}, Ugur and {Williams}, Mary and {Winkler}, Roland and {Wisotzki}, Lutz and {Ansorge}, Wolfgang R. and {Banerji}, Manda and {Gonzalez Solares}, Eduardo and {Irwin}, Mike and {Kennicutt}, Robert C. and {King}, Dave and {McMahon}, Richard G. and {Koposov}, Sergey and {Parry}, Ian R. and {Sun}, David and {Walton}, Nicholas A. and {Finger}, Gert and {Iwert}, Olaf and {Krumpe}, Mirko and {Lizon}, Jean-Louis and {Vincenzo}, Mainieri and {Amans}, Jean-Philippe and {Bonifacio}, Piercarlo and {Cohen}, Mathieu and {Francois}, Patrick and {Jagourel}, Pascal and {Mignot}, Shan B. and {Royer}, Fr{\'e}d{\'e}ric and {Sartoretti}, Paola and {Bender}, Ralf and {Grupp}, Frank and {Hess}, Hans-Joachim and {Lang-Bardl}, Florian and {Muschielok}, Bernard and {B{\"o}hringer}, Hans and {Boller}, Thomas and {Bongiorno}, Angela and {Brusa}, Marcella and {Dwelly}, Tom and {Merloni}, Andrea and {Nandra}, Kirpal and {Salvato}, Mara and {Pragt}, Johannes H. and {Navarro}, Ram{\'o}n and {Gerlofsma}, Gerrit and {Roelfsema}, Ronald and {Dalton}, Gavin B. and {Middleton}, Kevin F. and {Tosh}, Ian A. and {Boeche}, Corrado and {Caffau}, Elisabetta and {Christlieb}, Norbert and {Grebel}, Eva K. and {Hansen}, Camilla and {Koch}, Andreas and {Ludwig}, Hans-G. and {Quirrenbach}, Andreas and {Sbordone}, Luca and {Seifert}, Walter and {Thimm}, Guido and {Trifonov}, Trifon and {Helmi}, Amina and {Trager}, Scott C. and {Feltzing}, Sofia and {Korn}, Andreas and {Boland}, Wilfried},
        title = "{4MOST: 4-metre multi-object spectroscopic telescope}",
    booktitle = {Ground-based and Airborne Instrumentation for Astronomy IV},
         year = 2012,
       editor = {{McLean}, Ian S. and {Ramsay}, Suzanne K. and {Takami}, Hideki},
       series = {Society of Photo-Optical Instrumentation Engineers (SPIE) Conference Series},
       volume = {8446},
        month = sep,
          eid = {84460T},
        pages = {84460T},
          doi = {10.1117/12.926239},
archivePrefix = {arXiv},
       eprint = {1206.6885},
 primaryClass = {astro-ph.IM},
       adsurl = {https://ui.adsabs.harvard.edu/abs/2012SPIE.8446E..0TD}
}

@ARTICLE{2021A&A...649A...1G,
       author = {{Gaia Collaboration} and {Brown}, A.~G.~A. and {Vallenari}, A. and {Prusti}, T. and {de Bruijne}, J.~H.~J. and {Babusiaux}, C. and {Biermann}, M. and {Creevey}, O.~L. and {Evans}, D.~W. and {Eyer}, L. and {Hutton}, A. and {Jansen}, F. and {Jordi}, C. and {Klioner}, S.~A. and {Lammers}, U. and {Lindegren}, L. and {Luri}, X. and {Mignard}, F. and {Panem}, C. and {Pourbaix}, D. and {Randich}, S. and {Sartoretti}, P. and {Soubiran}, C. and {Walton}, N.~A. and {Arenou}, F. and {Bailer-Jones}, C.~A.~L. and {Bastian}, U. and {Cropper}, M. and {Drimmel}, R. and {Katz}, D. and {Lattanzi}, M.~G. and {van Leeuwen}, F. and {Bakker}, J. and {Cacciari}, C. and {Casta{\~n}eda}, J. and {De Angeli}, F. and {Ducourant}, C. and {Fabricius}, C. and {Fouesneau}, M. and {Fr{\'e}mat}, Y. and {Guerra}, R. and {Guerrier}, A. and {Guiraud}, J. and {Jean-Antoine Piccolo}, A. and {Masana}, E. and {Messineo}, R. and {Mowlavi}, N. and {Nicolas}, C. and {Nienartowicz}, K. and {Pailler}, F. and {Panuzzo}, P. and {Riclet}, F. and {Roux}, W. and {Seabroke}, G.~M. and {Sordo}, R. and {Tanga}, P. and {Th{\'e}venin}, F. and {Gracia-Abril}, G. and {Portell}, J. and {Teyssier}, D. and {Altmann}, M. and {Andrae}, R. and {Bellas-Velidis}, I. and {Benson}, K. and {Berthier}, J. and {Blomme}, R. and {Brugaletta}, E. and {Burgess}, P.~W. and {Busso}, G. and {Carry}, B. and {Cellino}, A. and {Cheek}, N. and {Clementini}, G. and {Damerdji}, Y. and {Davidson}, M. and {Delchambre}, L. and {Dell'Oro}, A. and {Fern{\'a}ndez-Hern{\'a}ndez}, J. and {Galluccio}, L. and {Garc{\'\i}a-Lario}, P. and {Garcia-Reinaldos}, M. and {Gonz{\'a}lez-N{\'u}{\~n}ez}, J. and {Gosset}, E. and {Haigron}, R. and {Halbwachs}, J. -L. and {Hambly}, N.~C. and {Harrison}, D.~L. and {Hatzidimitriou}, D. and {Heiter}, U. and {Hern{\'a}ndez}, J. and {Hestroffer}, D. and {Hodgkin}, S.~T. and {Holl}, B. and {Jan{\ss}en}, K. and {Jevardat de Fombelle}, G. and {Jordan}, S. and {Krone-Martins}, A. and {Lanzafame}, A.~C. and {L{\"o}ffler}, W. and {Lorca}, A. and {Manteiga}, M. and {Marchal}, O. and {Marrese}, P.~M. and {Moitinho}, A. and {Mora}, A. and {Muinonen}, K. and {Osborne}, P. and {Pancino}, E. and {Pauwels}, T. and {Petit}, J. -M. and {Recio-Blanco}, A. and {Richards}, P.~J. and {Riello}, M. and {Rimoldini}, L. and {Robin}, A.~C. and {Roegiers}, T. and {Rybizki}, J. and {Sarro}, L.~M. and {Siopis}, C. and {Smith}, M. and {Sozzetti}, A. and {Ulla}, A. and {Utrilla}, E. and {van Leeuwen}, M. and {van Reeven}, W. and {Abbas}, U. and {Abreu Aramburu}, A. and {Accart}, S. and {Aerts}, C. and {Aguado}, J.~J. and {Ajaj}, M. and {Altavilla}, G. and {{\'A}lvarez}, M.~A. and {{\'A}lvarez Cid-Fuentes}, J. and {Alves}, J. and {Anderson}, R.~I. and {Anglada Varela}, E. and {Antoja}, T. and {Audard}, M. and {Baines}, D. and {Baker}, S.~G. and {Balaguer-N{\'u}{\~n}ez}, L. and {Balbinot}, E. and {Balog}, Z. and {Barache}, C. and {Barbato}, D. and {Barros}, M. and {Barstow}, M.~A. and {Bartolom{\'e}}, S. and {Bassilana}, J. -L. and {Bauchet}, N. and {Baudesson-Stella}, A. and {Becciani}, U. and {Bellazzini}, M. and {Bernet}, M. and {Bertone}, S. and {Bianchi}, L. and {Blanco-Cuaresma}, S. and {Boch}, T. and {Bombrun}, A. and {Bossini}, D. and {Bouquillon}, S. and {Bragaglia}, A. and {Bramante}, L. and {Breedt}, E. and {Bressan}, A. and {Brouillet}, N. and {Bucciarelli}, B. and {Burlacu}, A. and {Busonero}, D. and {Butkevich}, A.~G. and {Buzzi}, R. and {Caffau}, E. and {Cancelliere}, R. and {C{\'a}novas}, H. and {Cantat-Gaudin}, T. and {Carballo}, R. and {Carlucci}, T. and {Carnerero}, M.~I. and {Carrasco}, J.~M. and {Casamiquela}, L. and {Castellani}, M. and {Castro-Ginard}, A. and {Castro Sampol}, P. and {Chaoul}, L. and {Charlot}, P. and {Chemin}, L. and {Chiavassa}, A. and {Cioni}, M. -R.~L. and {Comoretto}, G. and {Cooper}, W.~J. and {Cornez}, T. and {Cowell}, S. and {Crifo}, F. and {Crosta}, M. and {Crowley}, C. and {Dafonte}, C. and {Dapergolas}, A. and {David}, M. and {David}, P.},
        title = "{Gaia Early Data Release 3. Summary of the contents and survey properties}",
      journal = {\aap},
         year = 2021,
        month = may,
       volume = {649},
          eid = {A1},
        pages = {A1},
          doi = {10.1051/0004-6361/202039657},
archivePrefix = {arXiv},
       eprint = {2012.01533},
 primaryClass = {astro-ph.GA},
       adsurl = {https://ui.adsabs.harvard.edu/abs/2021A&A...649A...1G}
}

@INPROCEEDINGS{2005ASSL..327...41C,
       author = {{Chabrier}, Gilles},
        title = "{The Initial Mass Function: From Salpeter 1955 to 2005}",
    booktitle = {The Initial Mass Function 50 Years Later},
         year = 2005,
       editor = {{Corbelli}, E. and {Palla}, F. and {Zinnecker}, H.},
       series = {Astrophysics and Space Science Library},
       volume = {327},
        month = jan,
        pages = {41},
          doi = {10.1007/978-1-4020-3407-7_5},
archivePrefix = {arXiv},
       eprint = {astro-ph/0409465},
 primaryClass = {astro-ph},
       adsurl = {https://ui.adsabs.harvard.edu/abs/2005ASSL..327...41C}
}

@ARTICLE{2025arXiv250907060S,
       author = {{Sun}, Guochao and {Nguyen}, Tri and {Faucher-Gigu{\`e}re}, Claude-Andr{\'e} and {Lidz}, Adam and {Starkenburg}, Tjitske and {Scott}, Bryan R. and {Chang}, Tzu-Ching and {Furlanetto}, Steven R.},
        title = "{LIMFAST. IV. Learning High-Redshift Galaxy Formation from Multiline Intensity Mapping with Implicit Likelihood Inference}",
      journal = {arXiv e-prints},
         year = 2025,
        month = sep,
          eid = {arXiv:2509.07060},
        pages = {arXiv:2509.07060},
          doi = {10.48550/arXiv.2509.07060},
archivePrefix = {arXiv},
       eprint = {2509.07060},
 primaryClass = {astro-ph.GA},
       adsurl = {https://ui.adsabs.harvard.edu/abs/2025arXiv250907060S}
}

@ARTICLE{2022ApJ...933..236Z,
       author = {{Zhao}, Xiaosheng and {Mao}, Yi and {Wandelt}, Benjamin D.},
        title = "{Implicit Likelihood Inference of Reionization Parameters from the 21 cm Power Spectrum}",
      journal = {\apj},
         year = 2022,
        month = jul,
       volume = {933},
       number = {2},
          eid = {236},
        pages = {236},
          doi = {10.3847/1538-4357/ac778e},
archivePrefix = {arXiv},
       eprint = {2203.15734},
 primaryClass = {astro-ph.CO},
       adsurl = {https://ui.adsabs.harvard.edu/abs/2022ApJ...933..236Z}
}

@ARTICLE{1992EL.....19..451M,
       author = {{Marinari}, E. and {Parisi}, G.},
        title = "{Simulated tempering: a new Monte Carlo scheme}",
      journal = {EPL (Europhysics Letters)},
         year = 1992,
        month = jul,
       volume = {19},
        pages = {451},
          doi = {10.1209/0295-5075/19/6/002},
archivePrefix = {arXiv},
       eprint = {hep-lat/9205018},
 primaryClass = {hep-lat},
       adsurl = {https://ui.adsabs.harvard.edu/abs/1992EL.....19..451M}
}

@ARTICLE{2024NatAs...8.1457H,
       author = {{Hahn}, ChangHoon and {Lemos}, Pablo and {Parker}, Liam and {R{\'e}galdo-Saint Blancard}, Bruno and {Eickenberg}, Michael and {Ho}, Shirley and {Hou}, Jiamin and {Massara}, Elena and {Modi}, Chirag and {Moradinezhad Dizgah}, Azadeh and {Spergel}, David},
        title = "{Cosmological constraints from non-Gaussian and nonlinear galaxy clustering using the SIMBIG inference framework}",
      journal = {Nature Astronomy},
         year = 2024,
        month = nov,
       volume = {8},
        pages = {1457-1467},
          doi = {10.1038/s41550-024-02344-2},
       adsurl = {https://ui.adsabs.harvard.edu/abs/2024NatAs...8.1457H}
}
\bibliographystyle{JHEP}

\end{document}